\documentclass[fleqn,usenatbib,useAMS]{mnras}

\usepackage{subfiles}
\usepackage{amsmath}	
\usepackage{amssymb}	
\usepackage{multicol}        
\usepackage{bm}		
\usepackage{pdflscape}	
\usepackage{microtype}      
\usepackage{float}
\usepackage{graphicx}
\usepackage[skip=0pt]{caption}
\usepackage{natbib}
\usepackage{lipsum}
\usepackage{calc}  
\usepackage{enumitem}  
\usepackage[T1]{fontenc}
\usepackage{ae,aecompl}
\usepackage{stackengine}
\usepackage{booktabs,makecell}
\usepackage{siunitx}
\usepackage{tabularx}
\usepackage{cleveref}
\usepackage{multirow}
\usepackage{boldline}
\usepackage{array}
\usepackage{xfrac}
\usepackage{subcaption}
\usepackage{caption}
\usepackage{xcolor}

\title[Automated strong lensing]{Automated galaxy-galaxy strong lens modelling:\\no lens left behind}

\author[Etherington et al.]{\parbox{\textwidth}{Amy Etherington$^{1,2}$\thanks{Contact e-mail: \href{mailto:amy.etherington@durham.ac.uk}{amy.etherington@durham.ac.uk}}, 
James W.\ Nightingale$^{1,2}$,
Richard Massey$^{1,2}$,
XiaoYue Cao$^{3,4}$, 
Andrew Robertson$^{5}$,
Nicola C.\ Amorisco$^{1}$, 
Aristeidis Amvrosiadis$^{2}$, 
Shaun Cole$^{2}$,
Carlos S.\ Frenk$^{2}$,
Qiuhan He$^{2}$,
Ran Li$^{3,4}$, \& 
Sut-Ieng Tam$^{6}$ \\}
\\
$^{1}$Department of Physics, Centre for Extragalactic Astronomy, Durham University, South Rd, Durham, DH1 3LE \\
$^{2}$Department of Physics, Institute for Computational Cosmology, Durham University, South Road, Durham DH1 3LE, UK \\
$^{3}$National Astronomical Observatories, Chinese Academy of Sciences, 20A Datun Road, Chaoyang District, Beijing 100012, China\\
$^{4}$School of Astronomy and Space Science, University of Chinese Academy of Sciences, Beijing 100049, China\\
$^{5}$Jet Propulsion Laboratory, California Institute of Technology, 4800 Oak Grove Drive, Pasadena, CA 91109, USA\\
$^{6}$Academia Sinica Institute of Astronomy and Astrophysics (ASIAA), No.\ 1, Sec.\ 4, Roosevelt Road, Taipei 10617, Taiwan}

\date{}

\pubyear{2022}

\begin{document}
\label{firstpage}
\pagerange{\pageref{firstpage}--\pageref{lastpage}}
\maketitle

\begin{abstract}
The distribution of dark and luminous matter can be mapped around galaxies that gravitationally lens background objects into arcs or Einstein rings. New surveys will soon observe hundreds of thousands of galaxy lenses, and current, labour-intensive analysis methods will not scale up to this challenge. We develop an automatic, Bayesian method which we use to fit a sample of 59 lenses imaged by the Hubble Space Telescope. We set out to \textit{leave no lens behind} and focus on ways in which automated fits fail in a small handful of lenses, describing adjustments to the pipeline that ultimately allows us to infer accurate lens models for {\em all} 59 lenses. A high success rate is key to avoid catastrophic outliers that would bias large samples with small statistical errors. We establish the two most difficult steps to be subtracting foreground lens light and initialising a first, approximate lens model. After that, increasing model complexity is straightforward. We put forward a likelihood cap method to avoid the underestimation of errors due to pixel discretization noise inherent to pixel-based methods. With this new approach to error estimation, we find a mean $\sim1\%$ fractional uncertainty on the Einstein radius measurement which does not degrade with redshift up to at least $z=0.7$. This is in stark contrast to measurables from other techniques, like stellar dynamics, and demonstrates the power of lensing for studies of galaxy evolution. Our \texttt{PyAutoLens} software is open source, and is installed in the Science Data Centres of the ESA Euclid mission.

\end{abstract}


\section{Introduction}\label{Introduction}

Galaxy-scale strong lensing is the distortion of light rays from a background source into multiple images, by the gravitational field of a foreground galaxy along the same line of sight. From the apparent position, shape and flux of those multiple images, it is possible to infer both the intrinsic morphology of the background galaxy at magnified resolution, and the distribution of (all gravitating) mass in the foreground lens. 

In combination with kinematic measurements, lensing methods have inferred the mean total density profile of massive elliptical galaxies and how that evolves with redshift \citep{Gavazzi2007, Koopmans2009, Auger2010, Sonnenfeld2013b, Bolton2012}, and put constraints on their dark matter content, stellar mass-to-light ratio, and inner structure \citep{Sonnenfeld2012, Oldham2018, Nightingale2019, Shu2015, Shu2016c}. If the background source is variable and the mass model known, measurements of time delays between multiple images can constrain the value of the Hubble constant \citep{Suyu2017, Wong2019}. If the lens galaxy contains small substructures, they also perturb the multiple images, and provide a clean test of the nature of dark matter \citep{Vegetti2010, Li2016, Li2017, Hezaveh2016, Ritondale2019, Despali2019, Amorisco2021, He2021}.

Currently, a couple of hundred strong lensing systems have been observed, by dedicated surveys such as the Sloan Lens ACS (SLACS) \citep{Bolton2006,Auger2010}, BOSS Emission Line Lens (BELLS) \citep{Brownstein2012}, Strong Lensing Legacy (SL2S) \citep{Gavazzi2012} surveys, BELLS GALaxy-Ly$\alpha$ EmitteR sYstems (BELLS GALLERY) \citep{Shu2016a, Shu2016b}, the SLACS Survey for the Masses (S4TM) Survey \citep{Shu2017}, LEnSed laeS in the EBOSS suRvey (LESSER) \citep{Cao2021}, and the Spectroscopic Identification of Lensing Objects \citep{Talbot2018,Talbot2021}. 

During the next decade, a couple of hundred \textit{thousand} strong lenses will be discovered by wide-field surveys including Euclid, LSST, and SKA \citep{Collett2015}. Such large samples of strong lenses will contain rare `golden' systems such as double or triple source plane systems \citep{Collett2014, Collett2016a, Collett2020}, and unlock considerable scientific potential through vastly improved statistics \citep[e.g.][]{Birrer2020, Sonnenfeld2021, Sonnenfeld2021a, Cao2021, OrbanDeXivry2009}. To tackle the forthcoming thousand-fold increase in data volume, model inference must be automated, and made robust without human intervention. 

Convolutional Neural Networks (CNNs) are a fast approach that have recently been shown to be successful at lens modelling. \cite{Hezaveh2017} and \cite{Levasseur2017} modelled nine lens systems observed by the Hubble Space Telescope (HST). However, this approach requires a large, and significantly varied and unbiased training set of mock lenses to learn from. These are requirements that can be difficult to guarantee, which could be problematic for source galaxies with irregular morphologies. Using a different method, \cite{Shajib2021} used the \texttt{DOLPHIN} software to model 23 lenses from an initial sample of 50 SLACS lenses.

We use the \texttt{PyAutoLens} software (\citealt{Nightingale2015}, hereafter N15; \citealt{Nightingale2018}, hereafter N18), an open-source Bayesian forward-modelling project designed specifically with automation in mind. 
We develop an automated data analysis pipeline that models the distribution of foreground light and mass as a sum of smooth analytic functions, and the background light as either another sum of analytic functions \citep[e.g.][]{Tessore2016a}, or as a pixellated image \citep{Warren2003, Suyu2006, Dye2005, Vegetti2009, Joseph2019, Galan2021}. 
By fitting a mock sample of $\sim 500$ lenses we further show that previous versions of \texttt{PyAutoLens} (like many lens fitting algorithms) underestimated the statistical uncertainty of lens model parameters. A major component of this is a discretization effect inherent to pixel-based source reconstructions --- for which we provide a solution.

We apply our automated lens modelling pipeline to a uniform sample of 59 SLACS and BELLS GALLERY lenses that were observed with the Hubble Space Telescope. Our goal is to model every single lens and therefore \textit{leave no lens behind}: if we were analysing $\sim100,000$ lenses, even a low rate of (unflagged) failures would require unfeasible human intervention, and would bias the increasingly tight statistical precision of subsequent scientific analysis. Our first, `blind' analysis achieves a promising success rate of $85\%$. We then emphasise trying to understand \textit{why} some lenses are not well fit, and improve our pipeline until they are. This mirrors the kind of methodology that will be possible with future large samples: a fairly fixed initial framework, that is adapted after early results. In this paper we are trying to establish that first fixed framework. 

With the full sample modelled, we investigate the accuracy to which the Einstein radius is recovered. \cite{Cao2021} recently demonstrated the robustness of the measurement by comparing the Einstein radii of power-law fits to mock lenses with complex mass distributions, inferred from SDSS-MaNGA stellar dynamics data, to their true values. They showed that the Einstein radius was recovered to 0.1\% accuracy, taking into account both systematic and statistical sources of uncertainty. We examine how this compares to the statistical uncertainties we infer for the Einstein radii of the SLACS and GALLERY sample. Further, we compare to previous literature measurements \citep{Bolton2008a, Shu2016b} to verify our results and quantify how the uncertainty varies due to different methods and assumptions. Our work therefore provides an outlook on the accuracy to which we can anticipate measuring the Einstein radius in upcoming large samples of tens of thousands of lenses.

This paper is structured as follows. In Section~\ref{Theory and Models} we give a brief overview of lensing theory and provide the mass and light profile parameterisations we adopt. Section~\ref{Data} describes the sample selection and data reduction procedure for the data images of the SLACS and GALLERY samples. The method is then explained in detail in Section~\ref{Method} and applied to a sample of mock data in Section~\ref{lh_boosts} to investigate problems associated with pixelised source reconstructions. The results of applying the automated procedure to the SLACS and GALLERY samples are then presented in Section~\ref{Results}. Finally we discuss the implications for the future of automated analyses in Section ~\ref{Discussion} and summarise in Section~\ref{Summary}. Throughout this work we assume a Planck 2015 cosmological model \cite{Ade2016}. The results of every fit to the SLACS and GALLERY datasets can be found at the following link \url{https://zenodo.org/record/6104823}.

\section{Lens Modelling Theory}\label{Theory and Models}

\begin{figure*}
    \centering
    \includegraphics[width=\textwidth]{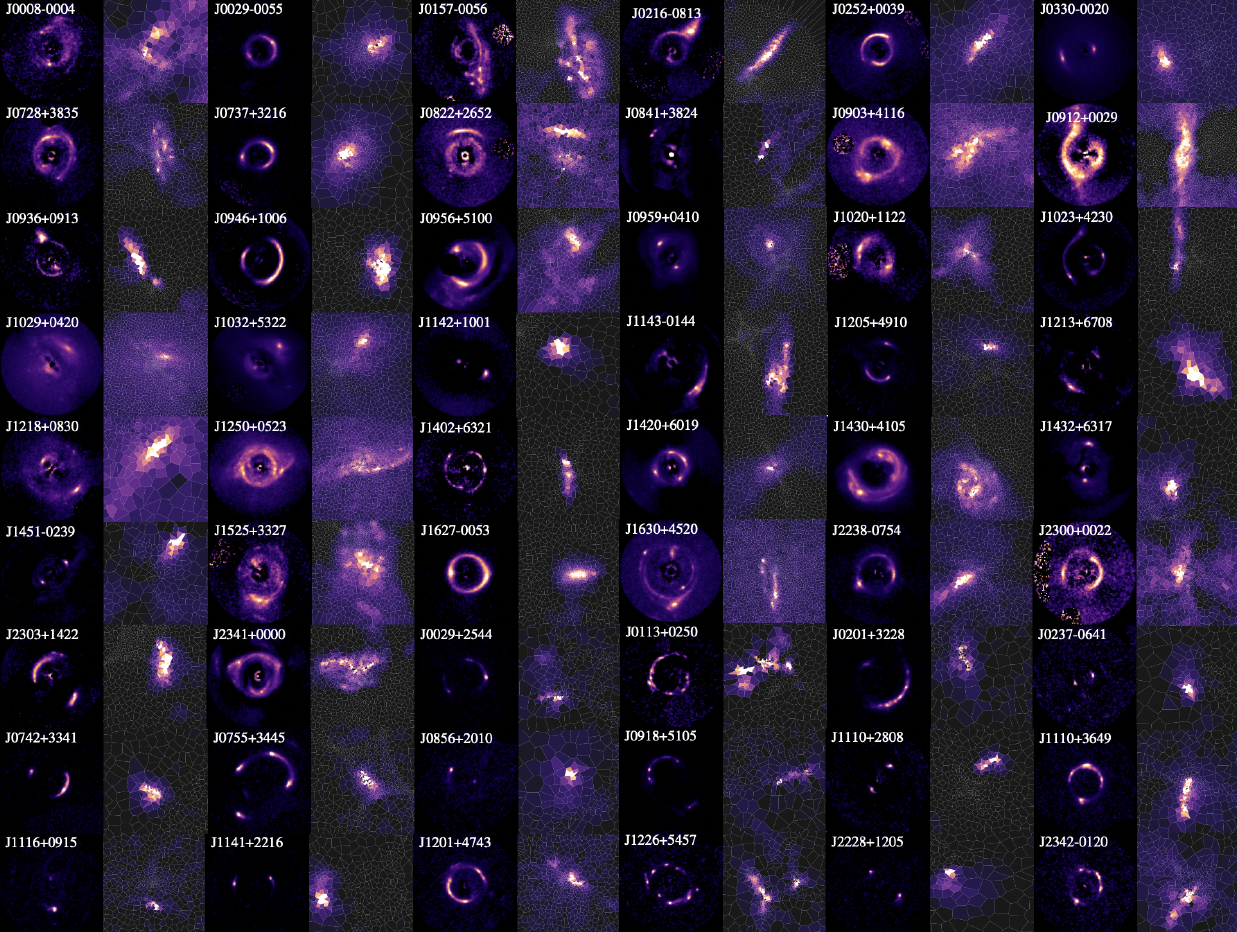}
    \caption{Lens subtracted data images (left) and their corresponding pixel-grid reconstructions (right) for the ``Gold'' sample of lens galaxies (see Section~\ref{automation} for a description of our classification process). Lenses are in order of Right Ascension, with SLACS lenses appearing first, followed by GALLERY lenses. The full model fits for these lenses, plotted with an indication of scale, are available in Appendix~\ref{model fits}.}
    \label{Figure:recons}
\end{figure*}

The aim of this study is to investigate the practicalities of automated extended source modelling to infer the mass distributions of a large sample of lenses. We give a brief overview of relevant theory for this analysis in Section~\ref{theory}, and describe our choice of mass and light profile parameterisations in Sections~\ref{Mass} and~\ref{Light}, respectively.

\subsection{Lensing theory}\label{theory}
Strong lensing occurs in and around regions where the surface mass density of the lens $\Sigma(R)$ exceeds the critical surface mass density for lensing,
\begin{equation}
    \Sigma_\mathrm{crit}=\frac{{\rm c}^2}{4{\rm \pi} {\rm G}}\frac{D_{\rm s}}{D_{\rm l} D_{\rm ls}},
    \label{eq: sigma crit}
\end{equation}
where $D_{\rm l}$, $D_{\rm s}$, and $D_{\rm ls}$ are respectively the angular diameter distances to the lens, to the source, and from the lens to the source, and ${\rm c}$ is the speed of light. Hence, assuming a cosmological model it is possible to fix the 3D geometry of the lens system using the observed redshifts of the foreground lens and background source galaxies. An extended distribution of matter can be described by its convergence, a dimensionless, 2D projected surface mass density defined as
\begin{equation}
    \kappa(x,y)=\frac{\Sigma(x,y)}{\Sigma_\mathrm{crit}}.
\end{equation}
The lensing properties of a galaxy with $\kappa(x,y)$ are characterised by the projected gravitational potential $\phi$ that satisfies the Poisson equation: $\nabla^2\phi=2\kappa$. The lens galaxy deflects light rays from the source galaxy by an amount described by the deflection angle field, $\bm{\alpha}=\nabla\phi$. The goal of lens modelling, then, is to solve the lens equation,
\begin{equation}
    \bm{\beta}=\bm{\theta} - \bm{\alpha}(\bm{\theta}), 
    \label{lens equation}
\end{equation}
which relates the observed image positions $\bm{\theta}=(\theta_1 , \theta_2)$ of deflected light rays in the image plane, from a source at position $\bm{\beta}=(\beta_1, \beta_2)$ in the source plane. Given a lensed image and (a model of) the distribution of foreground mass, one can invert Equation~\ref{lens equation} to recover the distribution of light in the source plane. In Figure~\ref{Figure:recons} the pixelised source plane reconstructions of the lenses fitted in this work are shown next to their lensed data image.

Gravitational lensing magnifies the background source, including an (infinitely thin) region of infinite magnification in the lens plane known as the tangential critical curve. Axisymmetric lenses have a circular critical curve known as the Einstein radius, $R_{\text{Ein}}$. The mean surface mass density inside $R_{\text{Ein}}$ is equal to the critical surface mass density $\Sigma_{\text{crit}}$ of the lens (Equation~\eqref{eq: sigma crit}). The Einstein radius and enclosed Einstein mass 
\begin{equation}
    M_\mathrm{Ein}=\mathrm{\pi} R_\mathrm{Ein}^2\Sigma_\mathrm{crit},
    \label{eq: einstein mass}
\end{equation}
are thus uniquely defined in the axisymmetric case, quantifying the size and efficiency of the lens. 

For asymmetric, irregular and realistic lenses, the definition of Einstein radius must be generalised. Several conventions are possible (see \citealt{Meneghetti2013} for a good overview) but we choose to use the \textit{effective} Einstein radius
\begin{equation}
    R_\mathrm{Ein,eff}=\sqrt{\frac{A}{\mathrm{\pi}}}~,
    \label{eq: einstein rad}
\end{equation}
where $A$ is the area enclosed by the tangential critical curve. This definition is self-consistent across different mass density profiles, and clearly recovers the definition of $R_{\text{Ein}}$ in the case of a circular critical curve. To calculate this in practice, we first obtain the set of points that defines the tangential critical curve contour from our lensing maps, using a marching squares algorithm, then compute the enclosed area using Green's theorem
\begin{equation}
    A = \iint dx\, dy = \oint x\, dy.
\end{equation}

\subsection{Mass profile parameterisation}\label{Mass}

We model the distribution of mass in the lens galaxy as a Power-Law Ellipsoidal Mass Distribution (PLEMD), assuming that this is able to capture the combined mass distribution of both baryonic and dark matter. The convergence is 
\begin{equation}
    \kappa(x,y) =
    \frac{\Sigma(x,y)}{\Sigma_\mathrm{crit}} =
    \frac{3-\gamma}{1+q}\bigg(\frac{b}{\sqrt{x^2+y^2/q^2}}\bigg)^{\gamma-1}
    \label{eq: kappa}
\end{equation}
\citep{Suyu2012}, where $\gamma$ is the logarithmic slope of the mass distribution in 3D, $1 \geq q > 0$ is the projected minor to major axis ratio of the elliptical isodensity contours, and $b\geq0$ is the angular scale length of the profile (referred to in some papers as the Einstein radius, but distinct from the more robust \textit{effective} Einstein radius in Equation~\ref{eq: einstein rad}). 
The profile has additional free parameters for the central coordinates $(x_{\rm c}, y_{\rm c})$ and position angle $\phi$, measured counterclockwise from the positive $x$-axis, and external shear. When varying the ellipticity, we actually sample from and adjust free parameters
\begin{align}
\varepsilon_{1} &=\frac{1-q}{1+q} \sin 2\phi~, &
\varepsilon_{2} &=\frac{1-q}{1+q} \cos 2\phi~.    
\label{eq: ellip}
\end{align}
because these are defined continuously in $-1<\varepsilon_{i}<1$, eliminating the periodic boundaries associated with angle $\phi$ and the discontinuity at $q=0$.  
We similarly parameterise the external lensing shear as components $\gamma_{1\text{ext}}$ and $\gamma_{2\text{ext}}$. The external shear magnitude $\gamma_{\text{ext}}$ and angle $\phi_{\text{ext}}$ are recovered from these parameters by
\begin{align}
    \gamma_{\text{ext}} &= \sqrt{\gamma_{1\text{ext}}^2+\gamma_{2\text{ext}}^2}~, &
    \tan{2\phi_{\text{ext}}}&=\frac{\gamma_{2\text{ext}}}{\gamma_{1\text{ext}}}~.
    \label{eq: shear}
\end{align}

The special case $\gamma=2$ recovers the Singular Isothermal Ellipse (SIE) mass distribution, in which the steady-state motions of particles have constant 1D velocity dispersion $\sigma_{\mathrm{SIE}}$, when projected along any line of sight. For this distribution of mass, the critical curve is the ellipse at $\kappa=1/2$. Our definition of effective Einstein radius (Equation~\ref{eq: einstein rad}) means that the ellipse is $R_\mathrm{Ein,eff} = qx^2 + y^2/q$, and the velocity dispersion is
\begin{equation}
    \sigma_{\mathrm{SIE}}=\mathrm{c}\sqrt{\frac{R_{\text{Ein}}D_\mathrm{s}}{4\mathrm{\pi} D_{\mathrm{l}}D_{\mathrm{ls}}}} ~.
\end{equation}

\subsection{Light profile parameterisation}\label{Light}

We model the foreground galaxy's light distribution as the sum of two S\'ersic  profiles with different ellipticities but a common centre. This replicates the bulge and disc components that constitute an Early-type Galaxy (ETG) \citep{Oh2017, Vika2014}, and significantly increased the Bayesian evidence compared to a single S\'ersic model, in a precursor study of three SLACS galaxies \citep{Nightingale2019}. The S\'ersic profile is 
\begin{equation}
    I(x, y) = I_\mathrm{{eff}}~ \mathrm{exp}\left\{-k_{\mathrm{eff}}\left[\left(\frac{\sqrt{qx^2+y^2/q}}{R_{\mathrm{eff}}}\right)^\frac{1}{n}-1\right]\right\},
    \label{eq: sersic}
\end{equation}
where $I_{\mathrm{eff}}$ is the surface brightness at the effective radius $R_{\mathrm{eff}}$, defined here in the intermediate axis normalisation\footnote{This definition keeps the area enclosed within a given isophote constant as $q$ is varied, and is distinct from `major axis normalisation' where the term $(qx^2 + y^2/q)$ would instead be $(x^2 + y^2 / q^2)$.}\!, $n$ is the S\'ersic index, and $k_{\mathrm{eff}}$ is a normalisation constant related to $n$ such that $R_{\mathrm{eff}}$ encloses half of the total light from the model \citep{Graham2005}. The axis ratio and position angle of each component are parameterized during the fitting process, using elliptical components as in Equation~\eqref{eq: ellip}.
Aside from the two components' common centre, all free parameters are fitted independently of each other to allow for more complex light distributions. For example, the flux ratio of the two S\'ersics is unconstrained, and the profiles may be elongated by different amounts and rotationally offset from one another.

We model the distribution of light in the source galaxy as either a single Se\'rsic profile or using a pixelated source reconstruction depending on the phase of the automated procedure, described in Section~\ref{sec: automated procedure}. The source galaxy is ultimately reconstructed on an adaptive Voronoi mesh, for which the procedure is described in detail in Section~\ref{sec: source recon}.

\section{Data}\label{Data} 

\subsection{Lens sample selection}\label{samples}

We analyse strong gravitational lenses around massive elliptical galaxies drawn from the SLACS \citep{Bolton2008} and BELLS GALLERY samples \citep{Shu2016b}. The SLACS sample were identified as lenses using SDSS spectroscopy to find higher redshift emission lines after subtracting a principle component model of the foreground galaxy spectrum \citep{Bolton2006}. This technique was modified for the GALLERY survey, to specifically select even higher redshift Ly$\alpha$-emitting (LAE) source galaxies \citep{Shu2016b}. Spectroscopic redshifts of the lens and source are known, and follow-up high resolution imaging has been carried out for all systems.

To keep the data quality reasonably uniform (as it would be for a large future survey), we restrict the SLACS sample to the 43 lenses imaged to at least 1-orbit depth in the HST Advanced Camera for Surveys (ACS) F814W band. We add the 17 grade-A confirmed LAE lenses from GALLERY, all of which have been observed to 1-orbit depth in the HST Wide Field Camera 3 (WFC3) F606W band. Several systems have second or third foreground lenses of low mass. However, for this first attempt at automation in which we shall try to fit only a single main lens, we have not considered GALLERY lens J0918+4518, which has two equally bright lens galaxies. We end up with a set of 59 lenses. 

\subsection{Data reduction}\label{reduction}

HST imaging of both the SLACS and GALLERY samples was reduced using custom pipelines. The procedure for the SLACS sample is described in \citet{Bolton2008a} and produces images with 0.05\arcsec pixels; the procedure for GALLERY is described in \citet{Brownstein2012} and \citet{Shu2016a}, and produces images with 0.04\arcsec pixels. The point spread function (PSF) was determined for both samples using the \texttt{Tiny Tim} software \cite{Krist1993}. The  aforementioned papers also describe an optional method to subtract the lens galaxy's light by fitting it with a b-spline. Our pipeline benefits from fitting the lens light simultaneously with its mass, so we shall generally not use the b-spline data. However, our pipeline struggles to automatically deblend the lens and source light of three systems, so we shall try the b-spline data there.

\section{Method}\label{Method}
{\renewcommand{\arraystretch}{2.0}
\begin{table*}
\centering
    \begin{tabularx}{\textwidth}{@{}*{6}{c}{X}}
        \hlineB{1}
        \thead{\textbf{Pipeline}} & \thead{\textbf{Phase}} & \thead{\textbf{Galaxy Component}} & \thead{\textbf{Model}} & \thead{\textbf{Varied}} & \thead{\textbf{Prior info}} & \textbf{Phase Description}\\
        \hlineB{1}
        \multirow{6}{*}{\shortstack[l]{Source \\ Parametric}} & \textbf{SP}$^{1}$ & Lens light & S\'ersic + Exp & \checkmark & - & Fit only the lens light model and subtract it from the data image. \\
        \cline{2-7}
        &\multirow{2}{*}{\textbf{SP}$^{2}$} & Lens mass & SIE + shear & \checkmark  & - &\multirow{2}{\hsize}{Fit the lens mass model and source light profile, comparing the lensed source model to the lens light subtracted image from \textbf{SP}$^{1}$.}  \\
        & & Source light & S\'ersic & \checkmark & - \\
        \cline{2-7}
        &\multirow{3}{*}{\textbf{SP}$^{3}$} & Lens light & S\'ersic + Exp &  & -  & \multirow{3}{\hsize}{Refit the lens light model with default priors and fit the mass and source models with priors informed from \textbf{SP}$^{2}$.}\\
        & & Lens mass & SIE + shear & \checkmark  & \textbf{SP}$^{2}$\\
        & & Source light & S\'ersic & \checkmark & \textbf{SP}$^{2}$ \\
        \cline{1-7}
        \multirow{12}{*}{\shortstack[l]{Source \\ Inversion}}& \multirow{3}{*}{\textbf{SI$^1$}} & Lens light & S\'ersic + Exp &  & \textbf{SP}$^3$ & \multirow{3}{\hsize}{Fix lens light and mass parameters to those from the source parametric pipeline and fit pixelization and regularisation parameters on magnification adaptive pixel-grid.}\\
        & & Lens mass & SIE + shear &  & \textbf{SP}$^3$ \\
        & & Source light & MPR & \checkmark & - \\
        \cline{2-7}
        & \multirow{3}{*}{\textbf{SI}$^2$} & Lens light & S\'ersic + Exp &  & \textbf{SP}$^3$  & \multirow{3}{\hsize}{Refine the lens mass model parameters, keeping lens light and source-grid parameters fixed to those from previous phases.} \\
        & & Lens mass & SIE + shear & \checkmark  & \textbf{SP}$^3$  \\
         & & Source light & MPR &  & \textbf{SI}$^1$   \\
        \cline{2-7}
        & \multirow{3}{*}{\textbf{SI}$^3$} & Lens light & S\'ersic + Exp &  & \textbf{SP}$^3$  & \multirow{3}{\hsize}{Fit BPR pixelization and regularisation parameters, using the lensed source image from \textbf{SI}$^2$ to determine the source galaxy pixel centres. Lens light and mass parameters are fixed to those from previous phases.}\\
        & & Lens mass & SIE + shear &  & \textbf{SP}$^3$ \\
        & & Source light &BPR& \checkmark  & -  \\
        \cline{2-7}
        & \multirow{3}{*}{\textbf{SI}$^4$} & Lens light & S\'ersic + Exp &  & \textbf{SP}$^3$ & \multirow{3}{\hsize}{Refine lens mass model parameters on the BPR grid, keeping lens light and source-grid parameters fixed to those from previous phases.} \\
        & & Lens mass & SIE + shear & \checkmark & \textbf{SI}$^2$   \\
        & & Source light & BPR &  & \textbf{SI}$^3$   \\
        \hlineB{1}
        \multirow{3}{*}{\shortstack[l]{Light \\ Parametric}} & \multirow{3}{*}{\textbf{LP}$^1$} & Lens light & S\'ersic + S\'ersic & \checkmark  & - & \multirow{3}{\hsize}{Fit lens light parameters, with lens mass and source parameters fixed to the result of the source inversion pipeline.}\\
        & & Lens mass & SIE + shear &  & \textbf{SI}$^4$  \\
         & & Source light & BPR & & \textbf{SI}$^3$  \\
        \hlineB{1}
        \multirow{6}{*}{\shortstack[l]{Mass \\ Total}} & \multirow{3}{*}{\textbf{MT}$^1$} & Lens light & S\'ersic + S\'ersic &  & \textbf{LP}$^1$ & \multirow{3}{\hsize}{Fit the lens mass parameters, now with the slope of the density profile free to vary within the uniform prior [1.5-3.0], all other mass priors are informed from \textbf{SI}$^4$. The lens and source light are fixed to those from the \textbf{LP}$^1$ pipeline.}\\
        & & Lens mass & PLEMD + shear & \checkmark & \textbf{SI}$^{4}$  \\
        & & Source light & BPR &  & \textbf{SI}$^3$  \\
        \cline{2-7}
        &\multirow{3}{*}{\textbf{MT}$^1_{\rm ext}$} & Lens light & S\'ersic + S\'ersic &  & \textbf{LP}$^1$ & \multirow{3}{\hsize}{An extension of the \textbf{MT}$^1$ phase to ensure robust error inference on parameters. The lens mass parameters are re-fitted, capping likelihood evaluations to a fixed value (See Section~\ref{lh_boosts}\ for details.)}\\
        & & Lens mass & PLEMD + shear & \checkmark & \textbf{MT}$^{1}$  \\
         & & Source light & BPR & \checkmark & \textbf{MT}$^1$  \\
         \hlineB{1}
    \end{tabularx}
    \caption{Composition of the pipelines that make up our uniform analysis. Where prior info is not passed from previous pipelines see Table~\ref{table:Priors} for the specific priors used on each model parameter.} 
    \label{table: pipelines}
\end{table*}}

\subsection{Overview}

Our strong lens analysis is carried out using the software \texttt{PyAutoLens}\footnote[2]{The \texttt{PyAutoLens} software is open source and available from \url{https://github.com/Jammy2211/PyAutoLens}}, which is described in N18, building on the works of \citet[][hereafter WD03]{Warren2003}, \citet[][hereafter S06]{Suyu2006} and N15. 

To fit a lens model to an image, {\tt PyAutoLens} first assumes a parameterisation for the distribution of light and mass in the lens, and the distribution of light in the source, using the parametric profiles described in Sections~\ref{Mass} and~\ref{Light}. The parameterised intensity $I$ of the lens light is evaluated at the centre of every image pixel, convolved with the instrumental PSF, and subtracted from the observed image. The mass model is then used to ray-trace image-pixels from their image-plane positions $\theta$ to source-plane positions $\beta$ (via the lens Equation~\ref{lens equation}). The source analysis finally follows, which {\tt PyAutoLens} performs using one of two approaches: (i) parametric profiles in the source-plane (e.g. the S\'ersic profile) are used to simply evaluate $I$ at every value of $\beta$; (ii) a pixelized source reconstruction is performed on an adaptive Voronoi mesh, where the values of $\beta$ are used to pair image-pixels to the Voronoi source pixels which reconstruct the source (see WD03, S06, N15 and N18 for a full description of lensing analyses with pixelized source reconstructions).

The following link (\url{https://github.com/Jammy2211/autolens_likelihood_function}) contains Jupyter notebooks that provide a visual step-by-step guide of the {\tt PyAutoLens} likelihood function used in this work. We have received feedback from readers of other papers using {\tt PyAutoLens} (who are less familiar with strong lens modelling) that they were unclear on the exact procedure that translates a strong lens model to a likelihood value. The notebooks aims to clarify this and provides links to all previous literature describing the {\tt PyAutoLens} likelihood function, alongside an explanation of the technical aspects of the linear algebra and Bayesian inference. We provide a brief description of the {\tt PyAutoLens} likelihood function below, but we recommend these notebooks to the interested reader if anything is unclear.

\subsection{Source Reconstruction}\label{sec: source recon}

After subtracting the foreground lens emission and ray-tracing coordinates to the source-plane via the mass model, the source is reconstructed in the source-plane using an adaptive Voronoi mesh which accounts for irregular or asymmetric source morphologies (see Figure~\ref{Figure:recons}). Our results use the \texttt{PyAutoLens} pixelisation \texttt{VoronoiBrightnessImage}, which adapts the centres of the Voronoi pixels to the reconstructed source morphology, such that more resolution is dedicated to its brighter central regions \citep{Nightingale2018}.

The reconstruction computes the linear superposition of PSF-smeared source pixel images which best fits the observed image. This uses the matrix $f_{\rm  ij}$, which maps the $j$th pixel of each lensed image to each source pixel $i$. Following the formalism of \citep[][WD03 hereafter]{Warren2003}, we define the vector $\vec{D}_{i} = \sum_{\rm  j=1}^{J}f_{ij}(d_{j} - b_{j})/\sigma_{j}^2$ and curvature matrix ${F}_{ik} = \sum_{\rm  j=1}^{J}f_{ij}f_{kj}/\sigma_{j}^2$, where $d_{j}$ are the observed image flux values with statistical uncertainties $\sigma{\rm _j}$ and $ b_{j}$ are the model lens light values. The source pixel surface brightnesses values are given by $s = F^{-1} D$ which are solved via a linear inversion that minimizes
\begin{equation}
\label{eqn:ChiSquared}
\chi^2 = \sum_{\rm  j=1}^{J} \bigg[ \frac{(\sum_{\rm  i=1}^{I} s_{i} f_{ij}) + b_{j} - d_{j}}{\sigma_{j}} \bigg] \, .
\end{equation}
The term $\sum_{\rm  i=1}^{I} s_{i} f_{ij}$ maps the reconstructed source back to the image-plane for comparison with the observed data.

This matrix inversion is ill-posed, therefore to avoid over-fitting noise the solution is regularized using a linear regularization matrix $H$ (see WD03). Regularization acts as a prior on the source reconstruction, penalizing solutions where the difference in reconstructed flux of these two neighboring Voronoi source pixels is large. Our results uses the  \texttt{PyAutoLens} regularization scheme \texttt{AdaptiveBrightness}, which adapts the degree of smoothing to the reconstructed source's luminous emission \citep{Nightingale2018}. This has three hyper parameters, the inner regularization coefficient, outer regularization coefficient and a parameter which controls how the outer and inner regions of the source plane are defined for regularization. The degree of smoothing is chosen objectively using the Bayesian formalism introduced by \citet{Suyu2006}. The likelihood function used in this work is taken from \citep{Dye2008} and is given by
\begin{eqnarray}
\label{eqn:evidence2}
-2 \,{  \mathrm{ln}} \, \epsilon &=& \chi^2 + s^{T}\textbf{H}s
+{ \mathrm{ln}} \, \left[ { \mathrm{det}} (\textbf{F}+\textbf{H})\right]
-{ \mathrm{ln}} \, \left[ { \mathrm{det}} (\textbf{H})\right]
\nonumber \\
& &
+ \sum_{\rm  j=1}^{J}
{ \mathrm{ln}} \left[2\pi (\sigma{_j})^2 \right]  \, .
\end{eqnarray}

\subsection{Automated Procedure}\label{sec: automated procedure}
\subsubsection{\texttt{PyAutoLens}}\label{AutoLens}

\texttt{PyAutoLens} is designed to approach lens modelling in a fully automated way \citep[N18, ][]{Nightingale2021}. This uses a technique we term `non-linear search chaining', which sequentially fits lens models of gradually increasing complexity. By initially fitting simpler lens models one can ensure that their corresponding non-linear parameter spaces are sampled in an efficient and robust manner that prevents local maxima being inferred. The resulting lens models then act as initialisation in subsequent model-fits which add more complexity to the lens model, guiding the non-linear search on where to look in parameter space for the highest likelihood lens models, ensuring the global maximum model has the highest chance of being inferred. Non-linear search chaining is performed using the probabilistic programming language \texttt{PyAutoFit} (\url{https://github.com/rhayes777/PyAutoFit}), a spin off project of \texttt{PyAutoLens} which generalises the statistical methods used to model strong lenses into a general purpose statistics library. 

To perform model-fitting \texttt{PyAutoLens} uses the nested sampling algorithm \texttt{dynesty} \citep{Speagle2020}, which obtains the posterior probability distributions for a given lens model's parameters. Nested sampling's ability to robustly sample from low dimensional (e.g.\ fewer than $\sim 30$ parameters), complex parameter space distributions makes it well suited to lens modelling. We use \texttt{dynesty}'s random walk sampling for every model-fit performed in this work, which we found gave a significant improvement over other sampling techniques owing to its better accounting of the covariance between lens model parameters. Since nested sampling starts by randomly sampling from the prior, the size and choice of priors directly impact the expected number of nested sampling iterations alongside how likely it is that a local maximum is incorrectly inferred. As such, using more informative priors will reduce the amount of time needed to integrate over the posterior and guide towards sampling the highest likelihood global maxima solutions. 

Non-linear search chaining allows us to construct informative priors from the results of one \texttt{dynesty} search and pass them to subsequent model-fits, thereby guiding them on where to sample parameter space. This uses a technique called prior passing (see N18), which sets the prior of each parameter as a Gaussian whose mean is that parameter's previously inferred median PDF (probability density function) value and its width is a customisable value specific to every lens model and parameter. The specific order of prior passing used in this study is given in Table~\ref{table: pipelines}. The prior widths have been carefully chosen to ensure they are broad enough not to omit valid lens model solutions, but sufficiently narrow to ensure the lens model does not inadvertently infer local maxima. More quantitatively, the prior widths are typically greater than $\sim 10$ times the errors we ultimately infer on each parameter, meaning it has negligible impact on the posterior (see Section~\ref{lh_boosts}). 

\subsubsection{User Setup}\label{Setup}
In this work, we use the standardised Source Light and Mass (SLaM) pipelines that are available, and fully customisable, in \texttt{PyAutoLens}. From these, we construct a pipeline that chains together a total of $11$ \texttt{dynesty} searches which we apply to every lens in our sample, which we describe in detail in Section~\ref{Pipeline}. Before we run the SLaM pipelines a few brief pre-processing steps must be carried out; we describe those here, as well as our chosen pipeline settings.

We define a circular mask centred on the lens galaxy that defines the image pixels we fit to. For the SLACS and GALLERY lenses we use a standard size of $3.5\arcsec$ and $3.0\arcsec$ radius, respectively. This is increased to $4.0\arcsec$ for the SLACS lenses J0912+0029 and J0216-0813, and for the GALLERY lens J0755+3445. All image pixels outside this mask are completely omitted from the analysis, meaning they are not traced to the source plane and included in the source reconstruction procedure.

We create scalable noise maps, unique to each lens, that define any regions inside the mask that we do not wish to fit (e.g.\ unrelated astronomical sources projected by chance along adjacent lines of sight). In these regions the image values are scaled to near zero and the noise-map values to large values such that the likelihood calculation effectively ignores them. Such areas of high flux would otherwise be indistinguishable from the source flux to the fitting procedure. We adopt this noise map approach, over the complete removal of such regions, since image-pixels are still traced to the source-plane and included in the source reconstruction procedure. This avoids creating discontinuities or `holes' in the source pixelisation which can degrade the quality of the overall reconstruction. The maps are produced in a graphical user interface (GUI) available in \texttt{PyAutoLens}, designed to reduce the human time necessary for creating each unique map ($\sim$1 minute per lens). We acknowledge this task is overly time-intensive when considering the incoming samples of tens of thousands of lenses and provide a discussion of possible routes to automation of this pre-processing step in Section~\ref{automation prospects}.

Finally, we store an array containing the coordinates of the pixels containing the peak surface brightness of each multiple image of the source galaxy, again selected by the user via a GUI. These coordinates are used to remove local maxima from the parameter spaces explored throughout the pipeline. In practise, this is done by discarding any models where the ray-traced points in the source plane are not within a positions threshold value of each other. This value is initially set to $0.7\arcsec$\footnote{This choice of arcsecond value reflects a low threshold for what we consider a plausible lens model, removing only extremely unphysical mass models. For example, without it the mass model could choose to be close to zero by fitting a source to only one multiple image with its centre aligned directly behind that image. We note this means we do not require the locations of the multiple images to be extremely accurate.}. Both the threshold and the positions themselves are then iteratively updated throughout the SLaM pipeline, by solving the lens equation using the maximum likelihood mass model estimated in a previous fit. For each iteration, the value is set to three times the separation of the positions found after solving the lens equation or a value of $0.2\arcsec$, whichever is largest. This ensures that, as we progress from parametric to pixelised source reconstructions, we avoid the under and over-magnified solutions that can be problematic for these methods \cite{Maresca2020}.

\subsubsection{Uniform Analysis}\label{Pipeline}
The uniform analysis ultimately aims to constrain the parameters describing the mass and light distributions. The lens galaxy's mass is parameterized as a PLEMD (Equation~\ref{eq: kappa}), while the lens light is modelled as a double Se\'rsic profile, which is a sum of two S\'ersic profiles (Equation~\ref{eq: sersic}) with a common centre. This is achieved by reconstructing the source galaxy's light distribution on an adaptive Brightness-based Pixelisation and Regularisation (BPR) grid. The uniform analysis is constructed from multiple pipelines that each focus on fitting a specific aspect of the lens model which we describe below. For an overview of the composition of the overall method see Table~\ref{table: pipelines}. A scaled down version of this pipeline was used by \citet{Cao2021} to model fifty simulated strong lenses.

We begin with the Source Parametric (SP) pipeline that fits the foreground lens galaxy's light profile, alongside a robust initialisation of less complex models for the mass distribution of the lens and light distribution of the source galaxy. The lens mass is modelled as an SIE (Equation~\ref{eq: kappa} with $\gamma=2$) plus external shear. The lens light is modelled by the sum of a S\'ersic and Exponential (Equation~\ref{eq: sersic} with n=1) profile. The source galaxy's light is described by a single S\'ersic profile; this is key to the initialisation of the model using the SP pipeline, as it allows us to compute an initial estimate of the mass profile without \texttt{dynesty} getting stuck in a local maximum (as methods with a pixelised source frequently do; N18, \citealt{Maresca2020}).

The Source Inversion (SI) pipeline then refines the lens galaxy's mass distribution by modelling the source galaxy using an adaptive pixelisation. This allows more realistic morphologies of the source galaxy to be recognised, which in turn improves the model for the lens galaxy's mass. 
The pixelisation and it's pixel-to-pixel regularisation are described by a set of hyper-parameters (see Section~\ref{sec: source recon} for more details), that are fitted for as free parameters in the fit, these are first initialised using a Magnification based Pixelisation and Regularisation (MPR) grid. The source model from this fit is then used to inform the the BPR pixelisation that adapts to the surface brightness of the source galaxy, thereby reconstructing areas of high flux with higher resolution and lower regularisation relative to areas of low flux. 

The Light Parametric (LP) pipeline re-fits the lens galaxy's light profile. This produces a more accurate estimate of the lens galaxy's light than previously, because the lensed source galaxy's light is now reconstructed using the Voronoi pixelisation, thereby reducing residuals which otherwise impact the lens light model fit. The lens light model is now composed of two S\'ersic profiles (the second component now has a free S\'ersic index). This fit is performed using broad uninformative priors and thus does not use any information about the lens galaxy's light profile estimated by the previous pipelines. 

Finally, the Mass Total (MT) pipeline extends the complexity of the model of the lens galaxy's mass to that of the PLEMD (Equation~\ref{eq: kappa}), whereby the slope of the density profile ($\gamma$) is now a free parameter in the model. A uniform prior between 1.5 and 3 is assumed on the slope. To ensure robust error inference on the parameters of our final model, the MT phase is extended by re-running the same model with a `likelihood cap' applied (see Section~\ref{lh_boosts} for details). The term `Mass Total' is used to distinguish this pipeline from the `Mass Light Dark' SLaM pipeline which is not used in this work. Instead of fitting a mass model that represents the total mass distribution this pipeline fits one that separately models the light and dark mater \citep{Nightingale2019}.

\subsubsection{Results Database}

Upon completion of a uniform pipeline there are \texttt{dynesty} samples of $11$ different model-fits, alongside additional metadata describing quantities such as each parameter's estimate, their errors and the \texttt{PyAutoLens} settings. Across our sample of $59$ strong lenses this creates over $500$ lens modelling results, necessitating tools to automate their processing and inspection. \texttt{PyAutoFit} outputs all modelling results to a queryable SQLite database \citep{sqlite} such that they can be easily loaded into a Jupyter notebook or Python script post-analysis. By adopting \texttt{PyAutoFit}, all \texttt{PyAutoLens} results support this SQLite database which is the primary tool we use for analysing lens modelling results.

\section{Dealing with noise in likelihood evaluations}\label{lh_boosts}

N15 demonstrated that pixelised source reconstructions can be subject to a discretization bias that ultimately leads to the underestimation of errors calculated by a typical non-linear search (N15). This is a result of discrete jumps in the likelihood as the lens model parameters are smoothly varied, which hinders convergence and parameter marginalisation. N15 suggests this may be a common problem for methods that employ pixelised sources. Here, we investigate the effects of the bias further using a large sample of mock observations.

\subsection{Mock data sample}\label{Mock Data} 

\begin{figure*}
    \centering
    \includegraphics{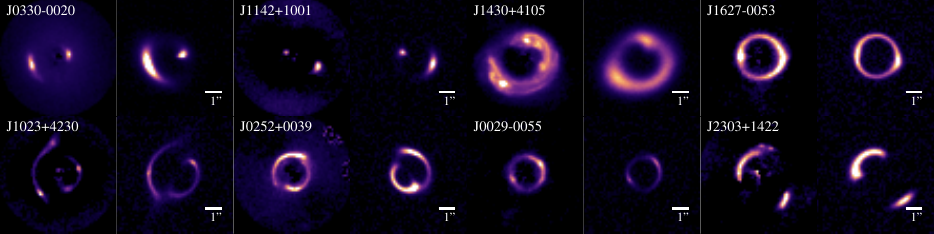}
    \caption{We create a sample of mock lenses that closely resemble each of the 59 SLACS and GALLERY lenses in our observed data sample, which we use for testing for data discretization bias. We show eight of these mock images (right panel) alongside the real data image they were simulated to represent (left panel with lens name). The mock images are simulated without light from the lens galaxy, as such we compare to the data images where the lens galaxy's (double Se\'rsic) light profile has been subtracted.}
    \label{Figure: mock data comp}
\end{figure*}

We create 59 synthetic lenses similar to our SLACS and GALLERY lenses, to approximately resemble the real data but with known truths. The mass distribution of each synthetic lens is a PLEMD; we set the radius $b$ and ellipticity parameters $\varepsilon_{1}$ and $\varepsilon_{2}$ to those of the SIE lens model measured in previous lensing analyses (see Table~5 of \cite{Bolton2008} and Table~2 of \cite{Shu2016a} for SLACS and GALLERY parameters, respectively). We set the slope $\gamma$ of the density profiles to the lensing and dynamics measurements of \cite{Auger2010} (SLACS) and \cite{Shu2016a} (GALLERY). Where the slope of the density profile is not available, we instead use the values inferred by preliminary runs of our own strong lensing-only analysis. The surface brightness of each source galaxy is simulated as an elliptical S\'ersic, the parameters of which are set to those inferred during preliminary runs of our Source Parametric Pipeline (see Section~\ref{Method} for more detail)\footnote[1]{The S\'ersic source parameters were optimised for an SIE mass profile but simulated with a PLEMD, leading to a difference in magnification of the source galaxy in the mock data. As a result, some lensed sources were simulated with lower signal-to-noise ratio (S/N) values than observed. In these cases we manually adjust their intensity value to give a peak S/N$\gtrsim$3.}. The redshifts of the lens and source are set to those known for the corresponding real strong lens.  

For every synthetic lens configuration, we create six mock observations, each with different realisations of observational noise. To mimic the HST observations the lensed image of the source is generated with a pixel scale of $0.05\arcsec$ (SLACS) and $0.04\arcsec$ (GALLERY) and convolved with the instrumental point spread function (PSF) modelled from the actual image of the strong lens we are simulating. The synthetic images include a flat sky background of 37.5 (SLACS) and 31.5 electrons per second (GALLERY) and six different realisations of Poisson noise. We choose not to simulate light from the lens galaxy since this has the potential to introduce systematic effects that we are not interested in investigating with this sample (see Section~\ref{lh_boosts}). Across the resulting suite of 354 synthetic observations, the S/N of the brightest pixel in each image ranges from 4 to 68. Figure~\ref{Figure: mock data comp} compares a subset of simulated mock lenses with their real data counterparts.

\subsection{The origin of discretization bias and error underestimation}

\begin{figure}
    \centering
    \includegraphics{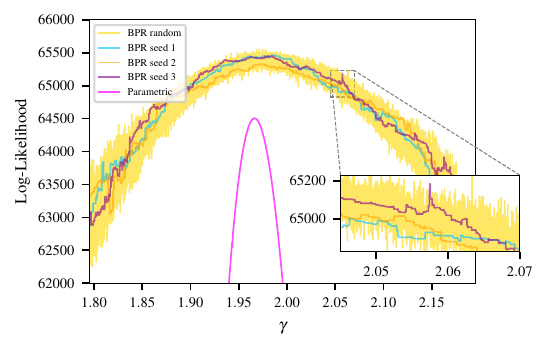}
    \caption{
    Comparison of the log likelihood as a function of density profile slope when using a parametric source (pink curve) or Brightness-based pixelisation and Regularisation (BPR) pixelisations to fit to mock data. All model parameters other than the slope are fixed to their true values. The yellow line reveals the full level of noise in the likelihood due to the particulars of the source plane pixelisation, by using a new random seed for the $k$-means algorithm that pixelates the source plane for every likelihood evaluation. The other three colours use fixed $k$-means seeds, as is done throughout the rest of this paper.}
    \label{Figure: lh surface grid comp}
\end{figure}

First, we investigate how discretization bias manifests in \texttt{PyAutoLens}, whose source pixelisation differs in its implementation from N15 and N18. This is illustrated in Figure~\ref{Figure: lh surface grid comp}, which plots the variation of the log likelihood of a lens model when changing only the slope parameter $\gamma$ of the mass distribution (fixing all other parameters to their true values). The parametric source model produces a smooth likelihood curve. The BPR pixelisation methods produce a higher likelihood, but one that is subject to seemingly random noise. These `spikes' in log likelihood occur over small ranges in the slope parameter; at least an order of magnitude smaller than the errors one infers for $\gamma$ when fitting this lens with a parametric source. This confuses the nested sampler, which converges to positive spikes in likelihood that are tiny volumes of the multi-dimensional parameter space, and thus significantly underestimate the total statistical uncertainty.

To perform a source reconstruction using a pixelised source, one must first define a procedure that determines the shape and locations of the source-plane pixels, its discretization. For example, in the case of \texttt{PyAutoLens}, one can alter the random seed that determines the centres of the Voronoi source pixels. This element of choice makes the likelihood ill-determined, as is demonstrated in Figure~\ref{Figure: lh surface grid comp} by the three different realisations of noise that are uncovered for the differently seeded grids (the only difference between the fits that produces the blue, orange, and purple likelihood surface is the choice of k-means seed that determines the source-pixel centres). If we choose to pass a random k-means seed to each individual fit (the yellow curve in Figure~\ref{Figure: lh surface grid comp}) the full scale of the noise due to different source discretizations is revealed, likelihood evaluations of almost identical lens models can yield very different likelihood values when the source pixelisation changes. Sampling the parameter space when using a random k-means seed is therefore prohibitively slow, ultimately leading to the non-linear search becoming stuck and being unable to converge.

\begin{figure}
    \centering
    \includegraphics{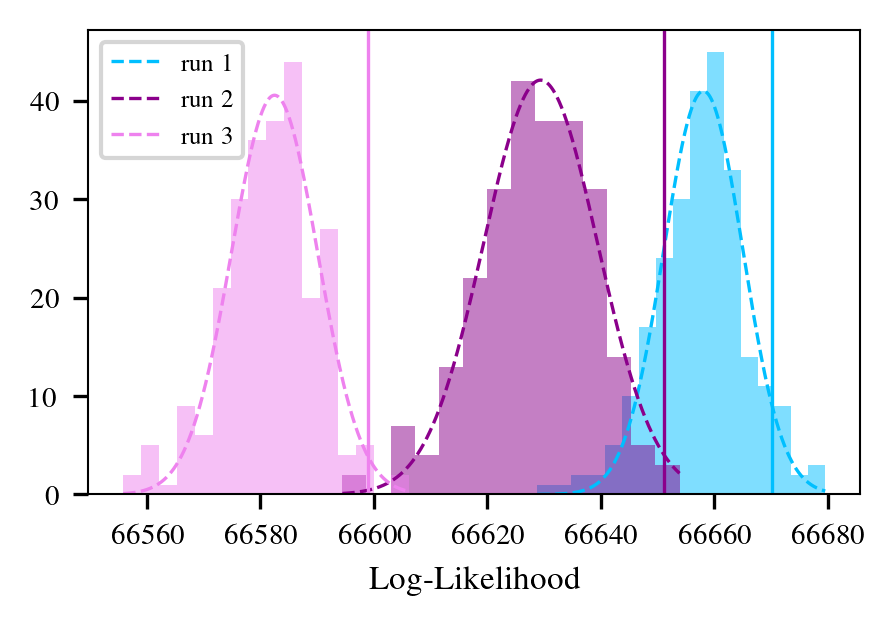}
    \caption{Histogram of log-likelihood values from re-fitting the best-fit model with a new $k$-means seed 500 times, while keeping the model parameters fixed. The dashed line is the fitted Gaussian curve to these values. The vertical line shows the maximum likelihood value of the best-fit parameters found without a likelihood cap, which is always boosted by noise to extremely high likelihood. For clarity we show three of the six distributions from different noise realisation images of the mock lens, the same behaviour is evident in the three distributions not shown here.}
    \label{Figure: lh histogram}
\end{figure}

In fact, repeat likelihood evaluations of an identical lens model also yield different likelihood values if the source pixelisation's discretization changes. Figure~\ref{Figure: lh histogram} shows the result of doing exactly this, where log likelihood values are computed using an identical lens model 500 times (we use the best fit lens model parameters from our fitting procedure to do this), with each computation using only a different Voronoi mesh to reconstruct the source. The three different coloured histograms show the results of this procedure for three of the six noise realisation images of a lens, that arrive at three different best fit lens models. In all cases, the histograms of log likelihood values show that changes in log likelihood of order $\sim 50$ are possible by just changing the source pixelisation. To perform parameter estimation, changes in log likelihood of order $\sim 10$ define how precisely a parameter is estimated at $\sim 3\sigma$ confidence. Thus, if our log likelihoods can fluctuate by of order $\sim 50$ in a seemingly arbitrary way, this will be detrimental to parameter and error estimation.

Why does the log likelihood vary when we change the source pixelisation? For a given lens model, there are certain source pixelisations where the centres of their Voronoi source pixels line up with the locations of the traced image-pixels mapped from the image data in a `preferable' way. Their alignment allows the source pixels to reconstruct the image data more accurately, in a way that is penalised less by regularisation (see S06). This produces what we consider an artificial `boost' in likelihood. Conversely, other pixelisations have a less fortuitous alignment, such that their reconstruction of the image data is worse and they are more heavily penalised by regularisation, producing an artificial `drop' in log likelihood. Figure \ref{Figure: lh histogram} shows that the distribution of log-likelihoods appears to be Gaussian, a property we will use when we put forward a solution to this problem.

We are now in a position to explain the spiky likelihood surface shown for the fixed seed BPR pixelisations in Figure~\ref{Figure: lh surface grid comp}. Let us first consider in more detail the BPR pixelisation implemented in \texttt{PyAutoLens}. To construct the source-pixel centres, the BPR pixelisation applies a weighted k-means algorithm in the image plane to determine a set of coordinates that are adapted to the lensed source's surface brightness. This k-means algorithm is seeded such that the same image-plane coordinates are inferred if the procedure (using the same inputs) is run multiple times (thus the completely random changes to the source pixelisation used to construct the histograms shown in Figure~\ref{Figure: lh histogram} cannot explain these likelihood spikes). These image-plane coordinates are then ray-traced via the mass model to the source-plane and are used as the centres of the source pixels of the Voronoi mesh. To produce the blue, orange, and purple curves shown in Figure~\ref{Figure: lh surface grid comp}, the coordinates that construct the source pixelisation are therefore fixed in the image-plane, but vary smoothly with the mass model in the source plane. The spiky likelihood surface can therefore be explained by how the continuous change in the positions of the source pixels generating the Voronoi pixelisation produces discrete changes in the Voronoi mesh (either creating new cells or changing the value of flux across cell boundaries - these changes may occur less frequently with interpolation of the source pixel grid). The reconstruction then receives boosts and drops in log likelihood as for certain mass models (values of $\gamma$) since the positions of the source pixels align more or less favourably with the data. 

\subsection{Testing for error underestimation in lens modelling}

In the context of a full non-linear search which varies every lens model parameter, we expect that likelihood spikes due to this preferable alignment of the source pixelisation with the data will be present, negatively impacting our inference on each parameter's PDF. To investigate this, we fit the full sample of 354 mock images (see section \ref{Mock Data}) with a uniform pipeline constructed from the SLaM pipelines in \texttt{PyAutoLens}. The pipeline is the equivalent of that described in Section~\ref{Pipeline} but created for fitting images without the lens galaxy's light distribution (see Appendix \ref{no light pipeline} for an overview of the pipeline). The pipeline, then, infers the mass parameters of the lens galaxy described by a PLEMD, while reconstructing the source galaxy on a BPR pixelisation. We choose not to fit for an external shear (which is not present in the lens models of the simulated data) in order to simplify our investigation of likelihood boosts. Our main goal, here, is to determine if the error estimates inferred by the non-linear search are being underestimated as a result of the data discretization bias.

\begin{figure}
    \centering
    \begin{subfigure}[b]{\columnwidth}
    \centering
    \includegraphics[width=0.82\textwidth]{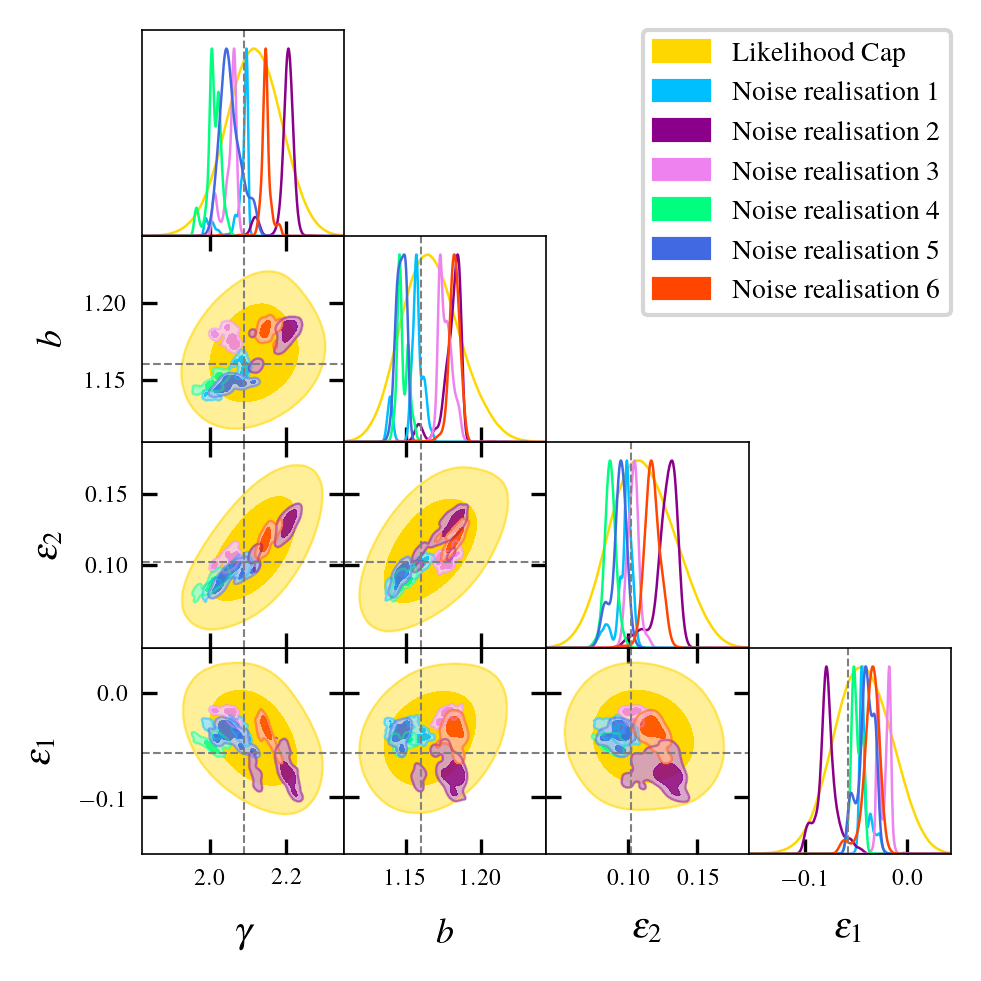}
    \end{subfigure}
    \vspace{-0.8\baselineskip}
    \begin{subfigure}[b]{\columnwidth}
    \centering
    \includegraphics[width=0.82\textwidth]{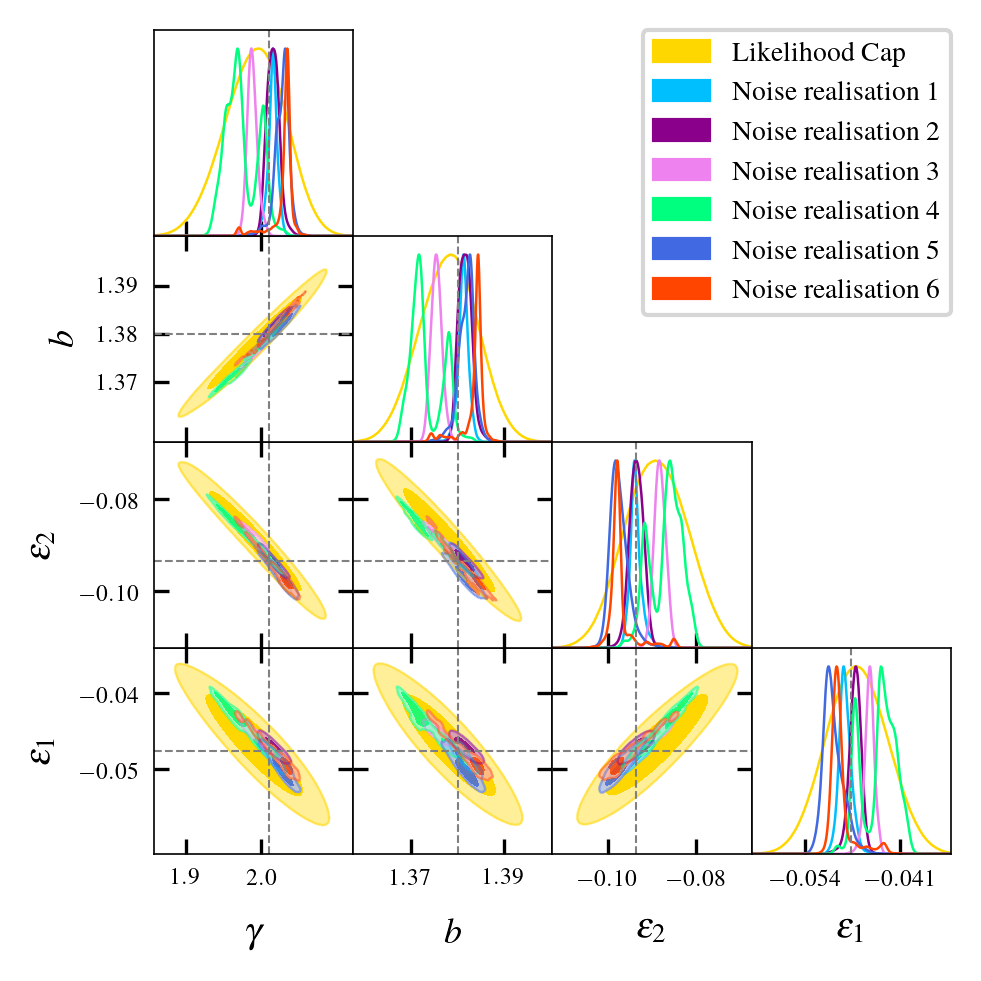}
    \end{subfigure}
    \vspace{-0.8\baselineskip}
    \begin{subfigure}[b]{\columnwidth}
    \centering
    \includegraphics[width=0.82\textwidth]{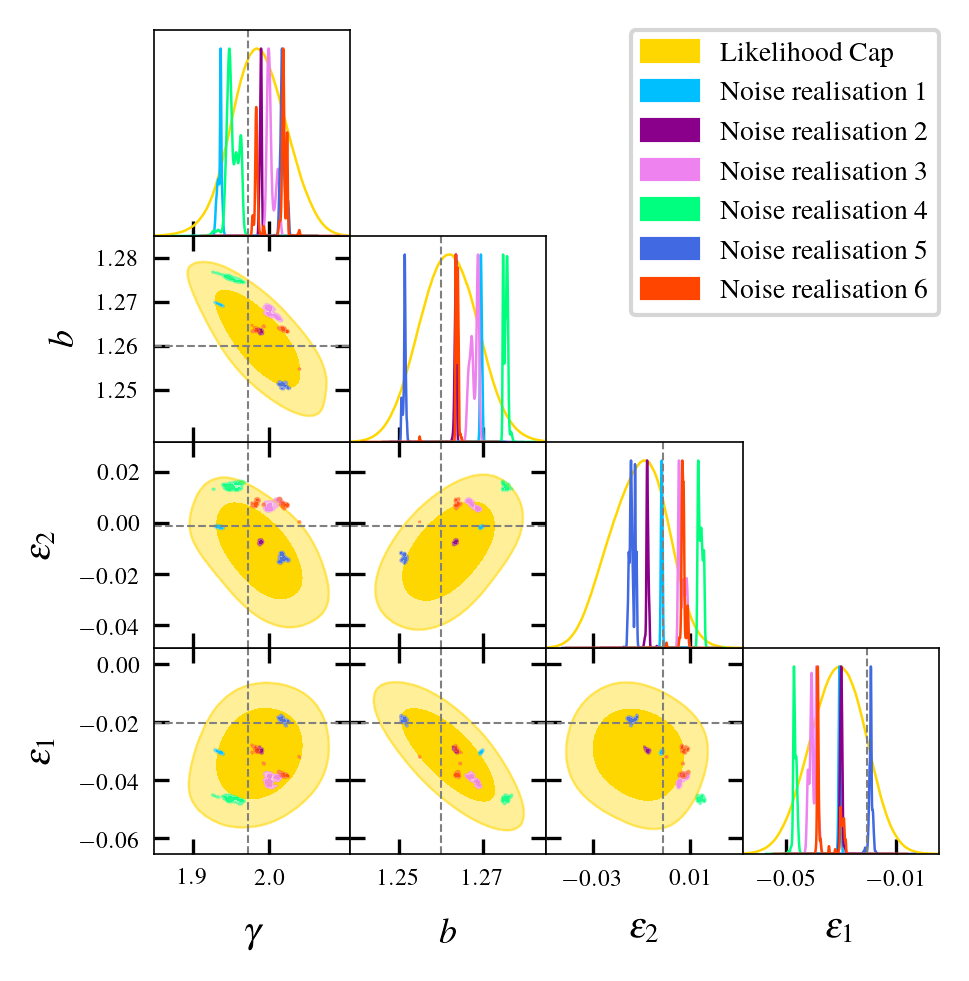}
    \end{subfigure}
    \caption{For three typical synthetic lenses, the posterior PDF of model parameters inferred from mock observations. With a likelihood cap (yellow), these PDFs have sufficient width to include the true value (crossed lines). Without a likelihood cap, the PDFs from mock data with different realisations of observational noise (six other colours) are too narrow because of noise in the likelihood evaluations. Fitted parameters shown are the mass-density slope ($\gamma$), mass normalisation ($R_{\text{al}}$), and two components of ellipticity ($\varepsilon_\text{1}$, $\varepsilon_\text{2}$); all other free parameters are marginalised over.}
    \label{Figure: pdf lh cap}
\end{figure}

Figure~\ref{Figure: pdf lh cap} shows the posterior PDFs obtained for individual runs of three lenses in our mock sample. For each lens, six realisations of the image data with different noise maps were simulated and fitted, which correspond to the six individual PDFs shown on each panel of Figure~\ref{Figure: pdf lh cap}. Not only do the PDF contours rarely contain the true parameter (represented by the grey dashed lines) they also rarely overlap with each other. To verify this is due to data discretization bias, for each of the 354 synthetic images we now produce 500 new likelihood evaluations --- fixing all lens and source model parameters to the best-fit values, but randomising the $k$-means seed used to pixelate the source plane. For 94.6\% of these 177,000 calculations, the new likelihood is lower than the best-fit model likelihood, indicating that the likelihood values inferred by \texttt{dynesty} were systematically boosted relative to the majority of possible source pixelisations. Figure~\ref{Figure: lh histogram} shows this for three example cases, where the solid lines show the maximum log likelihood model inferred via \texttt{dynesty} compared to a histogram of these 500 models draw using random $k$-means seeds.

\begin{figure}
    \centering
    \includegraphics[width=\columnwidth]{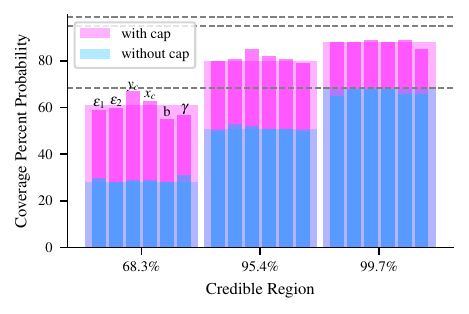}
    \caption{Coverage probabilities of the lens model parameters with (pink) and without (blue) a likelihood cap applied to the non-linear search. The thin bars give the coverage probabilities of individual lens model parameters as labelled, and the wider bars represent the average of these values.}
    \label{Figure: coverage}
\end{figure}

The likelihood boosted solutions inferred by \texttt{dynesty} occupy a tiny volume of parameter space, such that parameter marginalisation significantly underestimates the width of the posterior PDF. For each of the lens model parameters we calculate the percentage of the 354 model fits that recover the true parameter within their 1D marginalised 68.7\%,  95\%, and 99\% credible regions (blue bars in Figure~\ref{Figure: coverage}). On average for all lens model parameters the truth is recovered only $30\%$ or $50\%$ of the time at the  $68.7\%$ and $95\%$ credible regions, these coverage probabilities are significantly smaller than the percentage credible regions they were calculated for --- the reported uncertainties are too small.

\subsection{Likelihood Cap for improving sample statistics}

We now investigate the efficacy of placing a `log likelihood cap' on the non-linear search, where this cap is estimated in a way that seeks to smooth out likelihood spikes in parameter space. The cap is computed by taking the maximum likelihood lens model of the non-linear search inferred by the \textbf{MT}$^1_\textrm{ext}$ search in the SLaM pipeline and computing $500$ likelihood evaluations using this model but each with a different $k$-means seed. This process produces the histograms shown in figure~\ref{Figure: lh histogram}, which are fitted with a Gaussian whose mean then acts as the log likelihood cap. We then repeat the final \textbf{MT}$^1$ search of the pipeline (with identical parameters, hyper-parameters, $k$-means seed, etc.), but any log likelihood evaluation now returns no more than this value. If a log likelihood is computed above this cap, it is rounded down to the cap's value before it is returned to \texttt{dynesty}, we note that this assumes that \texttt{dynesty} has not converged on a local maxima in \textbf{MT}$^1$. The yellow shaded contours in Figure~\ref{Figure: pdf lh cap} show the PDFs inferred by \textbf{MT}$^1_\textrm{ext}$ using this log likelihood cap, which now appear larger, smoother, and do not have undesirable properties such as islands and discontinuities that are seen for the PDFs inferred without this cap. 

When performed on our 354 synthetic images, the final parameter estimation now converges more consistently for different realisations of noise (for the sake of visual clarity, Figure~\ref{Figure: pdf lh cap} only shows one PDF, but all six PDFs do now overlap for each dataset). The coverage probabilities for the 1D marginalised 68.7\% or 95\%  and 99\% credible regions have increased significantly for all lens model parameters with the use of the likelihood cap (see Figure~\ref{Figure: coverage} for the comparison with and without the likelihood cap). On average the true lens model parameters are recovered 61\% and 80\% of the time at the 68.7\% or 95\% credible regions, respectively. Although we do not obtain full coverage, this is a significant improvement in error estimation compared to not including the likelihood capped phase. Furthermore, for each lens model parameter we compare the mean of the best fit values of the six noise realisations, and find that these are recovered $74\%$ of the time at the 68.7\% credible region on average for all parameters. This suggests that the likelihood cap is producing errors that are consistent with the uncertainty due to random noise in the image, and that our posteriors recover the true values slightly less frequently than hoped due to systematic biases in particular lens configurations that offset the inferred parameters from the truth. 

Further testing is necessary to understand the systematics that result from the source discretization bias as well as any systematic offsets in inferred lens parameters in particular lens configurations. This would require a larger set of mocks than was simulated for this study (see Section~\ref{lens model testing} for more discussion) and is beyond the scope of this work. At present, it appears that the likelihood cap is effective at improving the coverage probability of the 68.7\% credible region (only 7\% shy of achieving coverage for lens models parameters on average). Since the mock data was simulated to be representative of the observed data, we assume this will be true of the errors on the data adopting the same approach. As such, all errors quoted in this work are those at the 68.7\% credible region of the PDFs inferred by the likelihood capped \textbf{MT}$^1_\textrm{ext}$ phase.
\section{Results}\label{Results}

\subsection{Automation}\label{automation}

We now inspect the results of our automated modelling procedure on the SLACS and BELLS GALLERY samples and quantify what fraction of lenses were fitted with a reliable lens model without human intervention. To facilitate this, we visually inspect every lens model, first after the \textbf{SP} pipeline and then again on completion of the uniform procedure. We label the final model of every lens in one of four categories:

\begin{itemize}
    \item \textbf{Gold (54/59):} The fit represents a physically plausible model of the lens and source.
    \item \textbf{Silver (4/59):} The fit represents a physically plausible model of the lens and source. However, achieving this required changes to data pre-processing that may not be easy to automate (e.g.\ masking, lens light subtraction), and may degrade the quality of the inferred lens model.
    \item \textbf{Bronze (1/59):} The fit represents a physically plausible model of the lens (with the correct number of multiple images), but other features in the data (e.g.\ residuals from lens light subtraction) visibly degrade the quality of the source model.
    \item \textbf{Failure (0/59):} The fit produces a physically implausible lens model (e.g.\ with an incorrect number of multiple images).
\end{itemize}

After a first blind run, we find 9 galaxies outside the ``Gold'' sample. In 8/9 cases, they went wrong during the first \textbf{SP} pipeline. We determine what went wrong, describe simple interventions, and rerun the pipeline. Our interventions successfully move all of these lenses into the ``Bronze'', ``Silver'' or ``Gold'' categories. Through this process, we suggest ways to reduce the failure rate in analyses of future large samples of lenses. For future analysis of large lens samples, one can anticipate undergoing this process on a subset of lenses before modelling the full sample.

If a lens ends up in the ``Gold'', ``Silver'', or ``Bronze'' categories, we consider its effective Einstein Radius $R_\mathrm{Ein, eff}$ to be measured accurately. If a lens is in the ``Gold'' or ``Silver'' categories, we also consider more detailed quantities of the mass model (e.g.\ the slope $\gamma$) to be reliable. Indeed, we shall find our best-fit models broadly consistent with those from previous literature, in Sections~\ref{section: r ein method comp} and \ref{lens model parameters}.

\subsubsection{Fully automated success}
{\renewcommand{\arraystretch}{1.4}
\begin{table*}
\centering
\begin{tabular}{llllllll}
\toprule
Class & Lens Name &         $R_{\rm Ein, eff}$ &                $\gamma$ &                       q &                 $\phi$ &             $\gamma^{\rm ext}$ &          $\phi^{\rm ext}$ \\
\midrule
\multirow{38}{*}{Gold} & J0008-0004 &   $1.15^{+0.007}_{-0.009}$ &  $2.08^{+0.08}_{-0.07}$ &  $0.72^{+0.03}_{-0.03}$ &     $42^{+5.1}_{-6.1}$ &  $0.023^{+0.018}_{-0.014}$ &      $96^{+15}_{-23}$ \\
       & J0029-0055 &  $0.934^{+0.007}_{-0.008}$ &  $2.32^{+0.13}_{-0.13}$ &  $0.78^{+0.06}_{-0.07}$ &      $22^{+9.3}_{-12}$ &  $0.013^{+0.019}_{-0.012}$ &      $11^{+32}_{-44}$ \\
       & J0157-0056 &  $0.912^{+0.013}_{-0.012}$ &  $2.23^{+0.08}_{-0.09}$ &  $0.56^{+0.07}_{-0.05}$ &    $112^{+4.8}_{-7.6}$ &  $0.182^{+0.021}_{-0.027}$ &   $102^{+3.0}_{-3.2}$ \\
       & J0216-0813 &  $1.183^{+0.014}_{-0.011}$ &  $1.99^{+0.05}_{-0.06}$ &   $0.8^{+0.04}_{-0.03}$ &     $75^{+9.2}_{-6.3}$ &  $0.009^{+0.012}_{-0.011}$ &       $2^{+37}_{-55}$ \\
       & J0252+0039 &  $1.024^{+0.004}_{-0.002}$ &  $1.92^{+0.08}_{-0.11}$ &  $0.89^{+0.02}_{-0.02}$ &    $111^{+3.5}_{-4.0}$ &  $0.024^{+0.005}_{-0.003}$ &   $117^{+6.1}_{-6.4}$ \\
       & J0330-0020 &  $1.088^{+0.009}_{-0.012}$ &  $2.15^{+0.02}_{-0.02}$ &  $0.79^{+0.07}_{-0.08}$ &       $94^{+11}_{-14}$ &  $0.041^{+0.021}_{-0.018}$ &      $54^{+12}_{-15}$ \\
       & J0728+3835 &  $1.244^{+0.012}_{-0.008}$ &   $1.99^{+0.12}_{-0.1}$ &  $0.68^{+0.05}_{-0.04}$ &     $65^{+3.4}_{-3.4}$ &  $0.068^{+0.015}_{-0.021}$ &    $61^{+5.5}_{-6.3}$ \\
       & J0737+3216 &  $0.976^{+0.003}_{-0.002}$ &  $2.28^{+0.07}_{-0.07}$ &  $0.86^{+0.04}_{-0.03}$ &     $96^{+5.4}_{-5.6}$ &  $0.109^{+0.007}_{-0.011}$ &    $10^{+1.7}_{-1.4}$ \\
       & J0822+2652 &  $1.129^{+0.011}_{-0.018}$ &   $2.1^{+0.08}_{-0.07}$ &  $0.54^{+0.04}_{-0.05}$ &     $75^{+4.9}_{-4.3}$ &    $0.1^{+0.021}_{-0.018}$ &    $72^{+6.6}_{-6.3}$ \\
       & J0841+3824 &  $0.956^{+0.096}_{-0.063}$ &   $2.27^{+0.2}_{-0.16}$ &  $0.69^{+0.15}_{-0.14}$ &      $117^{+24}_{-28}$ &  $0.144^{+0.047}_{-0.031}$ &   $117^{+6.8}_{-7.2}$ \\
       & J0903+4116 &  $1.261^{+0.005}_{-0.005}$ &  $2.23^{+0.05}_{-0.05}$ &  $0.88^{+0.04}_{-0.03}$ &      $52^{+12}_{-7.6}$ &   $0.062^{+0.01}_{-0.012}$ &    $63^{+5.6}_{-4.9}$ \\
       & J0912+0029 &  $1.393^{+0.011}_{-0.007}$ &  $2.14^{+0.05}_{-0.05}$ &  $0.79^{+0.04}_{-0.04}$ &     $27^{+8.1}_{-6.9}$ &  $0.033^{+0.011}_{-0.012}$ &    $126^{+12}_{-9.3}$ \\
       & J0936+0913 &  $1.081^{+0.004}_{-0.005}$ &  $2.13^{+0.08}_{-0.08}$ &  $0.79^{+0.05}_{-0.05}$ &    $134^{+4.9}_{-5.8}$ &  $0.061^{+0.013}_{-0.011}$ &   $105^{+6.6}_{-7.4}$ \\
       & J0946+1006 &  $1.409^{+0.001}_{-0.001}$ &  $2.06^{+0.03}_{-0.03}$ &    $0.9^{+0.0}_{-0.01}$ &     $68^{+1.8}_{-2.2}$ &   $0.09^{+0.004}_{-0.003}$ &  $68^{+0.79}_{-0.65}$ \\
       & J0956+5100 &  $1.314^{+0.002}_{-0.001}$ &  $2.05^{+0.02}_{-0.02}$ &  $0.79^{+0.01}_{-0.01}$ &    $143^{+1.0}_{-1.3}$ &  $0.066^{+0.003}_{-0.005}$ &   $53^{+1.4}_{-0.97}$ \\
       & J0959+0410 &  $0.985^{+0.014}_{-0.017}$ &  $2.08^{+0.07}_{-0.07}$ &   $0.52^{+0.07}_{-0.1}$ &     $59^{+5.3}_{-6.2}$ &  $0.038^{+0.024}_{-0.025}$ &      $60^{+18}_{-33}$ \\
       & J1020+1122 &  $1.065^{+0.011}_{-0.009}$ &  $2.15^{+0.11}_{-0.12}$ &  $0.54^{+0.04}_{-0.04}$ &    $131^{+3.3}_{-2.8}$ &  $0.159^{+0.023}_{-0.024}$ &   $131^{+3.2}_{-4.3}$ \\
       & J1023+4230 &  $1.411^{+0.009}_{-0.009}$ &  $1.95^{+0.16}_{-0.12}$ &  $0.92^{+0.05}_{-0.04}$ &      $177^{+14}_{-17}$ &   $0.023^{+0.01}_{-0.009}$ &      $68^{+18}_{-25}$ \\
       & J1029+0420 &    $0.947^{+0.01}_{-0.01}$ &  $1.43^{+0.05}_{-0.06}$ &  $0.62^{+0.02}_{-0.03}$ &    $111^{+3.4}_{-3.4}$ &    $0.152^{+0.02}_{-0.02}$ &   $100^{+2.3}_{-4.5}$ \\
       & J1032+5322 &   $1.03^{+0.011}_{-0.007}$ &  $2.11^{+0.02}_{-0.03}$ &  $0.69^{+0.07}_{-0.05}$ &    $143^{+5.5}_{-6.3}$ &  $0.039^{+0.019}_{-0.019}$ &     $167^{+13}_{-19}$ \\
       & J1142+1001 &  $0.908^{+0.024}_{-0.027}$ &    $2.03^{+0.1}_{-0.1}$ &  $0.49^{+0.11}_{-0.06}$ &    $144^{+4.7}_{-4.7}$ &     $0.21^{+0.04}_{-0.05}$ &   $148^{+5.0}_{-6.3}$ \\
       & J1143-0144 &  $1.611^{+0.013}_{-0.014}$ &  $2.15^{+0.03}_{-0.03}$ &  $0.73^{+0.04}_{-0.04}$ &    $116^{+5.6}_{-4.5}$ &    $0.038^{+0.01}_{-0.01}$ &    $166^{+13}_{-9.1}$ \\
       & J1205+4910 &  $1.218^{+0.008}_{-0.008}$ &  $1.92^{+0.07}_{-0.09}$ &  $0.74^{+0.08}_{-0.06}$ &    $149^{+6.8}_{-5.6}$ &  $0.019^{+0.011}_{-0.019}$ &      $99^{+33}_{-28}$ \\
       & J1213+6708 &  $1.322^{+0.018}_{-0.023}$ &   $2.8^{+0.07}_{-0.07}$ &  $0.92^{+0.07}_{-0.11}$ &        $3^{+74}_{-49}$ &  $0.045^{+0.018}_{-0.019}$ &       $9^{+15}_{-15}$ \\
       & J1218+0830 &   $1.217^{+0.01}_{-0.008}$ &  $2.35^{+0.07}_{-0.06}$ &  $0.35^{+0.03}_{-0.02}$ &    $144^{+1.4}_{-2.0}$ &  $0.353^{+0.011}_{-0.021}$ &  $140^{+1.3}_{-0.96}$ \\
       & J1250+0523 &  $1.144^{+0.006}_{-0.005}$ &  $1.84^{+0.04}_{-0.04}$ &  $0.91^{+0.03}_{-0.04}$ &    $129^{+7.2}_{-7.6}$ &   $0.024^{+0.014}_{-0.01}$ &    $132^{+15}_{-9.9}$ \\
       & J1402+6321 &  $1.349^{+0.005}_{-0.007}$ &  $2.00^{+0.18}_{-0.13}$ &  $0.72^{+0.04}_{-0.04}$ &     $63^{+3.1}_{-2.6}$ &  $0.030^{+0.019}_{-0.014}$ &    $1^{+14}_{-9.9}$ \\
       & J1420+6019 &  $1.075^{+0.002}_{-0.002}$ &  $1.94^{+0.04}_{-0.04}$ &  $0.43^{+0.02}_{-0.02}$ &  $111^{+0.45}_{-0.60}$ &  $0.118^{+0.009}_{-0.009}$ &   $110^{+1.1}_{-1.0}$ \\
       & J1430+4105 &  $1.481^{+0.002}_{-0.002}$ &  $2.02^{+0.01}_{-0.01}$ &  $0.91^{+0.01}_{-0.01}$ &    $120^{+2.1}_{-1.9}$ &  $0.088^{+0.002}_{-0.002}$ &  $22^{+0.64}_{-0.53}$ \\
       & J1432+6317 &   $1.284^{+0.01}_{-0.009}$ &  $1.79^{+0.06}_{-0.04}$ &  $0.88^{+0.05}_{-0.05}$ &      $102^{+12}_{-11}$ &  $0.099^{+0.016}_{-0.016}$ &   $115^{+3.8}_{-5.0}$ \\
       & J1451-0239 &   $0.96^{+0.017}_{-0.015}$ &   $2.29^{+0.1}_{-0.11}$ &  $0.54^{+0.06}_{-0.07}$ &     $30^{+3.7}_{-4.5}$ &  $0.193^{+0.042}_{-0.025}$ &    $27^{+3.6}_{-3.1}$ \\
       & J1525+3327 &   $1.29^{+0.012}_{-0.007}$ &  $1.92^{+0.06}_{-0.05}$ &  $0.59^{+0.04}_{-0.04}$ &    $117^{+2.8}_{-3.1}$ &    $0.14^{+0.01}_{-0.011}$ &    $87^{+3.1}_{-3.2}$ \\
       & J1627-0053 &  $1.217^{+0.002}_{-0.002}$ &  $2.08^{+0.08}_{-0.09}$ &  $0.84^{+0.03}_{-0.01}$ &      $8^{+2.6}_{-3.2}$ &  $0.019^{+0.005}_{-0.004}$ &     $6^{+6.8}_{-7.9}$ \\
       & J1630+4520 &  $1.791^{+0.006}_{-0.004}$ &  $1.96^{+0.09}_{-0.08}$ &  $0.83^{+0.01}_{-0.01}$ &     $70^{+2.7}_{-2.5}$ &  $0.023^{+0.006}_{-0.004}$ &    $59^{+7.6}_{-9.8}$ \\
       & J2238-0754 &  $1.268^{+0.004}_{-0.003}$ &  $2.07^{+0.09}_{-0.07}$ &  $0.83^{+0.03}_{-0.04}$ &    $137^{+5.8}_{-6.0}$ &  $0.004^{+0.007}_{-0.006}$ &       $3^{+52}_{-35}$ \\
       & J2300+0022 &  $1.219^{+0.008}_{-0.005}$ &  $2.55^{+0.07}_{-0.16}$ &  $0.62^{+0.05}_{-0.04}$ &     $74^{+3.8}_{-3.7}$ &  $0.094^{+0.012}_{-0.018}$ &     $9^{+4.0}_{-3.3}$ \\
       & J2303+1422 &  $1.628^{+0.007}_{-0.005}$ &  $2.09^{+0.04}_{-0.04}$ &  $0.53^{+0.05}_{-0.05}$ &     $34^{+1.6}_{-1.4}$ &  $0.002^{+0.007}_{-0.005}$ &     $171^{+44}_{-49}$ \\
       & J2341+0000 &  $1.338^{+0.009}_{-0.005}$ &  $2.12^{+0.06}_{-0.05}$ &   $0.8^{+0.03}_{-0.03}$ &     $81^{+3.7}_{-3.5}$ &  $0.027^{+0.009}_{-0.014}$ &   $167^{+8.9}_{-10.}$ \\
\cline{1-8}
\multirow{4}{*}{Silver} & J0959+4416 &   $0.972^{+0.023}_{-0.02}$ &   $2.5^{+0.19}_{-0.23}$ &   $0.67^{+0.14}_{-0.1}$ &      $83^{+16}_{-9.8}$ &  $0.027^{+0.037}_{-0.026}$ &      $88^{+57}_{-29}$ \\
       & J1016+3859 &   $1.004^{+0.026}_{-0.02}$ &   $2.23^{+0.15}_{-0.2}$ &  $0.56^{+0.13}_{-0.13}$ &      $92^{+11}_{-10.}$ &   $0.217^{+0.05}_{-0.042}$ &   $113^{+6.1}_{-4.9}$ \\
       & J1153+4612 &  $1.029^{+0.007}_{-0.005}$ &   $1.72^{+0.08}_{-0.1}$ &  $0.61^{+0.03}_{-0.03}$ &    $104^{+1.9}_{-1.6}$ &  $0.181^{+0.013}_{-0.013}$ &   $101^{+1.9}_{-2.1}$ \\
       & J1416+5136 &  $1.246^{+0.014}_{-0.018}$ &   $2.0^{+0.01}_{-0.01}$ &  $0.73^{+0.11}_{-0.09}$ &    $103^{+7.9}_{-3.9}$ &  $0.152^{+0.025}_{-0.032}$ &   $108^{+4.7}_{-2.8}$ \\
\cline{1-8}
Bronze & J1103+5322 &  $1.065^{+0.007}_{-0.007}$ &  $1.79^{+0.01}_{-0.01}$ &  $0.53^{+0.04}_{-0.04}$ &     $49^{+1.8}_{-1.5}$ &  $0.103^{+0.013}_{-0.009}$ &     $0^{+3.2}_{-2.2}$ \\
\bottomrule
\end{tabular}
    \caption{Best-fit physical parameters for SLACS lenses. These are derived quantities, obtained from the varied parameters of the lens mass model (Table~\ref{table: SLACS model fit params}). Lens light model parameters are presented in (Table~\ref{table: SLACS light parameter}).} 
    \label{table: SLACS mass}
\end{table*}}

{\renewcommand{\arraystretch}{1.4}
\begin{table*}
\centering

\begin{tabular}{llllllll}
\toprule
Class & Lens Name &                   $R_{\rm Ein, efff}$ &                $\gamma$ &                       q &               $\phi$ &             $\gamma^{\rm ext}$ &           $\phi^{\rm ext}$ \\
\midrule
\multirow{16}{*}{Gold} & J0029+2544 &  $1.347^{+0.014}_{-0.012}$ &  $2.05^{+0.12}_{-0.15}$ &  $0.65^{+0.07}_{-0.08}$ &  $128^{+6.7}_{-7.6}$ &  $0.029^{+0.033}_{-0.018}$ &     $149^{+40.}_{-31}$ \\
     & J0113+0250 &  $1.329^{+0.006}_{-0.005}$ &  $1.77^{+0.15}_{-0.11}$ &  $0.75^{+0.02}_{-0.02}$ &  $178^{+2.8}_{-4.4}$ &  $0.079^{+0.014}_{-0.012}$ &     $15^{+3.9}_{-5.9}$ \\
     & J0201+3228 &  $1.713^{+0.011}_{-0.005}$ &   $2.09^{+0.09}_{-0.1}$ &  $0.78^{+0.03}_{-0.02}$ &  $125^{+5.4}_{-3.6}$ &  $0.063^{+0.016}_{-0.014}$ &     $53^{+4.7}_{-5.7}$ \\
     & J0237-0641 &   $0.619^{+0.02}_{-0.025}$ &   $1.91^{+0.18}_{-0.1}$ &  $0.79^{+0.12}_{-0.09}$ &   $131^{+30.}_{-17}$ &  $0.027^{+0.044}_{-0.033}$ &        $6^{+36}_{-55}$ \\
     & J0742+3341 &   $1.241^{+0.01}_{-0.013}$ &  $2.21^{+0.06}_{-0.08}$ &  $0.29^{+0.04}_{-0.04}$ &   $56^{+3.4}_{-3.0}$ &  $0.107^{+0.016}_{-0.023}$ &     $44^{+6.4}_{-8.7}$ \\
     & J0755+3445 &  $2.073^{+0.005}_{-0.004}$ &  $1.77^{+0.08}_{-0.05}$ &  $0.53^{+0.01}_{-0.01}$ &   $15^{+1.9}_{-1.5}$ &   $0.24^{+0.006}_{-0.006}$ &     $28^{+1.6}_{-1.1}$ \\
     & J0856+2010 &   $0.951^{+0.035}_{-0.04}$ &  $2.23^{+0.08}_{-0.09}$ &  $0.36^{+0.09}_{-0.06}$ &   $45^{+5.9}_{-4.3}$ &   $0.153^{+0.023}_{-0.03}$ &     $93^{+6.5}_{-7.2}$ \\
     & J0918+5105 &  $1.645^{+0.005}_{-0.009}$ &  $2.38^{+0.16}_{-0.18}$ &  $0.78^{+0.04}_{-0.06}$ &    $95^{+18}_{-9.4}$ &   $0.259^{+0.034}_{-0.02}$ &   $125^{+0.60}_{-1.3}$ \\
     & J1110+2808 &  $0.904^{+0.027}_{-0.026}$ &  $2.03^{+0.09}_{-0.07}$ &  $0.82^{+0.08}_{-0.07}$ &     $77^{+22}_{-16}$ &  $0.123^{+0.043}_{-0.028}$ &     $55^{+7.1}_{-6.3}$ \\
     & J1110+3649 &  $1.151^{+0.001}_{-0.001}$ &  $2.23^{+0.07}_{-0.08}$ &  $0.77^{+0.02}_{-0.02}$ &  $174^{+1.4}_{-1.5}$ &  $0.025^{+0.005}_{-0.005}$ &     $64^{+5.6}_{-5.8}$ \\
     & J1116+0915 &  $0.811^{+0.053}_{-0.054}$ &  $2.22^{+0.16}_{-0.17}$ &  $0.21^{+0.05}_{-0.05}$ &   $86^{+4.3}_{-3.5}$ &  $0.393^{+0.053}_{-0.046}$ &     $88^{+3.6}_{-2.8}$ \\
     & J1141+2216 &  $1.283^{+0.027}_{-0.019}$ &  $2.13^{+0.09}_{-0.11}$ &  $0.58^{+0.09}_{-0.09}$ &   $57^{+5.5}_{-7.6}$ &  $0.043^{+0.026}_{-0.022}$ &      $38^{+23}_{-30.}$ \\
     & J1201+4743 &  $1.171^{+0.004}_{-0.002}$ &  $2.74^{+0.05}_{-0.21}$ &  $0.82^{+0.06}_{-0.06}$ &   $130^{+14}_{-7.8}$ &   $0.069^{+0.004}_{-0.01}$ &     $42^{+3.7}_{-3.4}$ \\
     & J1226+5457 &  $1.398^{+0.004}_{-0.003}$ &   $2.24^{+0.07}_{-0.1}$ &  $0.86^{+0.02}_{-0.02}$ &  $130^{+6.6}_{-7.9}$ &  $0.189^{+0.012}_{-0.012}$ &  $156^{+0.76}_{-0.75}$ \\
     & J2228+1205 &   $1.21^{+0.024}_{-0.024}$ &    $2.2^{+0.14}_{-0.1}$ &   $0.51^{+0.1}_{-0.05}$ &  $116^{+5.7}_{-8.0}$ &  $0.202^{+0.026}_{-0.023}$ &    $141^{+5.7}_{-3.3}$ \\
     & J2342-0120 &  $1.091^{+0.006}_{-0.004}$ &  $2.34^{+0.07}_{-0.09}$ &  $0.44^{+0.05}_{-0.03}$ &  $114^{+3.6}_{-2.5}$ &   $0.13^{+0.009}_{-0.016}$ &     $94^{+4.4}_{-2.6}$ \\
\bottomrule
\end{tabular}
    \caption{Best-fit physical parameters for BELLS GALLERY lenses. These are derived quantities, obtained from the varied parameters of the lens mass model (Table~\ref{table: GALLERY model fit params}). Lens light model parameters are presented in (Table~\ref{table: Gallery light parameter}).} 
    \label{table: GALLERY mass}
\end{table*}}

We immediately place 50/59 lenses (85\%) in the ``Gold'' sample after the first, blind run of our uniform pipeline. These models show low levels of residuals and physically plausible source galaxy morphologies. Best-fit model parameters are listed in Tables~\ref{table: SLACS mass} (SLACS) and~\ref{table: GALLERY mass} (GALLERY), and reconstructions are shown in Appendix~\ref{model fits}. 

\subsubsection{Semi-automated success}

\begin{figure*}
\centering
    \begin{subfigure}[b]{\columnwidth}
    \centering
    \includegraphics[width=\textwidth]{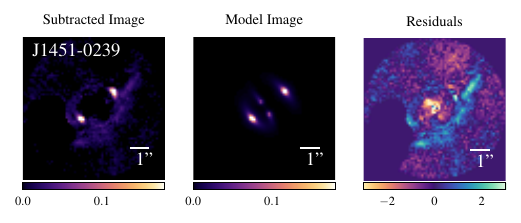}
    \caption{Unsuccessful model fit in the Source Parametric pipeline.}
    \end{subfigure}
    \begin{subfigure}[b]{\columnwidth}
    \centering
    \includegraphics[width=\textwidth]{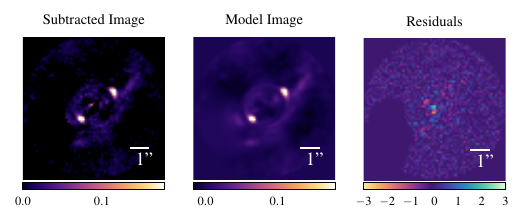}
    \caption{Successful model fit on completion of the pipeline.}
    \end{subfigure}
    \caption{(a) Model fits for the system that misses the counter image. (b) After tightening the prior on the elliptical components of the mass distribution to $\varepsilon_I\in$[$-0.2$, $0.2$], the system is fitted successfully, and is classified as a ``Gold'' model.}
    \label{Figure: 1451}
\end{figure*}

\begin{figure*}
\centering
    \begin{subfigure}[b]{\columnwidth}
    \centering
    \includegraphics[width=0.97\textwidth]{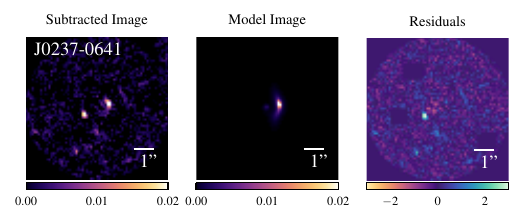}
    \end{subfigure}
    \vspace{-0.8\baselineskip}
    \begin{subfigure}[b]{\columnwidth}
    \centering
    \includegraphics[width=0.97\textwidth]{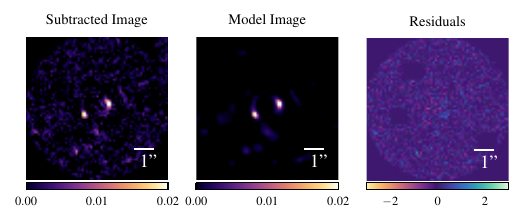}
    \end{subfigure}
    \vspace{-0.8\baselineskip}
    \begin{subfigure}[b]{\columnwidth}
    \centering
    \includegraphics[width=0.97\textwidth]{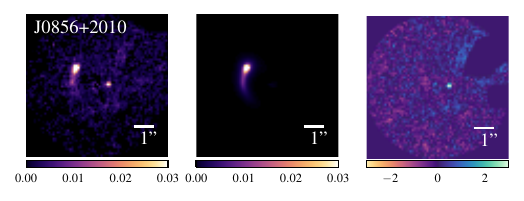}
    \caption{Unsuccessful model fit in the Source Parametric pipeline.}
    \end{subfigure}
    \begin{subfigure}[b]{\columnwidth}
    \centering
    \includegraphics[width=0.97\textwidth]{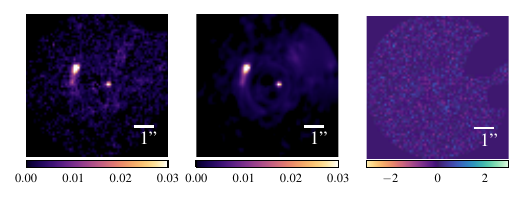}
    \caption{Successful model fit on completion of the pipeline.}
    \end{subfigure}
    \caption{(a) Model fits for the systems that fail to fit the counter image in the Source Parametric phase. (b) After tightening the prior on the elliptical components of the mass distribution to $\varepsilon_I\in$[$-0.2$, $0.2$], the systems are fitted successfully, and are classified as ``Gold'' models.}
    \label{Figure: singles}
\end{figure*}

\begin{figure*}
\centering
    \begin{subfigure}[b]{\columnwidth}
    \centering
    \includegraphics[width=\textwidth]{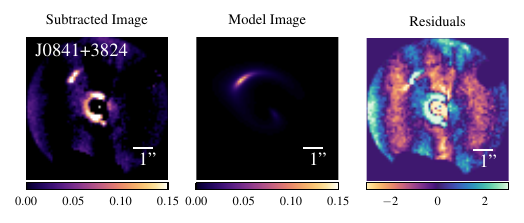}
    \caption{Unsuccessful model fit in the Source Parametric pipeline.}
    \end{subfigure}
    \begin{subfigure}[b]{\columnwidth}
    \centering
    \includegraphics[width=\textwidth]{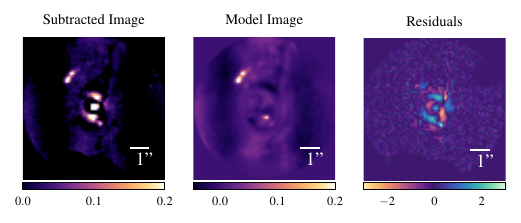}
    \caption{Successful model fit on completion of the pipeline.}
    \end{subfigure}
    \caption{(a) Model fits for the system that misses the counter image. (b) After tightening the prior on the elliptical components of the mass distribution to $\varepsilon_I\in$[$-0.2$, $0.2$], the system is fitted successfully, and is classified as a ``Gold'' model.}
    \label{Figure: 0841}
\end{figure*}

Fits to 4/59 lens systems converge to models with the wrong number of lensed images. In all four cases, the fits incorrectly converge to a highly elliptical mass distribution early in the \textbf{SP} pipeline, and could not recover the better solution in the \textbf{SI} or subsequent pipelines. The model of J1451$-$0239 fits 4 images to what is (by eye) a 2 image system (Figure~\ref{Figure: 1451}). Fits to J0237$-$0239 and J0856+2010 converge to single-image models, each missing a central counter-image that is close to the centre of the lens galaxy and therefore difficult to disentangle from the lens galaxy's light (Figure~\ref{Figure: singles}). The model of J0841+3824 is multiply imaged, but its very faint counter image is in the wrong location (Figure~\ref{Figure: 0841}).  

We fixed this by rerunning the pipeline for these lenses, but restricting the $\textbf{SP}^2$ phase to more circular mass models, via a uniform prior $\varepsilon_i\in$[$-0.2$, $0.2$], instead of a Gaussian with $\sigma=0.3$. To better find the global maximum likelihood solution for lenses J0237$-$0239, J0841+3824, and J0856+2010, we also increased the number of \texttt{dynesty} live points to 600 from 200 in $\textbf{SP}^2$ (this was not necessary for J1451-0239, where a change has no consequences other than increased runtime). With these settings, the automated modelling procedure is a success and the models (also shown in Figure~\ref{Figure:recons}) are moved into the ``Gold'' sample. 

These fits can be easily fixed by a more restrictive (or an all-round better) early initialisation. Our solution of forcing fairly circular models works well for early-type galaxy lenses, but would need to be rethought if the sample could include late-type galaxies with (edge-on) discs. Since spectroscopic lens detection techniques also identify the lens galaxy type, a different prior could be used for each. 

For now, we conclude that the biggest challenge of scaling up lens modelling to large samples is fitting an initial, physically plausible lens model. Once a simple lens model is correctly initialised, nothing prevents subsequent convergence of increasingly complex distributions of source light and lens mass. We shall discuss this further in Section~\ref{automation prospects}.

\subsubsection{Success with human intervention}

\begin{figure*}
\centering
    \begin{subfigure}[b]{\columnwidth}
    \centering
    \includegraphics[width=0.97\textwidth]{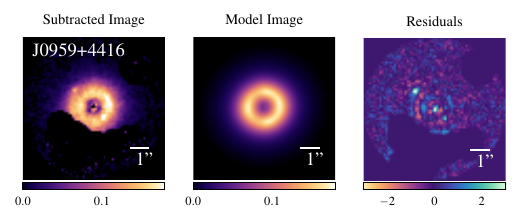}
    \end{subfigure}
    \vspace{-0.8\baselineskip}
    \begin{subfigure}[b]{\columnwidth}
    \centering
    \includegraphics[width=0.97\textwidth]{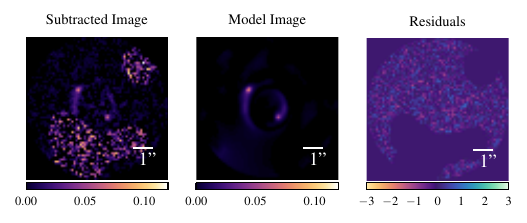}
    \end{subfigure}
    \vspace{-0.8\baselineskip}
    \begin{subfigure}[b]{\columnwidth}
    \centering
    \includegraphics[width=0.97\textwidth]{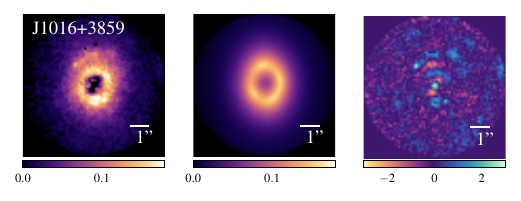}
    \end{subfigure}
    \vspace{-0.1\baselineskip}
    \begin{subfigure}[b]{\columnwidth}
    \centering
    \includegraphics[width=0.97\textwidth]{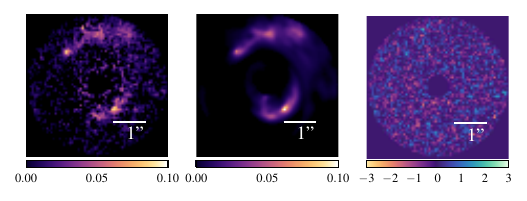}
    \end{subfigure}
    \vspace{-0.1\baselineskip}
    \begin{subfigure}[b]{\columnwidth}
    \centering
    \includegraphics[width=0.97\textwidth]{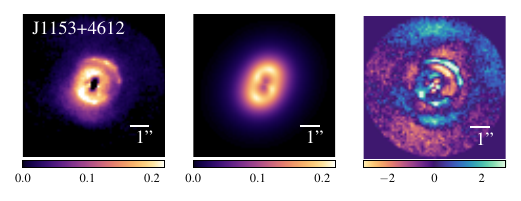}
    \caption{Unsuccessful model fit in the Source Parametric pipeline.}
    \end{subfigure}
    \begin{subfigure}[b]{\columnwidth}
    \centering
    \includegraphics[width=0.94\textwidth]{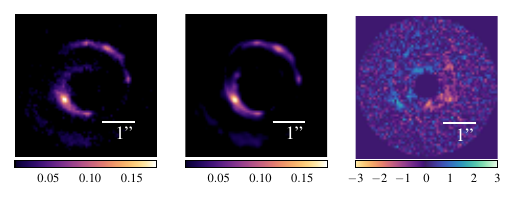}
    \caption{Successful model fit on completion of the pipeline.}
    \end{subfigure}
    \caption{(a) Model fits for the lens systems that fail to fit successful models in the Source Parametric pipeline as a result of bad lens light subtractions. The model reproduces lens light emission that remains in the subtracted image and significant residuals can be seen where the source emission is being ignored by the model. (b) For these systems we replace the data with b-spline subtracted data and use custom masks to arrive at successful model fits classified as ``Silver'' models.}
    \label{Figure:Bad Lens Subs}
\end{figure*}

Fits to 3/59 lens systems converge to a model in which imperfect lens light subtraction has left a spurious, residual ring of lens light that becomes considered part of the source. This again happens during the early \textbf{SP} pipeline, after which the S\'ersic model of the source is too large (Figure~\ref{Figure:Bad Lens Subs}a). Subsequent pixelised source models also include the residual lens light. Unlike the previous failure modes, we could not find small changes to the automated pipeline that fix these model fits.

For lenses J1153+4612, J1016+3859, and J0959+4416, we instead use the b-spline subtracted data (Section~\ref{reduction}). These versions pre-subtract the lens galaxy's light more cleanly than our double S\'ersic fit. Even then, we mask small remaining residuals near the centre of J1153+4612 and J1016+3859. We finally refit all three lenses using the version of the pipeline (which was also for the mock data) that does not fit the lens light. This results in successful models, as assessed by our visual inspection criteria (Figure~\ref{Figure:Bad Lens Subs}b). Although we arrive at successful model fits, we categorise these lenses in the ``Silver'' sample, because the lens light was not fitted in a Bayesian manner.

\begin{figure*}
\centering
    \begin{subfigure}[b]{\columnwidth}
    \centering
    \includegraphics[width=\textwidth]{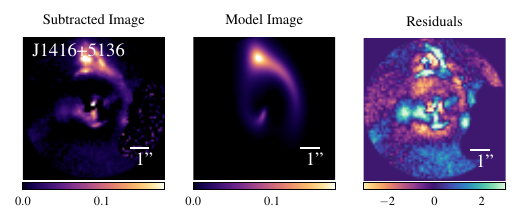}
    \caption{Unsuccessful model fit in the Source Parametric pipeline.}
    \end{subfigure}
    \begin{subfigure}[b]{\columnwidth}
    \centering
    \includegraphics[width=\textwidth]{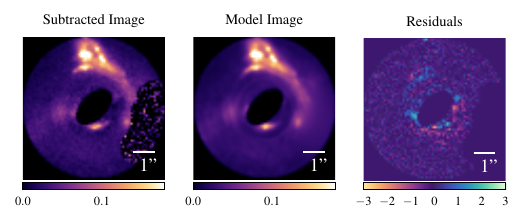}
    \caption{Successful model fit on completion of the pipeline.}
    \end{subfigure}
    \caption{(a) Model fits for the lens system that misses the counter image, instead fitting a counter image to lens light residuals. (b) The lens requires rerunning with our own double S\'ersic subtracted data using the without lens light pipeline, as well as a custom mask, to arrive at the successful ``Silver'' model fit.}
    \label{Figure:Bad Counter Image}
\end{figure*}

The fit to 1/59 lens systems includes a counter-image that reproduces a residual knot of lens light emission instead of the adjacent but fainter true counter-image (Figure~\ref{Figure:Bad Counter Image}). It can be fixed by masking the knot of lens light and rerunning the pipeline. However, this process would be difficult to automate with monochromatic imaging, so we place J1416+5136 in the ``Silver'' sample.

\subsubsection{Remaining problematic lens}

\begin{figure*}
\centering
    \begin{subfigure}[b]{\columnwidth}
    \centering
    \includegraphics[width=\textwidth]{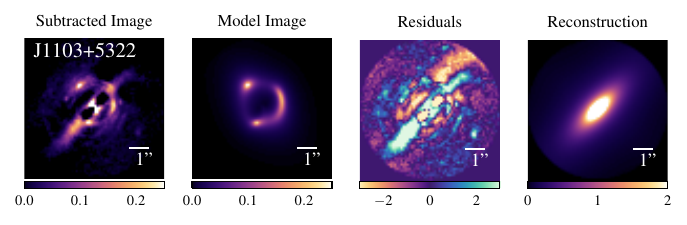}
    \caption{Successful model fit in the Source Parametric pipeline.}
    \end{subfigure}
    \begin{subfigure}[b]{\columnwidth}
    \centering
    \includegraphics[width=\textwidth]{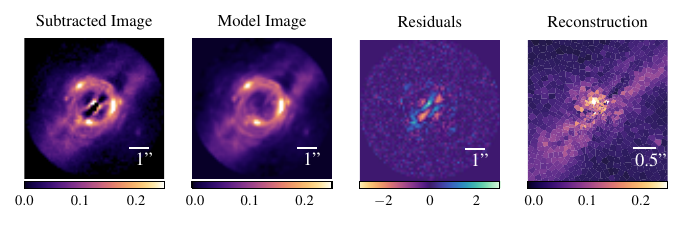}
    \caption{Unsuccessful model fit on completion of the pipeline.}
    \end{subfigure}
    \caption{(a) The single lens J1103+5322 is successful on completion of the Source Parametric pipeline, the parametric source avoids fitting to lens light residuals that remain in the subtracted image. (b) However, on completion of the pipeline the pixelised source reconstruction is unable to avoid fitting to these residuals, leading to this lens's classification of ``Bronze''. }
    \label{Figure: 1103}
\end{figure*}

The lens J1103+5322 is the only system that is unable to pass our visual inspection criteria on completion of the uniform pipeline. In the SP pipeline the model fits an appropriate model that fits the global lensed structure of the source, however significant residuals are present in the fit. The lens light subtraction leaves a quadrupole-like feature in the centre of the subtracted image as well as flux extending past the Einstein-ring feature. The SP pipeline is able to fit a model that fits solely to the source light, however continuation of the pipeline leads to a final model that reconstructs the lens light residual structure, which in Figure~\ref{Figure: 1103} can be seen to extend far beyond the emission from the source. This feature could impact the measurement of parameters which depend on the gradient of the flux in the lensed source, like the slope of the mass model. Replacing the data with the b-spline subtracted data resulted in similar residual lens light emission being reconstructed by the source galaxy. Nevertheless, we believe that this model estimates $R_\mathrm{Ein, eff}$ accurately, our measurement is within 5\% of previous literature measurements (see Section~\ref{section: r ein method comp} for a discussion on the expected uncertainty between these methods).  As a result, we place this lens in our ``Bronze'' sample.

\subsection{Statistical uncertainty on measurements}
\subsubsection{Effect of the Likelihood Cap}

\begin{figure}
\includegraphics[width=\columnwidth]{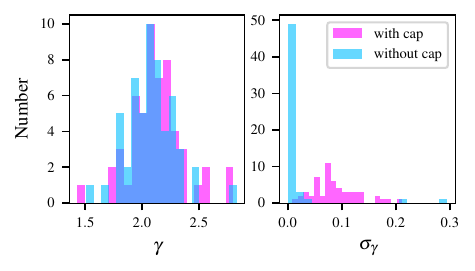}
\caption{Comparison of the distribution of inferred slopes (left) and their associated 1$\sigma$ credible region (right) with and without a likelihood cap applied to the non-linear search.}
\label{slope cap comp}
\end{figure}

\begin{table}
    \begin{tabularx}{\columnwidth}{c|c|c|c|c}
          \multirow{2}{*}{\shortstack[c]{\textbf{Model} \\ \textbf{Parameter}}} & \multicolumn{2}{c|}{\textbf{Mean Error}} & \multicolumn{2}{c}{\textbf{Median Error}} \\\cline{2-5}
           & cap & without cap & cap & without cap \\ \cline{1-5}
           b & 0.036 &  0.010& 0.027 & 0.005 \\
           $\gamma$ & 0.087 & 0.014 & 0.079 & 0.002 \\
           $\varepsilon_{1}$ & 0.039 & 0.010 & 0.028&  0.005 \\ 
           $\varepsilon_{2}$ & 0.038 & 0.025 & 0.031&  0.015 \\ 
           $\gamma_{{1}_{\rm ext}}$ & 0.018 & 0.005 & 0.016&  0.003 \\ 
           $\gamma_{{2}_{\rm ext}}$ & 0.019 & 0.009 & 0.017&  0.004 \\ 
           $x_{\rm c}$ & 0.016 & 0.006 & 0.014&  0.003 \\ 
           $y_{\rm c}$ & 0.016 & 0.004 & 0.013&  0.004 \\ 
    \end{tabularx}
    \caption{Summary of the average 68\% credible region errors inferred for all mass model parameters with and without a likelihood cap applied to the non-linear search.}
    \label{table:data errors}
\end{table}
In~Section~\ref{lh_boosts} we demonstrated the necessity of a likelihood capped phase (\textbf{MT}$^1_\textrm{ext}$) to increase the formal statistical errors inferred by the non-linear search such that they better recovered the true parameters on mock data. We now quantify the effect this phase has on the uncertainties inferred on real data (see Figure~\ref{slope cap comp} for its affect on the density profile slope errors). On average, we find that this approach has increased the inferred non-linear search errors by a factor of $\sim5$, as assessed by the median of individual factor increases for all mass model parameters. We quote the median increase to avoid bias from 5 lenses whose errors increase by a factor of over 1000 upon introduction of the log likelihood cap. On investigation, we found these lenses correspond to those with the largest difference between the likelihood inferred in \textbf{MT}$^1$ and the likelihood cap applied to \textbf{MT}$^1_\textrm{ext}$ (defined as the mean of 500 repeated likelihood evaluations with the same mass model, but different data discretizations). Hence, these lenses are the ones that were in the most ``likelihood-boosted'' regions of parameter space and as a result significantly underestimated the error. In the most extreme example, J0755+3445, the error inferred on the slope parameter with a likelihood cap is 64453 times larger than that inferred without a cap \cite[see][for a discussion of this lens]{Ritondale2019}. This highlights the scale at which the certainty of parameters can be incorrectly inferred  without consideration of the source discretization bias. Further quantification of the average errors inferred at the 68$\%$ credible region for each mass model parameter with and without a likelihood cap is given in Table~\ref{table:data errors}. 

\begin{figure}
\includegraphics[width=\columnwidth]{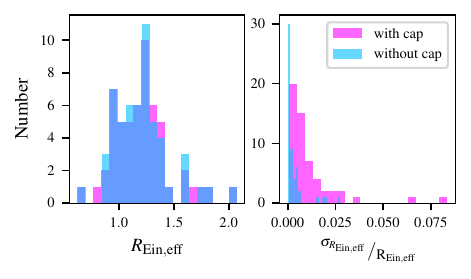}
\caption{Comparison of the distribution of inferred Einstein radii (left) and their associated percentage error at the 1$\sigma$ credible region (right), with and without a likelihood cap applied to the non-linear search.}
\label{Rein cap comp}
\end{figure}

Of all the mass model parameters, the likelihood cap has the largest effect on the density profile slope. The median factor increase in the error size before and after the cap is 24. The distribution of the 68\% credible region errors with and without the cap are plotted in the right panel of Figure~\ref{slope cap comp}. Notably there are two extreme outliers in the distribution of errors inferred without a cap, that are the two largest errors inferred across both distributions. For the lenses J1016+3859 and J0959+4416, both of which were replaced with b-spline subtracted data as an intervention to achieve model fits, the error actually decreases when the likelihood cap is applied. Although the uncertainty on the slope measurement is in general, as expected, significantly increased in \textbf{MT}$^1_\textrm{ext}$ relative to \textbf{MT}$^1$, the distribution of slopes inferred does not change significantly (left panel of Figure~\ref{slope cap comp}). The mean increases from 2.08 to 2.12, and the standard deviation increases from 0.21 to 0.24.

We derive errors on the effective Einstein radius by calculating a posterior PDF from all possible effective Einstein radii given the accepted non-linear search samples and their weights. We find the inclusion of the likelihood cap increases the mean 68\% credible region error on the effective Einstein radius from 0.3\% to 1.1\%, and does not affect the distribution of $R_\mathrm{Ein, eff}$ we infer (see Figure~\ref{Rein cap comp}). This suggests that, on average, the Einstein radius can be measured to $\sim1\%$ uncertainty, taking into account uncertainties in the noise and source discretization. We note that this does not account for any systematic error that would result from discrepancies between the assumed mass model and the underlying mass distribution. 
However, although the mean uncertainty on $R_\mathrm{Ein, eff}$ is low, two lenses (J0841+3824 and J1116+0915) have anomalously large uncertainties of 8.6\% and 6.6\% respectively. Hence, for some lens configurations it appears the Einstein radius can not be determined with such certainty. This may be an indication that the underlying mass distribution for these lenses is more complex than the PLEMD that we assume in our model fits. This seems reasonable for these two lenses since J0841+3824 is one of the few disky galaxies in the sample, with obvious extended spiral features in the data, and J1116+0915 contains a visible mass clump to the North of the lens that we do not fit for with our uniform approach.

\subsubsection{What drives the precision of a lens model?}\label{uncertainty correlations}

\begin{figure}
\includegraphics[width=\columnwidth]{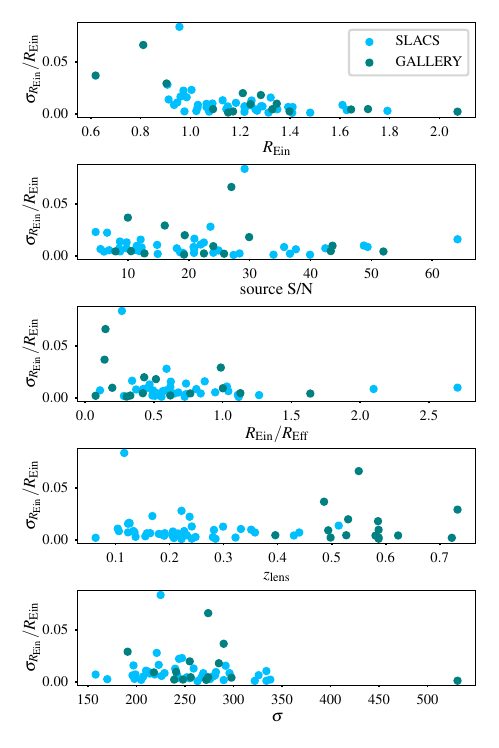}
\caption{Inferred percentage errors on the Einstein radii at the 68$\%$ credible region as a function of observable properties of the lens galaxy, and the S/N of the source. Parameters for linear fits to these data are given in Table~\ref{Table: correlations}.}
\label{err vs params}
\end{figure}

\begin{table}
\centering 
 \resizebox{\columnwidth}{!}{
\begin{tabular}{c c c} 
\hline\hline
Parameter & Gradient & Intercept  \\
\hline 
$R_{\rm Ein, eff} \left(\arcsec\right)$ & $-0.027\pm0.007$ & $0.044\pm0.008$\\ 
$R_{\rm Ein, eff}\left(\arcsec\right)\left[>5\% \textrm{ removed}\right]$ & $-0.017\pm0.004$ & $0.029\pm0.004$\\ 
peak source S/N & $\left(0.0\pm1.0\right)\times10^{-4}$  & $0.011\pm0.004$ \\
$R_{\rm Ein, eff}/R_{\rm Eff}$ & $\left(-6.4\pm4.0\right)\times10^{-4}$ & $0.015\pm0.004$\\
$z_{\rm lens}$ & $0.0\pm0.01$ & $0.010\pm0.004$\\
$\sigma \left(kms^{-1}\right)$ & $\left(-3.4\pm3.7\right)\times10^{-5}$ & $0.020\pm0.010$ \\
\hline 
\end{tabular}}
\caption{Linear fit results for the correlations with the uncertainty on the Einstein radius. Errors quoted on the gradient and intercept are the $1\sigma$ confidence intervals.}\label{Table: correlations}
\end{table}

To investigate what properties of the lens or data (if any) drive the precision of the lens model, we measure correlation coefficients between statistical uncertainty on the effective Einstein radius and observable properties of the lens galaxy: including the Einstein radius itself, the ratio of the Einstein radius to the effective radius, the lens redshift, the velocity dispersion of the lens, and the peak S/N of the source (Figure~\ref{err vs params}). Linear fits show no clear trend with most of these parameters. The only non-negligible correlation (defined as a non-zero gradient with $>$3$\sigma$ significance) is with the Einstein radius. The correlation remains when we repeat the linear fit removing the two uncertainties that are larger than $5\%$ that could bias the relation, although the effect size does reduce by over a third (Table~\ref{Table: correlations}).
   
\section{Discussion}\label{Discussion}
\subsection{Can we truly leave no lens behind?}\label{automation prospects}

The success of our uniform pipeline makes us optimistic for the future of automated strong lens analysis. We initially fitted 50/59 (85\%) lenses in a blind run. We increased this to 54/59 (92\%) ``Gold'' lenses after tweaking model priors, 58/59 (98\%) ``Gold'' or ``Silver'' lenses with some pre-fitting and masking of lens light, and 59/59 (100\%) including one successful model of the lens whose model of the source includes poorly-subtracted residuals of lens light. With just one pipeline, we have inferred parameters for 59/59 lenses that measure the lens galaxy's Einstein radius and other mass distribution parameters (of the power-law profile with an external shear we assume) that depend on only the first derivative of the potential of the lens galaxy. For 58/59 systems, we measure parameters describing their mass (including the parameters that depend on the gradient of the source flux such as $\gamma$). As well as this, we reconstruct a de-lensed image of the source galaxy, enabling study of its morphology. For 54/59 systems, we measure parameters describing their mass distribution \textit{and} light distribution (as a double S\'ersic profile) as well as reconstructing a de-lensed image of the source galaxy.

The most challenging step in automating lens modelling is in the initial estimation of a simple lens model (in this work, we use an SIE plus shear). Notably, once our early \textbf{SP} phases arrived at a successful fit to this model, the rest of our pipeline always ran to completion, successfully increasing the model complexity. We therefore recommend that effort to further improve automation should focus on `lens model initialisation' and find ways to avoid or flag the problematic solutions that occur at early stages of the analysis. Provided that our sample of lenses is representative of the larger population of lenses that will be discovered by future surveys, this strategy will lead to a high success rate for even complex mass fits and reduce the need for visual inspection of the results. An obvious starting point to improve lens model initialisation by \texttt{PyAutoLens} would be to further simplify the non-linear parameter space of the \textbf{SP} pipeline, for example by assuming models for the lens and source light with fewer parameters \citep[e.g.][]{Shapelets, Birrer2015, Tagore2016, Shexplets}.

Convolutional Neural Networks (CNNs) have also been suggested as a fast method for automated lens fitting \citep{Hezaveh2017, Levasseur2017, Morningstar2018a}. They provide a particularly compelling solution to the problem of lens model initialisation. For example, \cite{Pearson2021} combined a CNN with \texttt{PyAutoLens}, using models from the CNN to initialise the mass model priors of a \texttt{PyAutoLens} model-fit. In the majority of cases tested on mock data, the authors argued that a combination of the two methods outperformed either method individually. Indeed, the strengths of a CNN (fast run-times, avoidance of unphysical solutions) complement the weaknesses of Bayesian inference approaches like \texttt{PyAutoLens}. It is conceivable that a CNN could replace \texttt{PyAutoLens}'s initial lens model fits altogether and allow the method to focus entirely on fitting more complex lens models with well characterised errors: a task better suited to \texttt{PyAutoLens}'s fully Bayesian approach than a CNN. At least, a CNN might be able to identify and isolate which lenses will eventually make the gold sample, and reduce manual intervention \cite{Maresca2020}. CNNs will also have an as-yet unquantified fraction of failures. If the lenses where a CNN fails are different to where traditional model-fitting approaches fail, combining the two may be key to maximising the success rate of lens model initialisation.

The second major challenge for automated lens modelling is deblending the foreground lens light. Within our sample, \texttt{PyAutoLens} could not deblend the lens and source light in 3/59 systems, and required visual inspection to recognise these bad fits. In these cases, we instead used b-spline fits that were created via a time-consuming manual process. This issue will be more prevalent in Euclid, owing to its lower spatial resolution than HST and lens samples with smaller Einstein radii \citep{Collett2015} --- both of which move the source's light closer to that of the lens. Furthermore, our analysis included pre-processing steps that manually removed the light of foreground stars and interloper galaxies via a GUI, a task which is overly time-intensive for an individual scientist to perform on larger samples of lenses. 

We propose two directions for future work that could improve automatic deblending. First, there are CNN architectures dedicated to the task of image deblending and segmentation (these architectures do not attempt to estimate the lens model parameters). These have been applied successfully on galaxy catalogues \citep{Burke2019, Hausen2020} and in studies of strong lenses \citep{Rojas2021}, with multi-wavelength imaging seen to improve debelending quality. Alternatively, this task seems well suited to citizen science \citep{Kung2015, Marshall2016, More2016}, whereby members of the public could use a GUI to mark-out regions of the data they believe correspond to the lens, source and other objects. The desired outputs of either approach are pixel-level masks describing where the source, lens and other objects are in the image data, which could be used for the automated removal or masking of contaminating light before lens modelling begins.

\subsection{Einstein radius measurements and uncertainty}\label{section: r ein method comp}

The statistical precision with which the Einstein radius can be measured is promising for many possible scientific studies. For example, \cite{Sonnenfeld2021}'s proposed method to constrain the population-level parameters of lens galaxies relies on being able to accurately measure the Einstein radii of the sample of galaxies. Previous studies have attempted to account for the very small formal statistical uncertainties on model parameters (in particular those inferred with parametric source methods) and associated systematic uncertainties by comparing the fractional difference of parameter estimates using different approaches. \citet{Bolton2008} and \citet{Sonnenfeld2013a} reported a typical expected systematic uncertainty on the Einstein radius of $\sim$2--3\%. This value of uncertainty is often adopted over (or combined with) those determined from the non-linear search. Furthermore, given the Einstein radius is expected to be a model-independent quantity \citep{Falco1985,Unruh2017,Cao2021}, it is typically assumed that this amount of uncertainty accounts for differences in the assumed parameterisation of the mass model.  

\subsubsection{Einstein radii compared to previous measurements}\label{R_ein comp}
\begin{figure}
\includegraphics[width=\columnwidth]{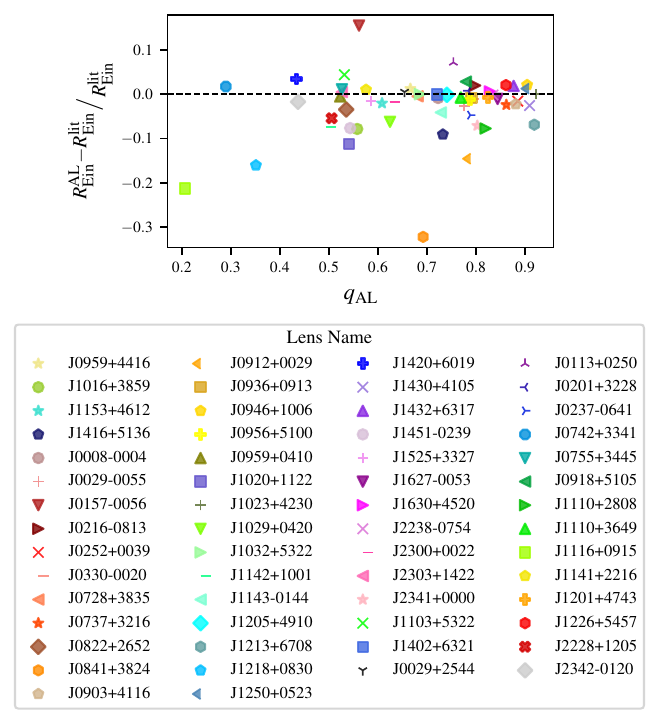}
\caption{The Einstein radii measured by \texttt{PyAutoLens} ($R_{\rm Ein}^{\rm AL}$) are generally consistent with those measured by previous analyses of the SLACS \citep{Bolton2008a} and GALLERY \citep{Shu2016a} lenses ($R_{\rm Ein}^{\rm lit}$). This shows the fractional difference between new and old measurements, as a function of PyAutoLens axis ratio, $q_{\rm AL}$.}
\label{Rein lit comp}
\end{figure}

In a similar fashion to \cite{Bolton2008} and \cite{Sonnenfeld2013b} we now compare our measurements of Einstein radii with those that exist in the literature (see Figure~\ref{Rein lit comp}) and estimate the uncertainty introduced as a result of the different methods. The full SLACS and GALLERY samples have previously been modelled with SIE profiles to measure the Einstein radii for supplementing a dynamical analysis of the lenses (SIE models of SLACS by \citealt{Bolton2008a} and SIE or SIE+shear models of GALLERY by \citealt{Shu2016a}). In this comparison, therefore, not only are the lensing methods very different, but we have also assumed the more complex PLEMD plus external shear (PL+ext) mass distribution for the lens galaxy. Compared to these previous measurements, we find Einstein radii with root mean square (RMS) fractional difference of 7.4\%. This is larger than the (empirically motivated) $\sim2$--$3$\% uncertainty that is typically assumed. 

Several differences between the methods could lead to variation between their inferred Einstein radii. \citet{Bolton2008a} and \citet{Shu2016a} model the background source using either a single or multiple S\'ersic ellipsoid components, and both choose different approaches to the lens light subtraction procedure than the one we adopt. While \citet{Bolton2008} and \citet{Sonnenfeld2013b} investigated differences like these, neither were concerned with differences in the assumed mass model. Indeed, for a subset of 14 lenses that were also analysed by \cite{Shajib2021} assuming a PL+ext model, the RMS fractional difference is only 1.6\%, it may be that the reduced uncertainty is a result of fitting the same mass model. Although, this is not of concern if the Einstein radius is indeed model-independent. For this \cite{Cao2021} provide good evidence, showing that the assumption of the PL+ext exhibits only $0.05\pm0.17$\% systematic error on the Einstein radius relative to complex ``MaNGA'' mock data.

Notably, though, we find that five of the six lenses whose $R_{\rm Ein}$ differ by over $10\%$ in the SLACS and GALLERY samples, are accompanied by extremely large values of external shear magnitude (ranging from 0.16 to 0.39) when fitted with our PL+ext models. If these high shear lenses are removed from the comparison, the RMS fractional difference drops to $4.2\%$. \cite{Cao2021} also demonstrated that the asymmetry in complex mass distributions can lead to the inference of spurious external shears. On average, they inferred an external shear magnitude of 0.015, despite the mock data being generated without external shear. In this work, we infer an average of 0.096 shear magnitude for the SLACS and GALLERY lenses. These large shear values may be partly a result of the additional complexity in the asymmetry of real lenses. \cite{Cao2021} required the Multiple Gaussian Expansion components, that represented the stellar mass, to share a common axis ratio and position angle --- this may not be a realistic representation of the angular structure of real lenses \citep{Nightingale2019}. Given that it is the lenses with high external shears that differ most from previous literature measurements of $R_{\rm Ein}$, we speculate that the assumption of a different mass model (in particular the assumption of external shear) may drive the larger fractional uncertainty. This would imply that the Einstein radius is less model-independent than is often assumed. Further work to test this hypothesis would be valuable. 

\subsubsection{Statistical uncertainty on Einstein radii}

We now consider the size of the errors we measure on the Einstein radius, based entirely on our own PL+ext models. Our likelihood cap method (Section~\ref{lh_boosts}) addressed the small formal statistical uncertainties on the mass model parameters and allows us to infer uncertainties that account for differences in possible noise realisations and the choice of data discretization. Moreover, since pixel-grid methods have the flexibility to reconstruct the source with as much complexity as the data needs, they are not subject to the overfitting that occurs in parametric source methods due to overly simplistic source assumptions. With this approach, we measure a mean uncertainty on the inferred Einstein radius of $\sim$1\%, albeit with a wide range of outliers, and 2/59 lens configurations exceeding 5\%. Adopting a uniform uncertainty could therefore be problematic for some statistical inferences. 

For example, determining the population level parameters of hundreds of thousands of lenses, as described by \cite{Sonnenfeld2021, Sonnenfeld2021a, Sonnenfeld2013b} might suffer from such inaccurate individual posteriors as those with up to 5\% uncertainty on the Einstein radius. The increase in the width of the posteriors inferred as a result of the likelihood cap approach demonstrated in this work should avoid biases in the population level parameters constrained in studies such as these. However, they will in turn increase the amount of lenses required to be able to make such a constraint. Moreover, the coverage probabilities of the lens model parameters with a likelihood cap (see Figure~\ref{Figure: coverage}) did not quite reach the expected level, potentially indicating an under confidence in the posterior. Under confidence in the posterior could lead to biases in estimates of the population parameters such as an overestimate in the scatter of the population \citep{Wagner-Carena2021}. We discuss the importance of further testing of the confidence of the individual posteriors further in Section~\ref{lens model testing}.

For comparison, \cite{Cao2021} inferred an average of 0.01\% statistical uncertainty on the Einstein radius when fitting to mock data simulated using ``MaNGA'' galaxies without the use of a likelihood cap. This order of magnitude difference from the uncertainties inferred in this study is likely a combination of the use of the likelihood cap increasing the errors in this work, and differences in the quality of the data. \cite{Cao2021}'s mock lensed sources are simulated with S/N of 50 and have visible extended arcs (or complete Einstein rings) that the lenses with the largest errors on $R_{\rm Ein}$ inferred in this work do not, often appearing closer to point-like. Furthermore, they do not include the lens galaxy's light, a component which we have shown in this study can hinder the lens model fitting procedure. 

\begin{figure}
\includegraphics[width=\columnwidth]{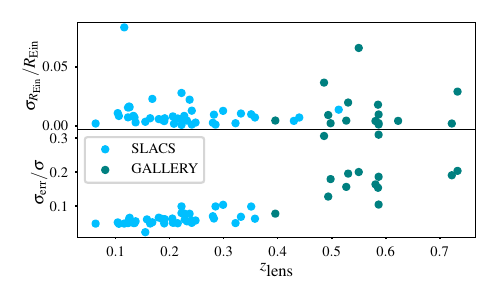}
\caption{The statistical uncertainty on a galaxy's total mass, when measured from its effective Einstein radius, does not degrade with lens redshift $z\lesssim0.7$ (top panel). This is in stark comparison to most astrophysical observables. For example, the uncertainty on a galaxy's total mass when measured from stellar dynamics (velocity dispersion) increases for more distant galaxies because of cosmological dimming and beam smearing (bottom panel).
}
\label{uncertainty dynamics comp}
\end{figure}

Based on the empirical relations we derived in Section~\ref{uncertainty correlations}, the certainty to which one can measure the Einstein radius is remarkably independent of a number of data properties and galaxy observables. For example, it might be expected that a higher S/N source galaxy image would tighten the constraints, however this does not appear to be the case for the Einstein radius measurement. This is encouraging for future surveys that will not achieve as high S/N as the HST data used in this study. 

The only parameter we investigate that exhibits a statistically significant correlation with measurement uncertainty on the Einstein radius is the Einstein radius itself. Measurements of the Einstein radius become less certain for small Einstein radii, and therefore low mass galaxies. This could also be relevant for surveys such as Euclid that are forecast to detect samples of lenses with smaller Einstein radii (typically $\sim$0.5\arcsec according to \citealt{Collett2015}). Interpolating from our empirical relationship, fitted to the sample excluding the two with anomalously large uncertainty, a lens with this Einstein radius should be measurable to $\sim2.1$\% accuracy. However, since the pixel-scale and PSF of the Euclid VIS instrument are roughly twice that of HST, this should be considered a lower limit.

\subsubsection{Implications for studies of galaxy evolution}

Notably, there does not appear to be a correlation between the lens redshift and measurement uncertainty on the Einstein radius. This highlights the power of strong lensing as a tool for investigations into galaxy evolution. If the lensing measurements do not degrade with redshift, then inferences of how galaxy properties evolve will be well constrained even to high redshift. This is in contrast to e.g.\ stellar dynamics data, where cosmological dimming effects reduce the certainty of the stellar velocity dispersion (and therefore dynamical mass) of distant galaxies. The increase in fractional uncertainty of the velocity dispersion, $\sigma_{\mathrm err} / \sigma$, within our SLACS and GALLERY samples is shown in Figure~\ref{uncertainty dynamics comp}. Within both samples $\sigma_{\rm err} / \sigma$ increases with redshift (the difference in fibre apertures used for SLACS and GALLERY means direct comparison of their errors is not straightforward, albeit they still highlight that in general higher redshift galaxy measurements are lower S/N). 

This creates an interesting dichotomy between using strong lensing to study galaxy evolution and other methods. In lensing, provided we are able to find the lenses at the highest redshifts (surveys such as Euclid and the Vera Rubin Observatory will observe lenses at redshift of up to $\sim2$ \citep{Collett2015}) we can anticipate that we will be able to measure their properties as well as those at lower redshifts. Issues that plague comparisons between the properties of low and high redshift galaxies via a technique like stellar dynamics, for example beam smearing \cite{Tiley2019a}, will therefore be less problematic. However, whilst comparing their properties may be more straightforward, strong lens samples will have complicated selection effects \cite{Sonnenfeld2022} that a carefully constructed dynamics sample can more easily mitigate. The reduced lensing efficiency of lower mass galaxies may also restrict the high redshift samples to only the most massive galaxies, albeit this is a limitation for most observing techniques. A strength of lensing, therefore, is that it offers a different means by which to study galaxy evolution that complements the strengths and weaknesses of other techniques.

\subsection{Measurements of other lens model parameters}\label{lens model parameters}

\begin{figure}
\includegraphics[width=\columnwidth]{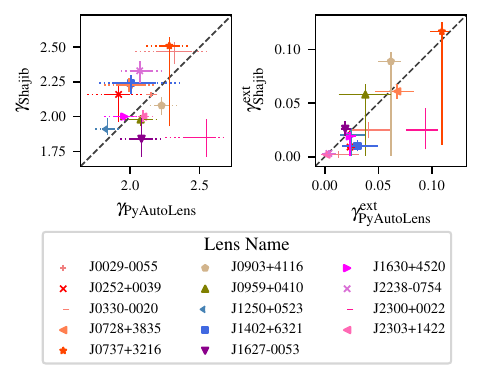}
\caption{Our measurements of the density profile slope (left) and the magnitude of external shear (right) in SLACS lenses, compared with previous, independent measurements by \citet{Shajib2021}.}
\label{slacs slope comp}
\end{figure}

In addition to the total mass enclosed within the Einstein radius, strong lensing information also constrains quantities like gradients of the distribution of mass, and the amount of external shear. This is captured in a model-dependent way via the parameters of our PL+ext mass model (see \citealt{Sonnenfeld2021} for a model-independent expression of this information). We shall now compare our measurements of radial density gradient $\gamma$ and shear magnitude $\gamma^{\rm ext}$, to measurements made using previous, independent analyses of overlapping sets of lenses. 

\citet{Shajib2021} modelled 23 SLACS lenses, including 14 in our sample. Like us, they used a uniform pipeline that simultaneously modelled the distribution of mass and light. They too described the lens galaxy's light as a double S\'ersic profile whose centres are aligned. However, unlike us, they fixed the S\'ersic index of each to values of $n=1$ and $n=4$ (the exponential and de Vaucouleurs profiles respectively) and join the axis ratios of the two profiles. A major difference in the two techniques lies in the source reconstruction; \citet{Shajib2021} reconstructed the source using a S\'ersic profile and shapelet basis functions. 

For all but one lens, \citet{Shajib2021} and our measurements of $\gamma$ and $\gamma^{\rm ext}$ are consistent (Figure~\ref{slacs slope comp}). For the discrepant lens J2300+0022, \texttt{PyAutoLens} infers $\gamma = 2.55$ and $\gamma^{\rm ext} = 0.08$, compared to \citet{Shajib2021}'s $\gamma = 1.85$ and $\gamma^{\rm ext} = 0.03$. We believe this discrepancy could be a result of the different order of operations in a model fit. \citet{Shajib2021} initialise their lens model assuming $\gamma^{\rm ext} = 0.0$ and relax this assumption once other components of the model are fit. In contrast, the first mass model we fit in our analysis assumes priors on the shear parameters that allow values up to $\gamma^{\rm ext} = 0.2$. Indeed, for J2300+0022 our search yields a best-fit shear of $\gamma^{\rm ext} = 0.07$. Discarding this lens, we find a mean difference of $-0.07\pm0.07$ between the slopes inferred by the two methods, where the error is propagated from the standard error on the means of the two samples. On average, \texttt{PyAutoLens} measures slightly shallower slopes than \cite{Shajib2021}, although this is not a significant difference -- the mean discrepancy for the sample is consistent with zero at the current uncertainty level. A larger sample of measurements may be able to discern if there are systematic differences introduced on the density slope as a result of the lensing technique. We note that we measure a scatter of 0.17 between the slope measurements suggesting there may be systematic uncertainty between the two methods.

\begin{figure}
\includegraphics[width=\columnwidth]{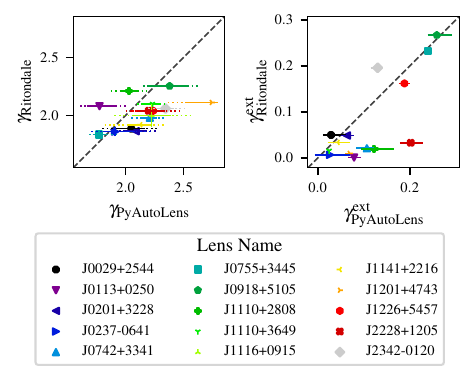}
\caption{
Our measurements of the density profile slope (left) and the magnitude of external shear (right) in BELLS GALLERY lenses, compared with previous, independent measurements by \citet{Ritondale2019}.
}
\label{gallery slope comp}
\end{figure}

\begin{figure}
\includegraphics[width=\columnwidth]{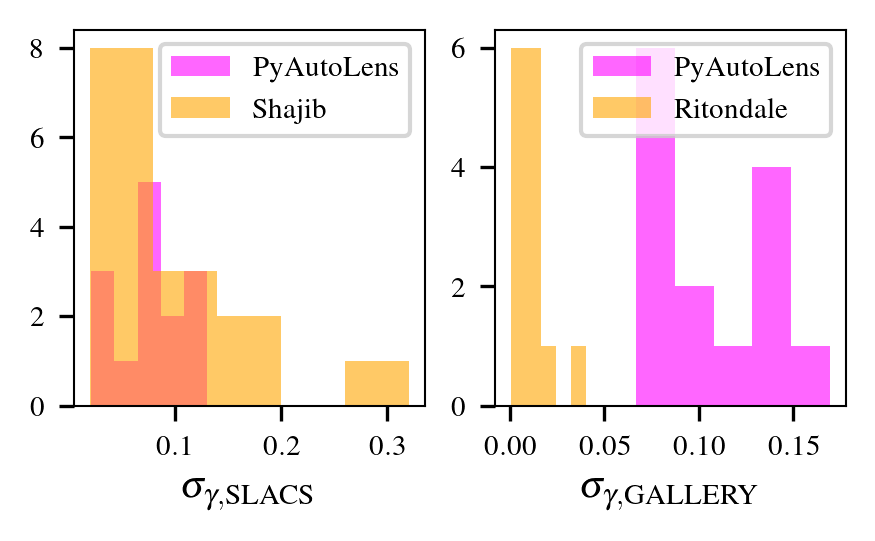}
\caption{The statistical uncertainty on measurements of the radial gradient of the total lens mass, reported by \texttt{PyAutoLens} are similar to those found by \citet{Shajib2021} for SLACS lenses (left). However, the uncertainty reported by \citet{Ritondale2019} for GALLERY lenses (right) is an order of magnitude smaller. That method uses a pixelised source, and may be subject to the source discretization bias that we discuss in Section~\ref{lh_boosts}.}
\label{pl err comp}
\end{figure}

\citet{Ritondale2019} modelled 17 GALLERY lenses, including 15 in our sample. Although they do not adopt a uniform analysis pipeline, their lens modelling technique more closely resembles ours, because they reconstruct the source galaxy using a pixelisation. On average, \texttt{PyAutoLens} measures a $0.13\pm0.21$ steeper density slope (Figure~\ref{gallery slope comp}). The scatter in this difference is comparable to the average uncertainty that we infer for the GALLERY lenses (0.11) but an order of magnitude larger than the average uncertainty inferred by \citet{Ritondale2019} (left panel of Figure~\ref{pl err comp}). In fact, the uncertainties inferred by \cite{Ritondale2019} more closely resemble those from \texttt{PyAutoLens} before we used a likelihood cap to avoid source discretization bias (Section~\ref{lh_boosts}). This suggests that discretization bias may also affect the pixelised-source method of \citet{Ritondale2019}. Conversely, the uncertainties derived by \citet{Shajib2021}, whose analytic approach to source reconstruction can not be affected by discretization bias, are similar to ours with the likelihood cap (right panel of Figure~\ref{pl err comp}).

It is reassuring that independent analyses yield results that are consistent in many ways. However, the relatively small number of lens systems in common to multiple analyses prevents much more detailed comparison between codes or modelling assumptions. The inconsistencies in other aspects of results highlights an urgent need for larger-scale tests.

\subsection{Large-scale tests of lens modelling}\label{lens model testing}

A vital but unintended consequence of this paper, is a solution to, and better understanding of the source discretization bias that previously caused parameter uncertainties to be underestimated. This occurred in both synthetic and real lenses, as a result of noise in the likelihood evaluations of methods using a pixelised source reconstruction (due to particular alignments of source pixels being arbitrarily more or less penalised by regularisation). Our likelihood cap solution successfully reduced noise and smoothed posterior PDFs. It increased the size of our uncertainties such that they had roughly the expected level of coverage, and improved the recovery of all parameters in our synthetic data. Although the likelihood cap was determined in an empirical way, the size of the inferred errors is inherently linked to this choice of likelihood cap. It may be that a different choice of likelihood cap could provide better coverage probabilities than the one we adopted. Further investigation would be warranted to understand at a deeper level what causes these spikes in likelihood in pixelised source reconstructions, as improvements may be possible by changing the approach to pixelising the source plane, or regularising the pixelised source.

Our work shows the importance of testing strong lens modelling methods on larger samples than previously attempted. Even our mock sample comprising 6 noise realisations of 59 lens configurations yields insufficient statistics to determine whether the inferred central values and statistical uncertainty on mass model parameters are consistent with the expectations of drawing each measurement from a normal distribution. Equally, whilst there is evidence for small systematic biases in the estimates of certain mass model parameters, we do not have enough unique lens configurations to determine the primary causes. Given that we are just a few years away from modelling samples of tens of thousands of lenses, tests of strong lens modelling methodology on synthetic data with complex mass distributions \citep[e.g.][]{Mukherjee2018,Enzi2020,Cao2022, He2022} must now scale up to ensure that error estimates are robust and systematic biases understood.

\subsection{Computational Costs}\label{CPU}

Every SLACS and GALLERY lens modelled in this work was analysed using a single 2x Intel Xeon Gold 5120 x @ 2.20GHz CPU, on the Distributed Research Utilising Advanced Computing (DiRAC) Data-Centric System on the COSMA7 machine at Durham University. Run times depend primarily on the number of image pixels fitted after masking, which due to the standard $3.5"$ circular mask used to fit most lenses is fixed. The lower resolution of SLACS lenses ($0.05$\,\arcsec\,pixel$^{-1}$) means they contain fewer image pixels than GALLERY lenses ($0.04$\,\arcsec\,pixel$^{-1}$) and the fits were therefore faster. For SLACS lenses, the source parametric pipeline takes between $10-24$ hours, the source inversion pipeline $10-36$ hours, the light pipeline $10-72$ hours and mass pipeline $6-48$ hours. GALLERY lenses take longer on average, where the source parametric pipeline takes between $10-36$ hours, the source inversion pipeline $10-48$ hours, the light pipeline $12-144$ hours and mass pipeline $12-72$ hours. The scatter in run times is due to many factors: lens galaxy S/N, source galaxy S/N, lens configuration, lens morphology, source morphology, etc.

Based on the longest GALLERY run times, the upper limit for the overall run time is $300$ CPU hours. For $100\,000$ strong lenses this would require $30\,000\,000$ CPU hours over the $5-10$ year lifetime of a survey like \textit{Euclid}, producing an upper limit of $\sim 6\,000\,000$ CPU hours per year. For the recent DiRAC resources allocation call, this amount of CPU time is a `small' project. We therefore anticipate that the analysis performed in this work will not be limited by CPU resources in the near future. Based on profiling of \texttt{PyAutoLens}, we anticipate the run time of a single lens will reduce by a factor of four or above when fitting lower resolution wide-field imaging data (e.g. the resolution of \textit{Euclid} data is $0.1$\,\arcsec\,pixel$^{-1}$). The $3.5"$ circular masks assumed throughout this work are also unnecessarily large for many lens systems, and reducing the mask size to $2.5"$ speeds up the analysis by factors of three and above.
\section{Summary}\label{Summary}

Tens of thousands of strong gravitational lenses will be imaged in the next few years, but current analysis techniques are labour-intensive. We use open source software \texttt{PyAutoLens} to develop a fully automated, Bayesian analysis of all 59 strong galaxy-galaxy lenses that have been observed by the Hubble Space Telescope (HST) under certain conditions. Adopting the open source software \texttt{PyAutoLens} provides an optimistic outlook for the future of automated analysis: for 54/59 lenses (92\%) we achieved successful model fits (determined via visual classification) with no human intervention. We illustrate why other fits initially went wrong, and present solutions that allowed us to infer accurate models for all 59/59 lenses (100\%) and recommend steps necessary for analysing the larger incoming samples. Notably, the difficulties primarily happen at the beginning of the analysis when attempting to determine an initial, approximate, lens model --- and often reflect confusion between light from the foreground lens and background source. Once a simple model is initialised, our pipeline worked flawlessly to automatically fit a sequence of more complex models that measure more detailed properties of the lens galaxy. We therefore discuss how combining a Convolutional Neural Network with a Bayesian approach like \texttt{PyAutoLens} could increase the automation success rate whilst extracting maximum physical information from each strong lens.

We also use synthetic observations of $\sim500$ lenses to explain and solve a problem common to pixel-based strong lensing methods that causes the statistical uncertainty on model parameters to be underestimated. This is fundamentally due to noise in likelihood evaluations, caused by discretization effects in pixelised reconstructions of the source galaxy. We implemented an empirical correction that `caps' the likelihood value to suppress noise. This significantly improves the measurement of the synthetic lens parameters, and leads to error estimates on different noise realisations of identical datasets that are more consistent with one another. On the real data we found this empirical correction (using the likelihood cap) gave a five fold average increase in the inferred uncertainties on model parameters. Comparing to previous literature results, we suggested this bias may be leading to uncertainty under estimation in other studies that use similar methods. Given the incoming samples of tens of thousands of strong gravitational lenses, we believe more detailed study of such systematics on larger mock samples is key.

Accurately knowing the systematic uncertainty on measurements of Einstein radius (total galaxy mass) will become vitally important for large samples of lenses, which beat down statistical uncertainty. Previous studies often assume a constant uncertainty of 2--3\%. We find substantial variation between lenses, with a mean of 1\% and 57/59 lenses with $<3\%$, but 2/59 lenses with $>5\%$. Future analysis of large samples, where careful control of systematics is paramount, must therefore adopt more rigorous errors. Our Einstein radii measurements assumed only a single type of parametric mass model and we do not investigate the degree of uncertainty that results from making different mass model assumptions. 

Notably, the uncertainty on our measurements of Einstein radii (and those of the lens models in general) do not increase with redshift. That is, we learn as much about the strong lenses at redshift $\sim 0.7$ as those at redshift $\sim 0.1$. This is in stark contrast to other astrophysical probes of a galaxy's structure (e.g.\ dynamics, morphology), where cosmological dimming effects and beam smearing degrade measurements of distant galaxies. Nor does uncertainty on Einstein radii depend strongly upon the signal-to-noise ratio of our data. This makes strong lensing a compelling technique to study galaxy evolution: once high redshift strong lenses are found, it should be straight forward to measure their properties. Of course, the technique has its own challenges, for example complicated selection effects, but it should nevertheless provide an invaluable tool for studies of galaxy evolution over the next decade.

\section*{Data Availability}

Text files and images of every model-fit performed in this work are available at \url{https://zenodo.org/record/6104823}. Full \texttt{dynesty} chains of every fit are available upon request.

\section*{Software Citations}

This work uses the following software packages:

\begin{itemize}

\item
\href{https://github.com/astropy/astropy}{{Astropy}}
\citep{astropy1, astropy2}

\item
\href{https://github.com/dfm/corner.py}{{Corner.py}}
\citep{corner}

\item
\href{https://github.com/joshspeagle/dynesty}{{Dynesty}}
\citep{dynesty}

\item
\href{https://github.com/matplotlib/matplotlib}{{Matplotlib}}
\citep{matplotlib}

\item
\href{numba` https://github.com/numba/numba}{{Numba}}
\citep{numba}

\item
\href{https://github.com/numpy/numpy}{{NumPy}}
\citep{numpy}

\item
\href{https://github.com/rhayes777/PyAutoFit}{{PyAutoFit}}
\citep{pyautofit}

\item
\href{https://github.com/Jammy2211/PyAutoLens}{{PyAutoLens}}
\citep{Nightingale2015, Nightingale2018, Nightingale2021}

\item
\href{https://www.python.org/}{{Python}}
\citep{python}

\item
\href{https://github.com/scikit-image/scikit-image}{{Scikit-image}}
\citep{scikit-image}

\item
\href{https://github.com/scikit-learn/scikit-learn}{{Scikit-learn}}
\citep{scikit-learn}

\item
\href{https://github.com/scipy/scipy}{{Scipy}}
\citep{scipy}

\item
\href{https://www.sqlite.org/index.html}{{SQLite}}
\citep{sqlite}

\end{itemize}

\section*{Acknowledgements}
We thank Yiping Shu for helping us undertake this study and providing helpful comments on this manuscript.

AE is supported by STFC via grants ST/R504725/1 and ST/T506047/1.  
JN and RM are supported by STFC via grant ST/T002565/1, and the UK Space Agency via grant ST/W002612/1. RL and XYC acknowledge support from the National Nature Science Foundation of China (Nos.\ 11988101, 11773032, 12022306), science research grants from the China Manned Space Project (Nos.\ CMS-CSST-2021-B01, CMS-CSST-2021-A01) and support from the K.C.Wong Education Foundation. AA, SMC, CSF and QH acknowledge support from the European Research Council (ERC) through Advanced Investigator grant DMIDAS (GA 786910). This work used both the Cambridge Service for Data Driven Discovery (CSD3) and the DiRAC Data-Centric system, which are operated by the University of Cambridge and Durham University on behalf of the STFC DiRAC HPC Facility (www.dirac.ac.uk). These were funded by BIS capital grant ST/K00042X/1, STFC capital grants ST/P002307/1, ST/R002452/1, ST/H008519/1, ST/K00087X/1, STFC Operations grants ST/K003267/1, ST/K003267/1, and Durham University. DiRAC is part of the UK National E-Infrastructure.

\bibliographystyle{mnras}
\typeout{}
\bibliography{references.bib, software.bib} 

\begin{thebibliography}{}
\makeatletter
\relax
\def\mn@urlcharsother{\let\do\@makeother \do\$\do\&\do\#\do\^\do\_\do\%\do\~}
\def\mn@doi{\begingroup\mn@urlcharsother \@ifnextchar [ {\mn@doi@}
  {\mn@doi@[]}}
\def\mn@doi@[#1]#2{\def\@tempa{#1}\ifx\@tempa\@empty \href
  {http://dx.doi.org/#2} {doi:#2}\else \href {http://dx.doi.org/#2} {#1}\fi
  \endgroup}
\def\mn@eprint#1#2{\mn@eprint@#1:#2::\@nil}
\def\mn@eprint@arXiv#1{\href {http://arxiv.org/abs/#1} {{\tt arXiv:#1}}}
\def\mn@eprint@dblp#1{\href {http://dblp.uni-trier.de/rec/bibtex/#1.xml}
  {dblp:#1}}
\def\mn@eprint@#1:#2:#3:#4\@nil{\def\@tempa {#1}\def\@tempb {#2}\def\@tempc
  {#3}\ifx \@tempc \@empty \let \@tempc \@tempb \let \@tempb \@tempa \fi \ifx
  \@tempb \@empty \def\@tempb {arXiv}\fi \@ifundefined
  {mn@eprint@\@tempb}{\@tempb:\@tempc}{\expandafter \expandafter \csname
  mn@eprint@\@tempb\endcsname \expandafter{\@tempc}}}

\bibitem[\protect\citeauthoryear{Ade et~al.,}{Ade et~al.}{2016}]{Ade2016}
Ade P.~A.,  et~al., 2016, \mn@doi [Astronomy and Astrophysics]
  {10.1051/0004-6361/201525830}, 594

\bibitem[\protect\citeauthoryear{Amorisco et~al.,}{Amorisco
  et~al.}{2022}]{Amorisco2021}
Amorisco N.~C.,  et~al., 2022, \mn@doi [Monthly Notices of the Royal
  Astronomical Society] {10.1093/mnras/stab3527}, 510, 2464

\bibitem[\protect\citeauthoryear{{Astropy Collaboration} et~al.,}{{Astropy
  Collaboration} et~al.}{2013}]{astropy1}
{Astropy Collaboration} et~al., 2013, \mn@doi [A\&A]
  {10.1051/0004-6361/201322068}, \href
  {http://adsabs.harvard.edu/abs/2013A%26A...558A..33A} {558, A33}

\bibitem[\protect\citeauthoryear{Auger, Treu, Bolton, Gavazzi, Koopmans,
  Marshall, Moustakas  \& Burles}{Auger et~al.}{2010}]{Auger2010}
Auger M.~W.,  Treu T.,  Bolton A.~S.,  Gavazzi R.,  Koopmans L.~V.,  Marshall
  P.~J.,  Moustakas L.~A.,   Burles S.,  2010, \mn@doi [Astrophysical Journal]
  {10.1088/0004-637X/724/1/511}, 724, 511

\bibitem[\protect\citeauthoryear{{Berg{\'e}}, {Massey}, {Baghi}  \&
  {Touboul}}{{Berg{\'e}} et~al.}{2019}]{Shexplets}
{Berg{\'e}} J.,  {Massey} R.,  {Baghi} Q.,   {Touboul} P.,  2019, \mn@doi
  [\mnras] {10.1093/mnras/stz787}, \href
  {https://ui.adsabs.harvard.edu/abs/2019MNRAS.486..544B} {486, 544}

\bibitem[\protect\citeauthoryear{Birrer, Amara  \& Refregier}{Birrer
  et~al.}{2015}]{Birrer2015}
Birrer S.,  Amara A.,   Refregier A.,  2015, \mn@doi [Astrophysical Journal]
  {10.1088/0004-637X/813/2/102}, 813

\bibitem[\protect\citeauthoryear{Birrer et~al.,}{Birrer
  et~al.}{2020}]{Birrer2020}
Birrer S.,  et~al., 2020, \mn@doi [Astronomy {\&} Astrophysics]
  {10.1051/0004-6361/202038861}, 1

\bibitem[\protect\citeauthoryear{Bolton, Burles, Koopmans, Treu  \&
  Moustakas}{Bolton et~al.}{2006}]{Bolton2006}
Bolton A.~S.,  Burles S.,  Koopmans L. V.~E.,  Treu T.,   Moustakas L.~A.,
  2006, \mn@doi [The Astrophysical Journal] {10.1086/498884}, 638, 703

\bibitem[\protect\citeauthoryear{Bolton, Burles, Koopmans, Treu, Gavazzi,
  Moustakas, Wayth  \& Schlegel}{Bolton et~al.}{2008a}]{Bolton2008a}
Bolton A.~S.,  Burles S.,  Koopmans L. V.~E.,  Treu T.,  Gavazzi R.,  Moustakas
  L.~A.,  Wayth R.,   Schlegel D.~J.,  2008a, \mn@doi [The Astrophysical
  Journal] {10.1086/589327}

\bibitem[\protect\citeauthoryear{Bolton, Treu, Koopmans, Gavazzi, Moustakas,
  Burles, Schlegel  \& Wayth}{Bolton et~al.}{2008b}]{Bolton2008}
Bolton A.~S.,  Treu T.,  Koopmans L. V.~E.,  Gavazzi R.,  Moustakas L.~A.,
  Burles S.,  Schlegel D.~J.,   Wayth R.,  2008b, \mn@doi [The Astrophysical
  Journal] {10.1086/589989}, 684, 248

\bibitem[\protect\citeauthoryear{Bolton et~al.,}{Bolton
  et~al.}{2012}]{Bolton2012}
Bolton A.~S.,  et~al., 2012, \mn@doi [Astrophysical Journal]
  {10.1088/0004-637X/757/1/82}, 757

\bibitem[\protect\citeauthoryear{Brownstein et~al.,}{Brownstein
  et~al.}{2012}]{Brownstein2012}
Brownstein J.~R.,  et~al., 2012, \mn@doi [Astrophysical Journal]
  {10.1088/0004-637X/744/1/41}, 744

\bibitem[\protect\citeauthoryear{Burke, Aleo, Chen, Liu, Peterson, Sembroski
  \& Lin}{Burke et~al.}{2019}]{Burke2019}
Burke C.~J.,  Aleo P.~D.,  Chen Y.~C.,  Liu X.,  Peterson J.~R.,  Sembroski
  G.~H.,   Lin J. Y.~Y.,  2019, \mn@doi [Monthly Notices of the Royal
  Astronomical Society] {10.1093/mnras/stz2845}, 490, 3952

\bibitem[\protect\citeauthoryear{Cao, Li, Shu, Mao, Kneib  \& Gao}{Cao
  et~al.}{2020}]{Cao2021}
Cao X.,  Li R.,  Shu Y.,  Mao S.,  Kneib J.~P.,   Gao L.,  2020, \mn@doi
  [Monthly Notices of the Royal Astronomical Society] {10.1093/mnras/staa3058},
  499, 3610

\bibitem[\protect\citeauthoryear{Cao et~al.,}{Cao et~al.}{2022}]{Cao2022}
Cao X.,  et~al., 2022, \mn@doi [Research in Astronomy and Astrophysics]
  {10.1088/1674-4527/ac3f2b}, 22

\bibitem[\protect\citeauthoryear{Collett}{Collett}{2015}]{Collett2015}
Collett T.~E.,  2015, \mn@doi [Astrophysical Journal]
  {10.1088/0004-637X/811/1/20}, 811, 20

\bibitem[\protect\citeauthoryear{Collett \& Auger}{Collett \&
  Auger}{2014}]{Collett2014}
Collett T.~E.,  Auger M.~W.,  2014, \mn@doi [Monthly Notices of the Royal
  Astronomical Society] {10.1093/mnras/stu1190}, 443, 969

\bibitem[\protect\citeauthoryear{Collett \& Bacon}{Collett \&
  Bacon}{2016}]{Collett2016a}
Collett T.~E.,  Bacon D.~J.,  2016, \mn@doi [Monthly Notices of the Royal
  Astronomical Society] {10.1093/mnras/stv2791}, 456, 2210

\bibitem[\protect\citeauthoryear{Collett \& Smith}{Collett \&
  Smith}{2020}]{Collett2020}
Collett T.~E.,  Smith R.~J.,  2020, \mn@doi [Monthly Notices of the Royal
  Astronomical Society] {10.1093/mnras/staa1804}, 497, 1654

\bibitem[\protect\citeauthoryear{Despali, Lovell, Vegetti, Crain  \&
  Oppenheimer}{Despali et~al.}{2019}]{Despali2019}
Despali G.,  Lovell M.,  Vegetti S.,  Crain R.~A.,   Oppenheimer B.~D.,  2019,
  \mn@doi [Monthly Notices of the Royal Astronomical Society]
  {10.1093/mnras/stz3068}, 17, 1

\bibitem[\protect\citeauthoryear{Dye \& Warren}{Dye \& Warren}{2005}]{Dye2005}
Dye S.,  Warren S.~J.,  2005, \mn@doi [The Astrophysical Journal]
  {10.1086/428340}, 623, 31

\bibitem[\protect\citeauthoryear{Dye, Evans, Belokurov, Warren  \& Hewett}{Dye
  et~al.}{2008}]{Dye2008}
Dye S.,  Evans N.~W.,  Belokurov V.,  Warren S.~J.,   Hewett P.,  2008, \mn@doi
  [Monthly Notices of the Royal Astronomical Society]
  {10.1111/j.1365-2966.2008.13401.x}, 388, 384

\bibitem[\protect\citeauthoryear{{E. E. Falco, M. V. Gorenstein} \&
  {I.I.Shapiro}}{{E. E. Falco, M. V. Gorenstein} \&
  {I.I.Shapiro}}{1985}]{Falco1985}
{E. E. Falco, M. V. Gorenstein} {I.I.Shapiro} 1985, American Astronomical
  Society, 21, 289

\bibitem[\protect\citeauthoryear{{Enzi}, {Vegetti}, {Despali}, {Hsueh}  \&
  {Metcalf}}{{Enzi} et~al.}{2020}]{Enzi2020}
{Enzi} W.,  {Vegetti} S.,  {Despali} G.,  {Hsueh} J.-W.,   {Metcalf} R.~B.,
  2020, \mn@doi [\mnras] {10.1093/mnras/staa1224}, \href
  {https://ui.adsabs.harvard.edu/abs/2020MNRAS.496.1718E} {496, 1718}

\bibitem[\protect\citeauthoryear{Foreman-Mackey}{Foreman-Mackey}{2016}]{corner}
Foreman-Mackey D.,  2016, \mn@doi [The J. Open Source Softw.]
  {10.21105/joss.00024}, 1, 24

\bibitem[\protect\citeauthoryear{Galan, Peel, Joseph, Courbin  \& Starck}{Galan
  et~al.}{2021}]{Galan2021}
Galan A.,  Peel A.,  Joseph R.,  Courbin F.,   Starck J.~L.,  2021, \mn@doi
  [Astronomy and Astrophysics] {10.1051/0004-6361/202039363}, 647

\bibitem[\protect\citeauthoryear{Gavazzi, Treu, Rhodes, Koopmans, Bolton,
  Burles, Massey  \& Moustakas}{Gavazzi et~al.}{2007}]{Gavazzi2007}
Gavazzi R.,  Treu T.,  Rhodes J.~D.,  Koopmans L. V.~E.,  Bolton A.~S.,  Burles
  S.,  Massey R.~J.,   Moustakas L.~A.,  2007, \mn@doi [The Astrophysical
  Journal] {10.1086/519237}, 667, 176

\bibitem[\protect\citeauthoryear{Gavazzi, Treu, Marshall, Brault  \&
  Ruff}{Gavazzi et~al.}{2012}]{Gavazzi2012}
Gavazzi R.,  Treu T.,  Marshall P.~J.,  Brault F.,   Ruff A.,  2012, \mn@doi
  [Astrophysical Journal] {10.1088/0004-637X/761/2/170}, 761

\bibitem[\protect\citeauthoryear{Graham \& Driver}{Graham \&
  Driver}{2005}]{Graham2005}
Graham A.~W.,  Driver S.~P.,  2005, \mn@doi [Publications of the Astronomical
  Society of Australia] {10.1071/AS05001}, 22, 118

\bibitem[\protect\citeauthoryear{Hausen \& Robertson}{Hausen \&
  Robertson}{2020}]{Hausen2020}
Hausen R.,  Robertson B.~E.,  2020, \mn@doi [The Astrophysical Journal
  Supplement Series] {10.3847/1538-4365/ab8868}, 248, 20

\bibitem[\protect\citeauthoryear{He et~al.,}{He et~al.}{2021}]{He2021}
He Q.,  et~al., 2021, \mn@doi [Monthly Notices of the Royal Astronomical
  Society] {10.1093/mnras/stac759}, 512, 5862

\bibitem[\protect\citeauthoryear{He et~al.,}{He et~al.}{2022}]{He2022}
He Q.,  et~al., 2022, 18, 1

\bibitem[\protect\citeauthoryear{Hezaveh et~al.,}{Hezaveh
  et~al.}{2016}]{Hezaveh2016}
Hezaveh Y.~D.,  et~al., 2016, \mn@doi [The Astrophysical Journal]
  {10.3847/0004-637x/823/1/37}, 823, 37

\bibitem[\protect\citeauthoryear{Hezaveh, Levasseur  \& Marshall}{Hezaveh
  et~al.}{2017}]{Hezaveh2017}
Hezaveh Y.~D.,  Levasseur L.~P.,   Marshall P.~J.,  2017, \mn@doi [Nature]
  {10.1038/nature23463}, 548, 555

\bibitem[\protect\citeauthoryear{Hipp}{Hipp}{2020}]{sqlite}
Hipp R.~D.,  2020, {SQLite}, \url {https://www.sqlite.org/index.html}

\bibitem[\protect\citeauthoryear{Hunter}{Hunter}{2007}]{matplotlib}
Hunter J.~D.,  2007, \mn@doi [Comput Sci Eng] {10.1109/MCSE.2007.55}, 9, 90

\bibitem[\protect\citeauthoryear{Joseph, Courbin, Starck  \& Birrer}{Joseph
  et~al.}{2019}]{Joseph2019}
Joseph R.,  Courbin F.,  Starck J.~L.,   Birrer S.,  2019, \mn@doi [Astronomy
  and Astrophysics] {10.1051/0004-6361/201731042}, 623

\bibitem[\protect\citeauthoryear{Koopmans et~al.,}{Koopmans
  et~al.}{2009}]{Koopmans2009}
Koopmans L.~V.,  et~al., 2009, \mn@doi [Astrophysical Journal]
  {10.1088/0004-637X/703/1/L51}, 703

\bibitem[\protect\citeauthoryear{Krist}{Krist}{1993}]{Krist1993}
Krist J.,  1993, ASP Conference Series, 52, 536

\bibitem[\protect\citeauthoryear{K{\"{u}}ng et~al.,}{K{\"{u}}ng
  et~al.}{2015}]{Kung2015}
K{\"{u}}ng R.,  et~al., 2015, \mn@doi [Monthly Notices of the Royal
  Astronomical Society] {10.1093/mnras/stu2554}, 447, 2170

\bibitem[\protect\citeauthoryear{Lam, Pitrou  \& Seibert}{Lam
  et~al.}{2015}]{numba}
Lam S.~K.,  Pitrou A.,   Seibert S.,  2015, \mn@doi [Proceedings of the Second
  Workshop on the LLVM Compiler Infrastructure in HPC - LLVM '15]
  {10.1145/2833157.2833162}, pp~1--6

\bibitem[\protect\citeauthoryear{Levasseur, Hezaveh  \& Wechsler}{Levasseur
  et~al.}{2017}]{Levasseur2017}
Levasseur L.~P.,  Hezaveh Y.~D.,   Wechsler R.~H.,  2017, \mn@doi
  [Astrophysical Journal Letters] {10.3847/2041-8213/aa9704}, 850

\bibitem[\protect\citeauthoryear{Li, Frenk, Cole, Gao, Bose  \& Hellwing}{Li
  et~al.}{2016}]{Li2016}
Li R.,  Frenk C.~S.,  Cole S.,  Gao L.,  Bose S.,   Hellwing W.~A.,  2016,
  \mn@doi [Monthly Notices of the Royal Astronomical Society]
  {10.1093/mnras/stw939}, 460, 363

\bibitem[\protect\citeauthoryear{Li, Frenk, Cole, Wang  \& Gao}{Li
  et~al.}{2017}]{Li2017}
Li R.,  Frenk C.~S.,  Cole S.,  Wang Q.,   Gao L.,  2017, \mn@doi [Monthly
  Notices of the Royal Astronomical Society] {10.1093/mnras/stx554}, 468, 1426

\bibitem[\protect\citeauthoryear{Maresca, Dye  \& Li}{Maresca
  et~al.}{2020}]{Maresca2020}
Maresca J.,  Dye S.,   Li N.,  2020, \mn@doi [Monthly Notices of the Royal
  Astronomical Society] {10.1093/mnras/stab387}, 13, 1

\bibitem[\protect\citeauthoryear{Marshall et~al.,}{Marshall
  et~al.}{2016}]{Marshall2016}
Marshall P.~J.,  et~al., 2016, \mn@doi [Monthly Notices of the Royal
  Astronomical Society] {10.1093/mnras/stv2009}, 455, 1171

\bibitem[\protect\citeauthoryear{{Massey} \& {Refregier}}{{Massey} \&
  {Refregier}}{2005}]{Shapelets}
{Massey} R.,  {Refregier} A.,  2005, \mn@doi [\mnras]
  {10.1111/j.1365-2966.2005.09453.x}, \href
  {https://ui.adsabs.harvard.edu/abs/2005MNRAS.363..197M} {363, 197}

\bibitem[\protect\citeauthoryear{Meneghetti, Bartelmann, Dahle  \&
  Limousin}{Meneghetti et~al.}{2013}]{Meneghetti2013}
Meneghetti M.,  Bartelmann M.,  Dahle H.,   Limousin M.,  2013, \mn@doi [Space
  Science Reviews] {10.1007/s11214-013-9981-x}, 177, 31

\bibitem[\protect\citeauthoryear{More et~al.,}{More et~al.}{2016}]{More2016}
More A.,  et~al., 2016, \mn@doi [Monthly Notices of the Royal Astronomical
  Society] {10.1093/mnras/stv1965}, 455, 1191

\bibitem[\protect\citeauthoryear{Morningstar, Hezaveh, Levasseur, Blandford,
  Marshall, Putzky  \& Wechsler}{Morningstar et~al.}{2018}]{Morningstar2018a}
Morningstar W.~R.,  Hezaveh Y.~D.,  Levasseur L.~P.,  Blandford R.~D.,
  Marshall P.~J.,  Putzky P.,   Wechsler R.~H.,  2018, \mn@doi [arXiv]
  {10.48550/arXiv.1808.00011}, pp~1--9

\bibitem[\protect\citeauthoryear{Mukherjee et~al.,}{Mukherjee
  et~al.}{2018}]{Mukherjee2018}
Mukherjee S.,  et~al., 2018, \mn@doi [Monthly Notices of the Royal Astronomical
  Society] {10.1093/mnras/sty1741}, 479, 4108

\bibitem[\protect\citeauthoryear{Nightingale \& Dye}{Nightingale \&
  Dye}{2015}]{Nightingale2015}
Nightingale J.~W.,  Dye S.,  2015, \mn@doi [Monthly Notices of the Royal
  Astronomical Society] {10.1093/mnras/stv1455}, 452, 2940

\bibitem[\protect\citeauthoryear{Nightingale, Dye  \& Massey}{Nightingale
  et~al.}{2018}]{Nightingale2018}
Nightingale J.,  Dye S.,   Massey R.,  2018, \mn@doi [Monthly Notices of the
  Royal Astronomical Society] {10.1093/mnras/sty1264}, 47, 1

\bibitem[\protect\citeauthoryear{Nightingale, Massey, Harvey, Cooper,
  Etherington, Tam  \& Hayes}{Nightingale et~al.}{2019}]{Nightingale2019}
Nightingale J.~W.,  Massey R.~J.,  Harvey D.~R.,  Cooper A.~P.,  Etherington
  A.,  Tam S.-I.,   Hayes R.~G.,  2019, \mn@doi [Monthly Notices of the Royal
  Astronomical Society] {10.1093/mnras/stz2220}

\bibitem[\protect\citeauthoryear{Nightingale, Hayes  \& Griffiths}{Nightingale
  et~al.}{2021a}]{pyautofit}
Nightingale J.~W.,  Hayes R.~G.,   Griffiths M.,  2021a, \mn@doi [J. Open
  Source Softw.] {10.21105/joss.02550}, 6, 2550

\bibitem[\protect\citeauthoryear{Nightingale et~al.,}{Nightingale
  et~al.}{2021b}]{Nightingale2021}
Nightingale J.,  et~al., 2021b, \mn@doi [Journal of Open Source Software]
  {10.21105/joss.02825}, 6, 2825

\bibitem[\protect\citeauthoryear{Oh, Greene  \& Lackner}{Oh
  et~al.}{2017}]{Oh2017}
Oh S.,  Greene J.~E.,   Lackner C.~N.,  2017, \mn@doi [The Astrophysical
  Journal] {10.3847/1538-4357/836/1/115}, 836, 115

\bibitem[\protect\citeauthoryear{Oldham \& Auger}{Oldham \&
  Auger}{2018}]{Oldham2018}
Oldham L.~J.,  Auger M.~W.,  2018, \mn@doi [Monthly Notices of the Royal
  Astronomical Society] {10.1093/mnras/sty065}, 476, 133

\bibitem[\protect\citeauthoryear{Orban De~Xivry \& Marshall}{Orban De~Xivry \&
  Marshall}{2009}]{OrbanDeXivry2009}
Orban De~Xivry G.,  Marshall P.,  2009, \mn@doi [Monthly Notices of the Royal
  Astronomical Society] {10.1111/j.1365-2966.2009.14925.x}, 399, 2

\bibitem[\protect\citeauthoryear{Pearson, Maresca, Li  \& Dye}{Pearson
  et~al.}{2021}]{Pearson2021}
Pearson J.,  Maresca J.,  Li N.,   Dye S.,  2021, \mn@doi [Monthly Notices of
  the Royal Astronomical Society] {10.1093/mnras/stab1547}, 505, 4362–4382

\bibitem[\protect\citeauthoryear{Pedregosa et~al.,}{Pedregosa
  et~al.}{2011}]{scikit-learn}
Pedregosa F.,  et~al., 2011, Journal of Machine Learning Research, 12, 2825

\bibitem[\protect\citeauthoryear{{Price-Whelan} et~al.,}{{Price-Whelan}
  et~al.}{2018}]{astropy2}
{Price-Whelan} A.~M.,  et~al., 2018, \mn@doi [AJ] {10.3847/1538-3881/aabc4f},
  \href {https://ui.adsabs.harvard.edu/#abs/2018AJ....156..123T} {156, 123}

\bibitem[\protect\citeauthoryear{Ritondale, Vegetti, Despali, Auger, Koopmans
  \& McKean}{Ritondale et~al.}{2019}]{Ritondale2019}
Ritondale E.,  Vegetti S.,  Despali G.,  Auger M.~W.,  Koopmans L.~V.,   McKean
  J.~P.,  2019, \mn@doi [Monthly Notices of the Royal Astronomical Society]
  {10.1093/mnras/stz464}, 485, 2179

\bibitem[\protect\citeauthoryear{Rojas et~al.,}{Rojas et~al.}{2021}]{Rojas2021}
Rojas K.,  et~al., 2021, \mn@doi [arXiv] {10.48550/arXiv.2109.00014}, pp 1--37

\bibitem[\protect\citeauthoryear{Shajib, Treu, Birrer  \& Sonnenfeld}{Shajib
  et~al.}{2021}]{Shajib2021}
Shajib A.~J.,  Treu T.,  Birrer S.,   Sonnenfeld A.,  2021, \mn@doi [Monthly
  Notices of the Royal Astronomical Society] {10.1093/mnras/stab536}, 503, 2380

\bibitem[\protect\citeauthoryear{Shu et~al.,}{Shu et~al.}{2015}]{Shu2015}
Shu Y.,  et~al., 2015, \mn@doi [Astrophysical Journal]
  {10.1088/0004-637X/803/2/71}, 803, 1

\bibitem[\protect\citeauthoryear{Shu, Bolton, Moustakas, Stern, Dey,
  Brownstein, Burles  \& Spinrad}{Shu et~al.}{2016a}]{Shu2016c}
Shu Y.,  Bolton A.~S.,  Moustakas L.~A.,  Stern D.,  Dey A.,  Brownstein J.~R.,
   Burles S.,   Spinrad H.,  2016a, \mn@doi [The Astrophysical Journal]
  {10.3847/0004-637x/820/1/43}, 820, 43

\bibitem[\protect\citeauthoryear{Shu et~al.,}{Shu et~al.}{2016b}]{Shu2016b}
Shu Y.,  et~al., 2016b, \mn@doi [The Astrophysical Journal]
  {10.3847/0004-637x/824/2/86}, 824, 86

\bibitem[\protect\citeauthoryear{Shu et~al.,}{Shu et~al.}{2016c}]{Shu2016a}
Shu Y.,  et~al., 2016c, \mn@doi [The Astrophysical Journal]
  {10.3847/1538-4357/833/2/264}, 833, 264

\bibitem[\protect\citeauthoryear{Shu et~al.,}{Shu et~al.}{2017}]{Shu2017}
Shu Y.,  et~al., 2017, \mn@doi [Astrophysical Journal]
  {10.3847/1538-4357/aa9794}

\bibitem[\protect\citeauthoryear{Sonnenfeld}{Sonnenfeld}{2021}]{Sonnenfeld2021a}
Sonnenfeld A.,  2021, Astronomy {\&} Astrophysics, pp 1--11

\bibitem[\protect\citeauthoryear{Sonnenfeld}{Sonnenfeld}{2022}]{Sonnenfeld2022}
Sonnenfeld A.,  2022, \mn@doi [Astronomy {\&} Astrophysics]
  {10.1051/0004-6361/202142467}, pp~1--9

\bibitem[\protect\citeauthoryear{Sonnenfeld \& Cautun}{Sonnenfeld \&
  Cautun}{2021}]{Sonnenfeld2021}
Sonnenfeld A.,  Cautun M.,  2021, Astronomy {\&} Astrophysics

\bibitem[\protect\citeauthoryear{Sonnenfeld, Treu, Gavazzi, Marshall, Auger,
  Suyu, Koopmans  \& Bolton}{Sonnenfeld et~al.}{2012}]{Sonnenfeld2012}
Sonnenfeld A.,  Treu T.,  Gavazzi R.,  Marshall P.~J.,  Auger M.~W.,  Suyu
  S.~H.,  Koopmans L.~V.,   Bolton A.~S.,  2012, \mn@doi [Astrophysical
  Journal] {10.1088/0004-637X/752/2/163}, 752

\bibitem[\protect\citeauthoryear{Sonnenfeld, Gavazzi, Suyu, Treu  \&
  Marshall}{Sonnenfeld et~al.}{2013b}]{Sonnenfeld2013a}
Sonnenfeld A.,  Gavazzi R.,  Suyu S.~H.,  Treu T.,   Marshall P.~J.,  2013b,
  \mn@doi [Astrophysical Journal] {10.1088/0004-637X/777/2/97}, 777

\bibitem[\protect\citeauthoryear{Sonnenfeld, Treu, Gavazzi, Suyu, Marshall,
  Auger  \& Nipoti}{Sonnenfeld et~al.}{2013a}]{Sonnenfeld2013b}
Sonnenfeld A.,  Treu T.,  Gavazzi R.,  Suyu S.~H.,  Marshall P.~J.,  Auger
  M.~W.,   Nipoti C.,  2013a, \mn@doi [Astrophysical Journal]
  {10.1088/0004-637X/777/2/98}, 777

\bibitem[\protect\citeauthoryear{Speagle}{Speagle}{2020a}]{Speagle2020}
Speagle J.~S.,  2020a, \mn@doi [Monthly Notices of the Royal Astronomical
  Society] {10.1093/mnras/staa278}, 493, 3132

\bibitem[\protect\citeauthoryear{Speagle}{Speagle}{2020b}]{dynesty}
Speagle J.~S.,  2020b, \mn@doi [MNRAS] {10.1093/mnras/staa278}, 493, 3132

\bibitem[\protect\citeauthoryear{Suyu}{Suyu}{2012}]{Suyu2012}
Suyu S.~H.,  2012, \mn@doi [Monthly Notices of the Royal Astronomical Society]
  {10.1111/j.1365-2966.2012.21661.x}, 426, 868

\bibitem[\protect\citeauthoryear{Suyu, Marshall, Hobson  \& Blandford}{Suyu
  et~al.}{2006}]{Suyu2006}
Suyu S.~H.,  Marshall P.~J.,  Hobson M.~P.,   Blandford R.~D.,  2006, \mn@doi
  [Monthly Notices of the Royal Astronomical Society]
  {10.1111/j.1365-2966.2006.10733.x}, 371, 983

\bibitem[\protect\citeauthoryear{Suyu et~al.,}{Suyu et~al.}{2017}]{Suyu2017}
Suyu S.~H.,  et~al., 2017, \mn@doi [Monthly Notices of the Royal Astronomical
  Society] {10.1093/mnras/stx483}, 468, 2590

\bibitem[\protect\citeauthoryear{Tagore \& Jackson}{Tagore \&
  Jackson}{2016}]{Tagore2016}
Tagore A.~S.,  Jackson N.,  2016, \mn@doi [Monthly Notices of the Royal
  Astronomical Society] {10.1093/mnras/stw057}, 457, 3066

\bibitem[\protect\citeauthoryear{Talbot et~al.,}{Talbot
  et~al.}{2018}]{Talbot2018}
Talbot M.~S.,  et~al., 2018, \mn@doi [Monthly Notices of the Royal Astronomical
  Society] {10.1093/mnras/sty653}, 477, 195

\bibitem[\protect\citeauthoryear{Talbot, Brownstein, Dawson, Kneib  \&
  Bautista}{Talbot et~al.}{2021}]{Talbot2021}
Talbot M.~S.,  Brownstein J.~R.,  Dawson K.~S.,  Kneib J.~P.,   Bautista J.,
  2021, \mn@doi [Monthly Notices of the Royal Astronomical Society]
  {10.1093/mnras/stab267}, 502, 4617

\bibitem[\protect\citeauthoryear{Tessore, Bellagamba  \& Metcalf}{Tessore
  et~al.}{2016}]{Tessore2016a}
Tessore N.,  Bellagamba F.,   Metcalf R.~B.,  2016, \mn@doi [Monthly Notices of
  the Royal Astronomical Society] {10.1093/mnras/stw2212}, 463, 3115

\bibitem[\protect\citeauthoryear{Tiley et~al.,}{Tiley
  et~al.}{2019}]{Tiley2019a}
Tiley A.~L.,  et~al., 2019, \mn@doi [Monthly Notices of the Royal Astronomical
  Society] {10.1093/mnras/stz428}, 485, 834

\bibitem[\protect\citeauthoryear{Unruh, Schneider  \& Sluse}{Unruh
  et~al.}{2017}]{Unruh2017}
Unruh S.,  Schneider P.,   Sluse D.,  2017, \mn@doi [Astronomy and
  Astrophysics] {10.1051/0004-6361/201629048}, 601, 1

\bibitem[\protect\citeauthoryear{Van~Rossum \& Drake}{Van~Rossum \&
  Drake}{2009}]{python}
Van~Rossum G.,  Drake F.~L.,  2009, Python 3 Reference Manual.
CreateSpace, Scotts Valley, CA

\bibitem[\protect\citeauthoryear{Van~der Walt, Sch{\"o}nberger, Nunez-Iglesias,
  Boulogne, Warner, Yager, Gouillart  \& Yu}{Van~der Walt
  et~al.}{2014}]{scikit-image}
Van~der Walt S.,  Sch{\"o}nberger J.~L.,  Nunez-Iglesias J.,  Boulogne F.,
  Warner J.~D.,  Yager N.,  Gouillart E.,   Yu T.,  2014, PeerJ, 2, e453

\bibitem[\protect\citeauthoryear{Vegetti \& Koopmans}{Vegetti \&
  Koopmans}{2009}]{Vegetti2009}
Vegetti S.,  Koopmans L.~V.,  2009, \mn@doi [Monthly Notices of the Royal
  Astronomical Society] {10.1111/j.1365-2966.2009.15559.x}, 400, 1583

\bibitem[\protect\citeauthoryear{Vegetti, Koopmans, Bolton, Treu  \&
  Gavazzi}{Vegetti et~al.}{2010}]{Vegetti2010}
Vegetti S.,  Koopmans L.~V.,  Bolton A.,  Treu T.,   Gavazzi R.,  2010, \mn@doi
  [Monthly Notices of the Royal Astronomical Society]
  {10.1111/j.1365-2966.2010.16865.x}, 408, 1969

\bibitem[\protect\citeauthoryear{Vika, Bamford, H{\"{a}}u{\ss}ler  \&
  Rojas}{Vika et~al.}{2014}]{Vika2014}
Vika M.,  Bamford S.~P.,  H{\"{a}}u{\ss}ler B.,   Rojas A.~L.,  2014, \mn@doi
  [Monthly Notices of the Royal Astronomical Society] {10.1093/mnras/stu1696},
  444, 3603

\bibitem[\protect\citeauthoryear{{Virtanen} et~al.,}{{Virtanen}
  et~al.}{2020}]{scipy}
{Virtanen} P.,  et~al., 2020, \mn@doi [Nature Methods]
  {10.1038/s41592-019-0686-2}, \href {https://rdcu.be/b08Wh} {17, 261}

\bibitem[\protect\citeauthoryear{Wagner-Carena, Park, Birrer, Marshall, Roodman
   \& Wechsler}{Wagner-Carena et~al.}{2021}]{Wagner-Carena2021}
Wagner-Carena S.,  Park J.~W.,  Birrer S.,  Marshall P.~J.,  Roodman A.,
  Wechsler R.~H.,  2021, \mn@doi [The Astrophysical Journal]
  {10.3847/1538-4357/abdf59}, 909, 187

\bibitem[\protect\citeauthoryear{Warren \& Dye}{Warren \&
  Dye}{2003}]{Warren2003}
Warren S.~J.,  Dye S.,  2003, \mn@doi [The Astrophysical Journal]
  {10.1086/375132}, 590, 673

\bibitem[\protect\citeauthoryear{Wong et~al.,}{Wong et~al.}{2019}]{Wong2019}
Wong K.~C.,  et~al., 2019, \mn@doi [Monthly Notices of the Royal Astronomical
  Society] {10.1093/mnras/stz3094}, 498, 1420–1439

\bibitem[\protect\citeauthoryear{{van der Walt}, {Colbert}  \&
  {Varoquaux}}{{van der Walt} et~al.}{2011}]{numpy}
{van der Walt} S.,  {Colbert} S.~C.,   {Varoquaux} G.,  2011, \mn@doi [Comput
  Sci Eng] {10.1109/MCSE.2011.37}, 13, 22

\makeatother
\end{thebibliography}

\appendix
\section{Without Lens Light Pipeline}
The pipelines that make up the uniform analysis for modelling a lensed image that does not contain the lens galaxy's light are presented in Table~\ref{no light pipeline}. This pipeline was used to analyse the mock data in this work. As well as this, a variation on this analysis, that also includes external shear in the mass model, was used to fit the four lenses that required lens subtracted data to arrive at successful model fits. The initial model fit priors, and those used when we choose not to inform priors with prior passing, are given in Table~\ref{table:Priors}.

{\renewcommand{\arraystretch}{2.0}
\begin{table*}
\centering
    \begin{tabularx}{\textwidth}{@{}*{6}{c}{X}}
          \hlineB{1}
          \thead{\textbf{Pipeline}} & \thead{\textbf{Phase}} & \thead{\textbf{Galaxy Component}} & \thead{\textbf{Model}} & \thead{\textbf{Varied}} & \thead{\textbf{Prior info}} & \textbf{Phase Description}\\
          \hlineB{1}
          \multirow{2}{*}{\shortstack[l]{Source \\ Parametric}} & \multirow{2}{*}{\textbf{SP}$^1$} & Lens mass & SIE & \checkmark & - &\multirow{2}{\hsize}{Fit the lens mass model and source light profile, comparing the lensed source model to mock image.}  \\
          & & Source light & S\'ersic & \checkmark & - \\
          \hlineB{1}
         \multirow{8}{*}{\shortstack[l]{Source \\ Inversion}}& \multirow{2}{*}{\textbf{SI$^1$}}& Lens mass & SIE & & \textbf{SP}$^3$ & \multirow{2}{\hsize}{Fix lens mass parameters to those from the source parametric pipeline and fit pixelization and regularisation parameters on magnification adaptive pixel-grid.}\\
         & & Source light & MPR & \checkmark & - \\
          \cline{2-7}
         & \multirow{2}{*}{\textbf{SI}$^2$} & Lens mass & SIE  & \checkmark & \textbf{SP}$^3$  & \multirow{2}{\hsize}{Refine the lens mass model parameters, keeping source-grid parameters fixed to those from previous phase.} \\
          & & Source light & MPR & & \textbf{SI}$^1$   \\
         \cline{2-7}
         & \multirow{2}{*}{\textbf{SI}$^3$} & Lens mass & SIE  & & \textbf{SP}$^3$ & \multirow{2}{\hsize}{Fit BPR pixelization and regularisation parameters, using the lensed source image from \textbf{SI}$^2$ to determine the source galaxy pixel centres. Lens mass parameters are fixed to those from previous phase.}\\
         & & Source light &BPR& \checkmark & -  \\
         \cline{2-7}
         & \multirow{2}{*}{\textbf{SI}$^4$} & Lens mass & SIE  & \checkmark & \textbf{SI}$^2$  & \multirow{2}{\hsize}{Refine lens mass model parameters on the BPR grid, keeping lens light and source-grid parameters fixed to those from previous phases.} \\
         & & Source light & BPR & & \textbf{SI}$^3$   \\
          \hlineB{1}
          \multirow{2}{*}{\shortstack[l]{Mass \\ Total}} & \multirow{2}{*}{\textbf{MT}$^1$} & Lens mass & PLEMD  & \checkmark & \textbf{SI}$^{4}$ & \multirow{2}{\hsize}{Fit the lens mass parameters, now with the slope of the density profile free to vary within the uniform prior [1.5-3.0], all other mass priors are informed from \textbf{SI}$^4$.}\\
          & & Source light & BPR &  & \textbf{SI}$^3$  \\
          \hlineB{1}
    \end{tabularx}
    \caption{Pipeline model components for the analysis which fits to a lensed image which does not contain emission from the lens galaxy.} 
    \label{no light pipeline}
\end{table*}}

\begin{table*}
\begin{tabular}{ l | l | l } 
\multicolumn{1}{p{1.5cm}|}{\centering \textbf{Model}} 
& \multicolumn{1}{p{1.5cm}|}{\centering \textbf{Parameter}} 
& \multicolumn{1}{p{1.1cm}|}{\centering \textbf{Prior}} 
\\ \hline
& & \\[-4pt]
 \textbf{Elliptical}     & $b\left(\arcsec\right)$          &  $\mathcal{U}(0,8)$  \\
 \textbf{Power-Law (PL)} & $\gamma$                         & $\mathcal{U}(1.5,3)$  \\     
                         & $\varepsilon_{1}$                & $\mathcal{N}(0,0.3)$ \\
                         & $\varepsilon_{2}$                & $\mathcal{N}(0,0.3)$ \\ 
                         & $x_{\rm c}\left(\arcsec\right)$  & $\mathcal{N}(0,0.05)$ \\
                         & $y_{\rm c}\left(\arcsec\right)$  & $\mathcal{N}(0,0.05)$ \\                        
& & \\[-4pt]
\hline
& & \\[-4pt]
\textbf{Sersic}          & $R_{\rm eff} (\arcsec)$           & $\mathcal{U}(0,30)$ \\
                         & $n$                               & $\mathcal{U}(0.5,5)$ \\
                         & $\log_{10} I_{0}$($e^{\rm -}s^{\rm -1}$)                          & $\mathcal{U}(-6,6)$ \\   
                         & $\varepsilon_{1}$                 & $\mathcal{N}(0,0.5)$ \\
                         & $\varepsilon_{2}$                 & $\mathcal{N}(0,0.5)$ \\
                         & $x_{\rm c}\left(\arcsec\right)$   & $\mathcal{N}(0,0.1)$ \\
                         & $y_{\rm c}\left(\arcsec\right)$   & $\mathcal{N}(0,0.1)$ \\
& & \\[-4pt]
\hline
& & \\[-4pt]                       
\textbf{Shear}          & $\gamma_{1_{\rm ext}}$ & $\mathcal{U}(-0.2,0.2)$ \\
                        & $\gamma_{2_{\rm ext}}$ & $\mathcal{U}(-0.2,0.2)$ \\                         
\end{tabular}
\caption{The initial priors on every parameter of every light and mass profile fitted in this work. Column 1 gives the model component name. Column 2 gives the parameter. Column 3 gives the prior, where $\mathcal{U}(a,b)$ is a uniform prior between $a$ and $b$, and $\mathcal{N}(\mu,\sigma)$ is a Gaussian prior with mean $\mu$ and variance $\sigma^2$. 
Note that due to prior passing (see Section \ref{Method}) the final priors used to fit a model, corresponding to the results given in this work, will be updated from the above values. The priors of every fit can be found at the following link \url{https://zenodo.org/record/6104823}.}
\label{table:Priors}
\end{table*}

\section{Inferred Model Parameters} \label{model parameters}
We present the best fit model parameters for all SLACS and GALLERY lenses. The PLEMD+ext mass model parameters are given in Tables~\ref{table: SLACS model fit params} (SLACS) and \ref{table: GALLERY model fit params} (GALLERY). The double S\'ersic light model parameters for the Gold sample are presented in Tables~\ref{table: SLACS light parameter} (SLACS), and \ref{table: Gallery light parameter} (GALLERY). We present the light parameters only for the ``Gold'' sample since the ``Silver'' and ``Bronze'' samples either do not fit the lens light or provide models we do not trust. All errors quoted are those derived from the 68\% credible region of the PDF output from \texttt{dynesty}. 

{\renewcommand{\arraystretch}{1.4}
\begin{table*}
\centering
 \resizebox{\textwidth}{!}{

\begin{tabular}{llllllllll}
\toprule
Class & Lens Name &              $b\left(\arcsec\right)$ &                              $\gamma$ &                          $\varepsilon_{1}$ &                          $\varepsilon_{2}$ &                          $\gamma_{1_{\rm ext}}$ &                          $\gamma_{2_{\rm ext}}$ &    $x_{\rm c}\left(\arcsec\right)$ &                                            $y_{\rm c} \left(\arcsec\right)$
  \\
\midrule
\multirow{10}{*}{Gold} & J0008-0004 &  $1.178^\textrm{+0.021}_\textrm{-0.015}$ &  $2.08^\textrm{+0.08}_\textrm{-0.07}$ &    $0.16^\textrm{+0.028}_\textrm{-0.025}$ &   $0.014^\textrm{+0.031}_\textrm{-0.033}$ &  $-0.006^\textrm{+0.011}_\textrm{-0.015}$ &  $-0.023^\textrm{+0.017}_\textrm{-0.017}$ &  $-0.015^\textrm{+0.012}_\textrm{-0.012}$ &   $0.034^\textrm{+0.012}_\textrm{-0.014}$ \\
     & J0029-0055 &  $0.971^\textrm{+0.026}_\textrm{-0.015}$ &  $2.32^\textrm{+0.13}_\textrm{-0.13}$ &    $0.089^\textrm{+0.05}_\textrm{-0.044}$ &    $0.089^\textrm{+0.062}_\textrm{-0.04}$ &   $0.005^\textrm{+0.021}_\textrm{-0.017}$ &   $0.012^\textrm{+0.019}_\textrm{-0.023}$ &   $-0.019^\textrm{+0.014}_\textrm{-0.02}$ &  $-0.017^\textrm{+0.014}_\textrm{-0.016}$ \\
     & J0157-0056 &  $0.999^\textrm{+0.033}_\textrm{-0.034}$ &  $2.23^\textrm{+0.08}_\textrm{-0.09}$ &  $-0.198^\textrm{+0.067}_\textrm{-0.049}$ &   $-0.199^\textrm{+0.054}_\textrm{-0.06}$ &   $-0.077^\textrm{+0.022}_\textrm{-0.02}$ &  $-0.165^\textrm{+0.023}_\textrm{-0.027}$ &   $-0.137^\textrm{+0.026}_\textrm{-0.02}$ &   $0.031^\textrm{+0.026}_\textrm{-0.026}$ \\
     & J0216-0813 &  $1.188^\textrm{+0.013}_\textrm{-0.013}$ &  $1.99^\textrm{+0.05}_\textrm{-0.06}$ &   $0.056^\textrm{+0.034}_\textrm{-0.035}$ &  $-0.097^\textrm{+0.024}_\textrm{-0.029}$ &    $0.001^\textrm{+0.02}_\textrm{-0.019}$ &   $0.009^\textrm{+0.016}_\textrm{-0.016}$ &   $0.009^\textrm{+0.008}_\textrm{-0.006}$ &   $0.011^\textrm{+0.008}_\textrm{-0.008}$ \\
     & J0252+0039 &  $1.021^\textrm{+0.005}_\textrm{-0.006}$ &  $1.92^\textrm{+0.08}_\textrm{-0.11}$ &    $-0.041^\textrm{+0.01}_\textrm{-0.01}$ &   $-0.045^\textrm{+0.01}_\textrm{-0.008}$ &   $-0.02^\textrm{+0.005}_\textrm{-0.005}$ &  $-0.013^\textrm{+0.005}_\textrm{-0.005}$ &     $0.0^\textrm{+0.006}_\textrm{-0.005}$ &  $-0.005^\textrm{+0.006}_\textrm{-0.006}$ \\
     & J0330-0020 &  $1.113^\textrm{+0.022}_\textrm{-0.022}$ &  $2.15^\textrm{+0.02}_\textrm{-0.02}$ &  $-0.017^\textrm{+0.052}_\textrm{-0.046}$ &  $-0.119^\textrm{+0.042}_\textrm{-0.043}$ &    $0.039^\textrm{+0.018}_\textrm{-0.02}$ &  $-0.013^\textrm{+0.019}_\textrm{-0.018}$ &  $-0.051^\textrm{+0.017}_\textrm{-0.026}$ &  $-0.021^\textrm{+0.017}_\textrm{-0.026}$ \\
     & J0728+3835 &  $1.274^\textrm{+0.029}_\textrm{-0.024}$ &   $1.99^\textrm{+0.12}_\textrm{-0.1}$ &    $0.145^\textrm{+0.03}_\textrm{-0.028}$ &  $-0.122^\textrm{+0.027}_\textrm{-0.027}$ &    $0.056^\textrm{+0.015}_\textrm{-0.02}$ &  $-0.037^\textrm{+0.012}_\textrm{-0.013}$ &  $-0.006^\textrm{+0.013}_\textrm{-0.012}$ &   $0.004^\textrm{+0.013}_\textrm{-0.012}$ \\
     & J0737+3216 &  $0.982^\textrm{+0.009}_\textrm{-0.008}$ &  $2.28^\textrm{+0.07}_\textrm{-0.07}$ &  $-0.017^\textrm{+0.017}_\textrm{-0.014}$ &   $-0.072^\textrm{+0.021}_\textrm{-0.02}$ &   $0.038^\textrm{+0.007}_\textrm{-0.007}$ &    $0.103^\textrm{+0.009}_\textrm{-0.01}$ &  $-0.008^\textrm{+0.004}_\textrm{-0.004}$ &  $-0.006^\textrm{+0.004}_\textrm{-0.004}$ \\
     & J0822+2652 &  $1.235^\textrm{+0.034}_\textrm{-0.035}$ &   $2.1^\textrm{+0.08}_\textrm{-0.07}$ &   $0.147^\textrm{+0.041}_\textrm{-0.043}$ &  $-0.264^\textrm{+0.039}_\textrm{-0.038}$ &   $0.057^\textrm{+0.025}_\textrm{-0.025}$ &  $-0.082^\textrm{+0.017}_\textrm{-0.024}$ &  $-0.014^\textrm{+0.022}_\textrm{-0.016}$ &  $-0.103^\textrm{+0.022}_\textrm{-0.028}$ \\
     & J0841+3824 &  $1.005^\textrm{+0.158}_\textrm{-0.139}$ &   $2.27^\textrm{+0.2}_\textrm{-0.16}$ &  $-0.149^\textrm{+0.089}_\textrm{-0.096}$ &   $-0.104^\textrm{+0.175}_\textrm{-0.12}$ &   $-0.118^\textrm{+0.043}_\textrm{-0.04}$ &  $-0.083^\textrm{+0.049}_\textrm{-0.052}$ &   $-0.25^\textrm{+0.046}_\textrm{-0.043}$ &  $-0.204^\textrm{+0.046}_\textrm{-0.034}$ \\
\bottomrule
\end{tabular}}

    \caption{Mass distribution model fit parameters for the first ten SLACS lenses. The full table is available online.} 
    \label{table: SLACS model fit params}
\end{table*}}

{\renewcommand{\arraystretch}{1.4}
\begin{table*}
\centering

 \resizebox{\textwidth}{!}{

\begin{tabular}{llllllllll}
\toprule
 Class & Lens Name &   b $\left(\arcsec\right)$ &                $\gamma$ &            $\varepsilon_{1}$ &            $\varepsilon_{2}$ &          $\gamma_{1_{\rm ext}}$ &                          $\gamma_{2_{\rm ext}}$   &                                         $x_{\rm c}\left(\arcsec\right)$ &                                            $y_{\rm c} \left(\arcsec\right)$ \\
\midrule
\multirow{16}{*}{Gold} & J0029+2544 &  $1.395^\textrm{+0.074}_\textrm{-0.039}$ &  $2.05^\textrm{+0.12}_\textrm{-0.15}$ &  $-0.203^\textrm{+0.055}_\textrm{-0.078}$ &   $0.014^\textrm{+0.045}_\textrm{-0.055}$ &   $-0.025^\textrm{+0.03}_\textrm{-0.031}$ &  $-0.049^\textrm{+0.025}_\textrm{-0.027}$ &    $0.09^\textrm{+0.023}_\textrm{-0.026}$ &   $0.036^\textrm{+0.023}_\textrm{-0.026}$ \\
     & J0113+0250 &  $1.293^\textrm{+0.037}_\textrm{-0.028}$ &  $1.77^\textrm{+0.15}_\textrm{-0.11}$ &  $-0.006^\textrm{+0.017}_\textrm{-0.019}$ &   $0.068^\textrm{+0.013}_\textrm{-0.017}$ &   $0.041^\textrm{+0.013}_\textrm{-0.013}$ &    $0.14^\textrm{+0.012}_\textrm{-0.012}$ &   $0.031^\textrm{+0.009}_\textrm{-0.011}$ &   $0.008^\textrm{+0.009}_\textrm{-0.008}$ \\
     & J0201+3228 &  $1.727^\textrm{+0.022}_\textrm{-0.018}$ &   $2.09^\textrm{+0.09}_\textrm{-0.1}$ &   $-0.114^\textrm{+0.026}_\textrm{-0.02}$ &   $-0.02^\textrm{+0.019}_\textrm{-0.012}$ &    $0.06^\textrm{+0.015}_\textrm{-0.012}$ &  $-0.039^\textrm{+0.016}_\textrm{-0.008}$ &   $0.002^\textrm{+0.016}_\textrm{-0.014}$ &   $0.026^\textrm{+0.016}_\textrm{-0.013}$ \\
     & J0237-0641 &  $0.615^\textrm{+0.117}_\textrm{-0.078}$ &   $1.91^\textrm{+0.18}_\textrm{-0.1}$ &  $-0.117^\textrm{+0.033}_\textrm{-0.018}$ &   $0.026^\textrm{+0.075}_\textrm{-0.101}$ &    $0.006^\textrm{+0.051}_\textrm{-0.06}$ &   $-0.015^\textrm{+0.068}_\textrm{-0.04}$ &   $0.146^\textrm{+0.033}_\textrm{-0.033}$ &   $0.021^\textrm{+0.033}_\textrm{-0.027}$ \\
     & J0742+3341 &  $1.684^\textrm{+0.134}_\textrm{-0.097}$ &  $2.21^\textrm{+0.06}_\textrm{-0.08}$ &   $0.506^\textrm{+0.053}_\textrm{-0.049}$ &   $0.001^\textrm{+0.061}_\textrm{-0.061}$ &   $0.107^\textrm{+0.018}_\textrm{-0.024}$ &  $-0.217^\textrm{+0.026}_\textrm{-0.025}$ &   $0.136^\textrm{+0.018}_\textrm{-0.048}$ &  $-0.062^\textrm{+0.018}_\textrm{-0.015}$ \\
     & J0755+3445 &    $2.0^\textrm{+0.071}_\textrm{-0.055}$ &  $1.77^\textrm{+0.08}_\textrm{-0.05}$ &   $0.156^\textrm{+0.016}_\textrm{-0.014}$ &   $0.131^\textrm{+0.016}_\textrm{-0.012}$ &   $0.201^\textrm{+0.008}_\textrm{-0.009}$ &   $0.268^\textrm{+0.007}_\textrm{-0.011}$ &   $0.069^\textrm{+0.004}_\textrm{-0.011}$ &  $-0.159^\textrm{+0.004}_\textrm{-0.005}$ \\
     & J0856+2010 &  $1.157^\textrm{+0.071}_\textrm{-0.087}$ &  $2.23^\textrm{+0.08}_\textrm{-0.09}$ &   $0.474^\textrm{+0.079}_\textrm{-0.079}$ &  $-0.151^\textrm{+0.069}_\textrm{-0.098}$ &  $-0.021^\textrm{+0.032}_\textrm{-0.032}$ &  $-0.014^\textrm{+0.027}_\textrm{-0.023}$ &   $0.171^\textrm{+0.021}_\textrm{-0.039}$ &  $-0.063^\textrm{+0.021}_\textrm{-0.018}$ \\
     & J0918+5105 &  $1.642^\textrm{+0.035}_\textrm{-0.037}$ &  $2.38^\textrm{+0.16}_\textrm{-0.18}$ &  $-0.024^\textrm{+0.057}_\textrm{-0.064}$ &  $-0.081^\textrm{+0.023}_\textrm{-0.041}$ &  $-0.246^\textrm{+0.023}_\textrm{-0.033}$ &  $-0.122^\textrm{+0.007}_\textrm{-0.007}$ &  $-0.019^\textrm{+0.027}_\textrm{-0.014}$ &   $0.004^\textrm{+0.027}_\textrm{-0.016}$ \\
     & J1110+2808 &  $0.902^\textrm{+0.029}_\textrm{-0.026}$ &  $2.03^\textrm{+0.09}_\textrm{-0.07}$ &   $0.041^\textrm{+0.053}_\textrm{-0.081}$ &  $-0.045^\textrm{+0.056}_\textrm{-0.058}$ &   $0.114^\textrm{+0.042}_\textrm{-0.037}$ &   $-0.09^\textrm{+0.025}_\textrm{-0.029}$ &   $-0.106^\textrm{+0.03}_\textrm{-0.023}$ &   $-0.141^\textrm{+0.03}_\textrm{-0.035}$ \\
     & J1110+3649 &  $1.188^\textrm{+0.011}_\textrm{-0.012}$ &  $2.23^\textrm{+0.07}_\textrm{-0.08}$ &  $-0.024^\textrm{+0.007}_\textrm{-0.008}$ &  $-0.016^\textrm{+0.013}_\textrm{-0.013}$ &   $0.019^\textrm{+0.003}_\textrm{-0.003}$ &   $0.129^\textrm{+0.007}_\textrm{-0.006}$ &    $-0.0^\textrm{+0.003}_\textrm{-0.003}$ &  $-0.009^\textrm{+0.003}_\textrm{-0.002}$ \\
     & J1116+0915 &  $1.247^\textrm{+0.188}_\textrm{-0.156}$ &  $2.22^\textrm{+0.16}_\textrm{-0.17}$ &   $0.071^\textrm{+0.097}_\textrm{-0.101}$ &  $-0.393^\textrm{+0.069}_\textrm{-0.072}$ &   $0.016^\textrm{+0.034}_\textrm{-0.045}$ &  $-0.653^\textrm{+0.041}_\textrm{-0.053}$ &  $-0.034^\textrm{+0.047}_\textrm{-0.032}$ &   $0.086^\textrm{+0.047}_\textrm{-0.044}$ \\
     & J1141+2216 &  $1.381^\textrm{+0.067}_\textrm{-0.071}$ &  $2.13^\textrm{+0.09}_\textrm{-0.11}$ &   $0.243^\textrm{+0.058}_\textrm{-0.069}$ &   $0.009^\textrm{+0.072}_\textrm{-0.085}$ &   $0.042^\textrm{+0.032}_\textrm{-0.033}$ &  $-0.117^\textrm{+0.032}_\textrm{-0.031}$ &    $0.088^\textrm{+0.04}_\textrm{-0.033}$ &   $-0.035^\textrm{+0.04}_\textrm{-0.028}$ \\
     & J1201+4743 &  $1.221^\textrm{+0.023}_\textrm{-0.018}$ &  $2.74^\textrm{+0.05}_\textrm{-0.21}$ &  $-0.095^\textrm{+0.019}_\textrm{-0.032}$ &   $0.007^\textrm{+0.035}_\textrm{-0.043}$ &     $0.069^\textrm{+0.0}_\textrm{-0.017}$ &  $-0.016^\textrm{+0.007}_\textrm{-0.009}$ &  $-0.046^\textrm{+0.001}_\textrm{-0.004}$ &   $0.025^\textrm{+0.001}_\textrm{-0.011}$ \\
     & J1226+5457 &  $1.385^\textrm{+0.008}_\textrm{-0.009}$ &   $2.24^\textrm{+0.07}_\textrm{-0.1}$ &  $-0.074^\textrm{+0.011}_\textrm{-0.015}$ &   $0.127^\textrm{+0.014}_\textrm{-0.016}$ &   $-0.139^\textrm{+0.01}_\textrm{-0.007}$ &   $-0.011^\textrm{+0.009}_\textrm{-0.01}$ &   $0.023^\textrm{+0.004}_\textrm{-0.004}$ &   $0.006^\textrm{+0.004}_\textrm{-0.003}$ \\
     & J2228+1205 &  $1.338^\textrm{+0.072}_\textrm{-0.063}$ &    $2.2^\textrm{+0.14}_\textrm{-0.1}$ &   $-0.262^\textrm{+0.049}_\textrm{-0.08}$ &   $0.048^\textrm{+0.079}_\textrm{-0.057}$ &   $-0.196^\textrm{+0.03}_\textrm{-0.035}$ &  $-0.198^\textrm{+0.033}_\textrm{-0.022}$ &  $-0.057^\textrm{+0.026}_\textrm{-0.026}$ &  $-0.005^\textrm{+0.026}_\textrm{-0.031}$ \\
     & J2342-0120 &  $1.313^\textrm{+0.048}_\textrm{-0.044}$ &  $2.34^\textrm{+0.07}_\textrm{-0.09}$ &   $-0.298^\textrm{+0.06}_\textrm{-0.034}$ &  $-0.129^\textrm{+0.036}_\textrm{-0.025}$ &  $-0.019^\textrm{+0.014}_\textrm{-0.021}$ &  $-0.254^\textrm{+0.015}_\textrm{-0.008}$ &    $0.051^\textrm{+0.013}_\textrm{-0.01}$ &    $0.02^\textrm{+0.013}_\textrm{-0.014}$ \\
\bottomrule
\end{tabular}}
    \caption{GALLERY mass distribution model fit parameters.} 
    \label{table: GALLERY model fit params}
\end{table*}}

{\renewcommand{\arraystretch}{1.3}
\begin{table*}
\centering
\small
 \resizebox{\textwidth}{!}{
\begin{tabular}{lllllllllllll}

\toprule
lens & noise & S\'ersic &                                           $R_{\rm eff} (\arcsec)$ &                                  $n$ &                               $I_{0}\left(\times10^{-3}\right)$ &                           $\phi$ &                                   $q$ &                     $\varepsilon_{1}$ &                     $\varepsilon_{2}$ &                                $x_{\rm c} \left(\times10^{-3}\arcsec\right)$ &                                $y_{\rm c}\left(\times10^{-3}\arcsec\right)$ \\
\midrule
\multirow{2}{*}{J0029+2544} & \multirow{2}{*}{$0.11^\textrm{+0.11}_\textrm{-0.045}$} & I &   $16.84^\textrm{+3.89}_\textrm{-6.68}$ &   $3.9^\textrm{+0.58}_\textrm{-0.5}$ &   $0.04^\textrm{+0.01}_\textrm{-0.02}$ &     $-2^\textrm{+4}_\textrm{-4}$ &  $0.33^\textrm{+0.09}_\textrm{-0.14}$ &  $-0.04^\textrm{+0.07}_\textrm{-0.07}$ &    $0.5^\textrm{+0.16}_\textrm{-0.18}$ &    $0.49^\textrm{+0.63}_\textrm{-0.8}$ &  $-0.71^\textrm{+0.51}_\textrm{-1.02}$ \\
           &                                  & II &    $0.59^\textrm{+0.04}_\textrm{-0.02}$ &  $3.5^\textrm{+0.04}_\textrm{-0.04}$ &  $17.93^\textrm{+0.61}_\textrm{-1.19}$ &    $-43^\textrm{+3}_\textrm{-3}$ &   $0.82^\textrm{+0.01}_\textrm{-0.0}$ &     $-0.1^\textrm{+0.0}_\textrm{-0.0}$ &   $0.01^\textrm{+0.01}_\textrm{-0.01}$ &                                        &                                        \\
\cline{1-12}
\cline{2-12}
\multirow{2}{*}{J0113+0250} & \multirow{2}{*}{$0.00032^\textrm{+0.0012}_\textrm{-0.00018}$} & I &     $2.43^\textrm{+0.82}_\textrm{-0.6}$ &  $1.0^\textrm{+0.28}_\textrm{-0.23}$ &    $1.5^\textrm{+0.36}_\textrm{-0.35}$ &    $-71^\textrm{+3}_\textrm{-3}$ &  $0.35^\textrm{+0.07}_\textrm{-0.08}$ &  $-0.29^\textrm{+0.06}_\textrm{-0.05}$ &  $-0.38^\textrm{+0.09}_\textrm{-0.09}$ &     $5.0^\textrm{+1.41}_\textrm{-1.5}$ &  $-2.03^\textrm{+1.76}_\textrm{-1.55}$ \\
           &                                  & II &     $1.72^\textrm{+0.24}_\textrm{-0.2}$ &  $3.9^\textrm{+0.21}_\textrm{-0.24}$ &    $2.15^\textrm{+0.5}_\textrm{-0.38}$ &     $-3^\textrm{+0}_\textrm{-0}$ &  $0.54^\textrm{+0.01}_\textrm{-0.02}$ &  $-0.04^\textrm{+0.01}_\textrm{-0.01}$ &   $0.29^\textrm{+0.01}_\textrm{-0.01}$ &                                        &                                        \\
\cline{1-12}
\cline{2-12}
\multirow{2}{*}{J0201+3228} & \multirow{2}{*}{$160^\textrm{+370}_\textrm{-130}$} & I &    $2.12^\textrm{+0.59}_\textrm{-0.35}$ &  $1.4^\textrm{+0.19}_\textrm{-0.18}$ &   $7.65^\textrm{+1.47}_\textrm{-1.69}$ &    $-47^\textrm{+2}_\textrm{-3}$ &  $0.79^\textrm{+0.02}_\textrm{-0.02}$ &  $-0.12^\textrm{+0.01}_\textrm{-0.01}$ &  $-0.01^\textrm{+0.01}_\textrm{-0.01}$ &   $0.68^\textrm{+0.63}_\textrm{-0.56}$ &    $3.2^\textrm{+0.63}_\textrm{-0.61}$ \\
           &                                  & II &    $1.09^\textrm{+0.08}_\textrm{-0.08}$ &  $4.9^\textrm{+0.06}_\textrm{-0.08}$ &   $9.03^\textrm{+0.75}_\textrm{-0.69}$ &    $-85^\textrm{+4}_\textrm{-4}$ &  $0.91^\textrm{+0.01}_\textrm{-0.01}$ &  $-0.01^\textrm{+0.01}_\textrm{-0.01}$ &  $-0.04^\textrm{+0.01}_\textrm{-0.01}$ &                                        &                                        \\
\cline{1-12}
\cline{2-12}
\multirow{2}{*}{J0237-0641} & \multirow{2}{*}{$1.4^\textrm{+15}_\textrm{-1.4}$} & I &  $10.62^\textrm{+12.78}_\textrm{-7.09}$ &    $3.5^\textrm{+1.3}_\textrm{-0.8}$ &   $0.15^\textrm{+0.45}_\textrm{-0.12}$ &    $3^\textrm{+26}_\textrm{-17}$ &  $0.64^\textrm{+0.31}_\textrm{-0.19}$ &    $0.03^\textrm{+0.2}_\textrm{-0.07}$ &   $0.22^\textrm{+0.12}_\textrm{-0.21}$ &    $0.48^\textrm{+0.7}_\textrm{-0.83}$ &  $-2.23^\textrm{+0.71}_\textrm{-0.73}$ \\
           &                                  & II &    $0.91^\textrm{+0.11}_\textrm{-0.14}$ &  $4.8^\textrm{+0.11}_\textrm{-0.47}$ &   $6.92^\textrm{+1.71}_\textrm{-0.82}$ &  $80^\textrm{+20}_\textrm{-316}$ &  $0.98^\textrm{+0.02}_\textrm{-0.05}$ &    $0.0^\textrm{+0.01}_\textrm{-0.01}$ &  $-0.01^\textrm{+0.01}_\textrm{-0.01}$ &                                        &                                        \\
\cline{1-12}
\cline{2-12}
\multirow{2}{*}{J0742+3341} & \multirow{2}{*}{$1.1^\textrm{+28}_\textrm{-1.1}$} & I &   $10.31^\textrm{+8.57}_\textrm{-8.83}$ &  $3.1^\textrm{+1.28}_\textrm{-0.95}$ &   $0.24^\textrm{+2.42}_\textrm{-0.14}$ &     $27^\textrm{+7}_\textrm{-7}$ &  $0.53^\textrm{+0.14}_\textrm{-0.11}$ &    $0.25^\textrm{+0.09}_\textrm{-0.1}$ &   $0.18^\textrm{+0.09}_\textrm{-0.13}$ &  $-0.19^\textrm{+0.58}_\textrm{-0.63}$ &   $1.35^\textrm{+0.59}_\textrm{-0.68}$ \\
           &                                  & II &    $1.04^\textrm{+0.25}_\textrm{-0.11}$ &  $4.6^\textrm{+0.16}_\textrm{-0.23}$ &   $9.12^\textrm{+1.38}_\textrm{-2.27}$ &     $62^\textrm{+1}_\textrm{-2}$ &  $0.71^\textrm{+0.01}_\textrm{-0.18}$ &   $0.14^\textrm{+0.01}_\textrm{-0.01}$ &   $-0.1^\textrm{+0.02}_\textrm{-0.01}$ &                                        &                                        \\
\bottomrule
\end{tabular}}
    \caption{Light model parameters for the first five GALLERY lenses in order of Right Ascension. The full table is available online.} 
    \label{table: Gallery light parameter}
\end{table*}}

{\renewcommand{\arraystretch}{1.3}
\begin{table*}
\centering
\small
 \resizebox{\textwidth}{!}{
\begin{tabular}{lllllllllllll}
\toprule
lens & noise & S\'ersic &                                           $R_{\rm eff} (\arcsec)$ &                                  $n$ &                               $I_{0}\left(\times10^{-3}\right)$ &                           $\phi$ &                                   $q$ &                     $\varepsilon_{1}$ &                     $\varepsilon_{2}$ &                                $x_{\rm c} \left(\times10^{-3}\arcsec\right)$ &                                $y_{\rm c}\left(\times10^{-3}\arcsec\right)$ \\
\midrule
\multirow{2}{*}{J0008-0004} & \multirow{2}{*}{$1500^\textrm{+210}_\textrm{-460}$} & I &   $27.25^\textrm{+1.2}_\textrm{-22.87}$ &  $2.2^\textrm{+2.65}_\textrm{-0.94}$ &        $0.01^\textrm{+0.0}_\textrm{-0.0}$ &   $46^\textrm{+3}_\textrm{-15}$ &  $0.63^\textrm{+0.11}_\textrm{-0.23}$ &    $0.23^\textrm{+0.2}_\textrm{-0.19}$ &  $-0.02^\textrm{+0.04}_\textrm{-0.03}$ &   $-3.25^\textrm{+0.0}_\textrm{-2.21}$ &   $4.05^\textrm{+0.0}_\textrm{-1.36}$ \\
           &                                   & II &    $1.69^\textrm{+0.69}_\textrm{-0.09}$ &   $4.3^\textrm{+0.28}_\textrm{-0.1}$ &     $27.93^\textrm{+0.0}_\textrm{-11.49}$ &    $26^\textrm{+3}_\textrm{-1}$ &    $0.9^\textrm{+0.0}_\textrm{-0.01}$ &    $0.04^\textrm{+0.01}_\textrm{-0.0}$ &     $0.03^\textrm{+0.0}_\textrm{-0.0}$ &                                        &                                       \\
\cline{1-12}
\cline{2-12}
\multirow{2}{*}{J0029-0055} & \multirow{2}{*}{$470^\textrm{+120}_\textrm{-100}$} & I &    $0.33^\textrm{+0.02}_\textrm{-0.02}$ &   $2.8^\textrm{+0.1}_\textrm{-0.09}$ &   $905.32^\textrm{+0.08}_\textrm{-80.08}$ &    $22^\textrm{+0}_\textrm{-0}$ &     $0.9^\textrm{+0.0}_\textrm{-0.0}$ &     $0.04^\textrm{+0.0}_\textrm{-0.0}$ &     $0.04^\textrm{+0.0}_\textrm{-0.0}$ &   $-5.57^\textrm{+0.0}_\textrm{-0.26}$ &   $1.92^\textrm{+0.0}_\textrm{-0.25}$ \\
           &                                   & II &     $3.0^\textrm{+0.22}_\textrm{-0.17}$ &  $1.6^\textrm{+0.11}_\textrm{-0.11}$ &       $49.4^\textrm{+0.0}_\textrm{-5.06}$ &    $27^\textrm{+1}_\textrm{-1}$ &  $0.79^\textrm{+0.01}_\textrm{-0.01}$ &   $0.09^\textrm{+0.01}_\textrm{-0.01}$ &   $0.07^\textrm{+0.01}_\textrm{-0.01}$ &                                        &                                       \\
\cline{1-12}
\cline{2-12}
\multirow{2}{*}{J0157-0056} & \multirow{2}{*}{$120^\textrm{+30.}_\textrm{-26}$} & I &    $1.86^\textrm{+0.56}_\textrm{-0.34}$ &  $0.8^\textrm{+0.28}_\textrm{-0.13}$ &        $7.3^\textrm{+0.0}_\textrm{-1.09}$ &   $-58^\textrm{+5}_\textrm{-6}$ &  $0.72^\textrm{+0.05}_\textrm{-0.07}$ &  $-0.15^\textrm{+0.04}_\textrm{-0.04}$ &  $-0.07^\textrm{+0.04}_\textrm{-0.04}$ &   $-5.09^\textrm{+0.0}_\textrm{-0.14}$ &   $1.65^\textrm{+0.0}_\textrm{-0.31}$ \\
           &                                   & II &    $1.04^\textrm{+0.02}_\textrm{-0.06}$ &  $4.9^\textrm{+0.04}_\textrm{-0.07}$ &     $66.77^\textrm{+0.01}_\textrm{-1.89}$ &    $68^\textrm{+0}_\textrm{-0}$ &    $0.67^\textrm{+0.0}_\textrm{-0.0}$ &     $0.13^\textrm{+0.0}_\textrm{-0.0}$ &    $-0.14^\textrm{+0.0}_\textrm{-0.0}$ &                                        &                                       \\
\cline{1-12}
\cline{2-12}
\multirow{2}{*}{J0216-0813} & \multirow{2}{*}{$850^\textrm{+52}_\textrm{-70.}$} & I &     $1.54^\textrm{+0.33}_\textrm{-0.2}$ &  $3.9^\textrm{+0.32}_\textrm{-0.24}$ &   $115.98^\textrm{+0.03}_\textrm{-28.97}$ &    $85^\textrm{+0}_\textrm{-0}$ &    $0.81^\textrm{+0.0}_\textrm{-0.0}$ &     $0.02^\textrm{+0.0}_\textrm{-0.0}$ &    $-0.11^\textrm{+0.0}_\textrm{-0.0}$ &   $-7.13^\textrm{+0.0}_\textrm{-0.39}$ &   $3.76^\textrm{+0.0}_\textrm{-0.44}$ \\
           &                                   & II &    $3.75^\textrm{+2.41}_\textrm{-0.49}$ &  $0.8^\textrm{+0.27}_\textrm{-0.13}$ &    $16.96^\textrm{+0.01}_\textrm{-10.44}$ &    $50^\textrm{+3}_\textrm{-5}$ &  $0.66^\textrm{+0.07}_\textrm{-0.13}$ &    $0.2^\textrm{+0.14}_\textrm{-0.05}$ &  $-0.04^\textrm{+0.04}_\textrm{-0.04}$ &                                        &                                       \\
\cline{1-12}
\cline{2-12}
\multirow{2}{*}{J0252+0039} & \multirow{2}{*}{$210^\textrm{+52}_\textrm{-43}$} & I &    $0.94^\textrm{+0.03}_\textrm{-0.03}$ &  $0.9^\textrm{+0.05}_\textrm{-0.05}$ &     $125.89^\textrm{+0.0}_\textrm{-3.86}$ &   $-65^\textrm{+1}_\textrm{-1}$ &  $0.77^\textrm{+0.01}_\textrm{-0.01}$ &   $-0.1^\textrm{+0.01}_\textrm{-0.01}$ &  $-0.09^\textrm{+0.01}_\textrm{-0.01}$ &   $-6.69^\textrm{+0.0}_\textrm{-0.38}$ &  $-1.61^\textrm{+0.0}_\textrm{-0.42}$ \\
           &                                   & II &    $0.62^\textrm{+0.05}_\textrm{-0.04}$ &   $4.9^\textrm{+0.06}_\textrm{-0.1}$ &   $132.34^\textrm{+0.01}_\textrm{-11.09}$ &  $54^\textrm{+20}_\textrm{-22}$ &  $0.99^\textrm{+0.01}_\textrm{-0.01}$ &      $0.0^\textrm{+0.0}_\textrm{-0.0}$ &     $-0.0^\textrm{+0.0}_\textrm{-0.0}$ &                                        &                                       \\
\bottomrule
\end{tabular}}
    \caption{Light model parameters for the first five SLACS lenses in order of Right Ascension. The full table is available online.} 
    \label{table: SLACS light parameter}
\end{table*}}

\section{Model fits}\label{model fits}
\begin{figure*}
    \centering
    \includegraphics[width=0.92\linewidth]{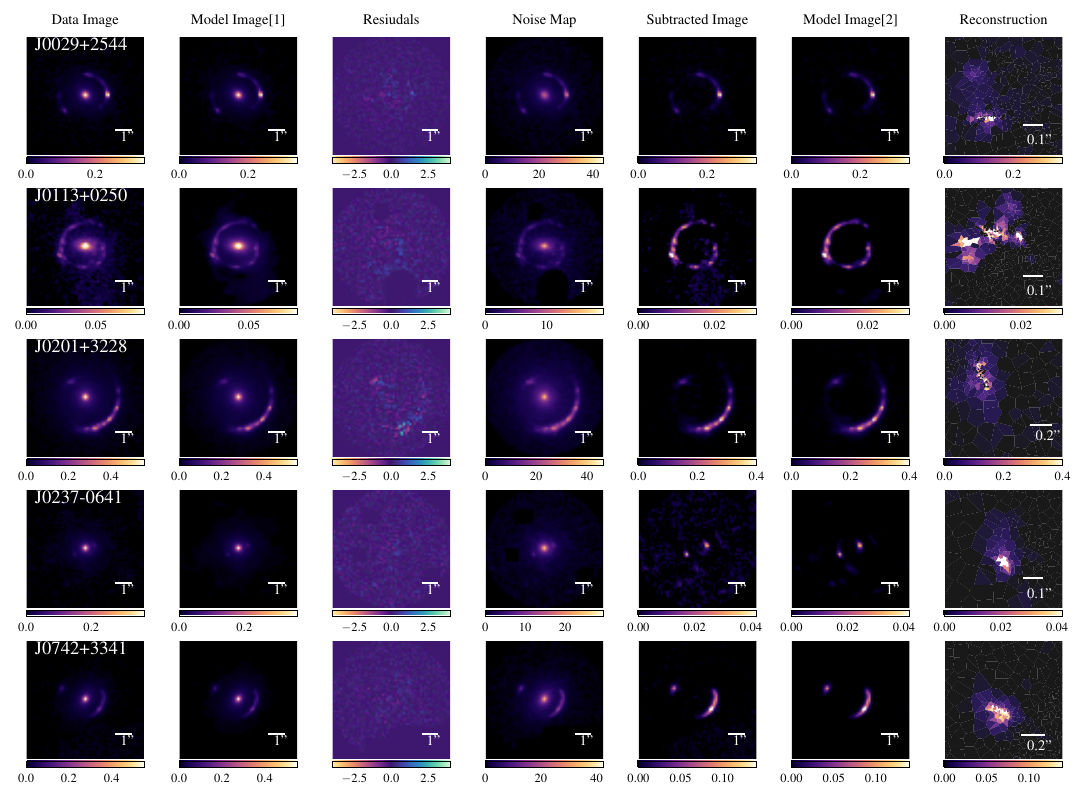}
    \caption{Model fits for the first five GALLERY lenses in order of Right Ascension. Model fits for the full sample of lenses are available online. Residuals are the normalised residuals.}
    \label{Figure: GALLERY modelfits}
\end{figure*}

\begin{figure*}
    \centering
    \includegraphics[width=0.92\linewidth]{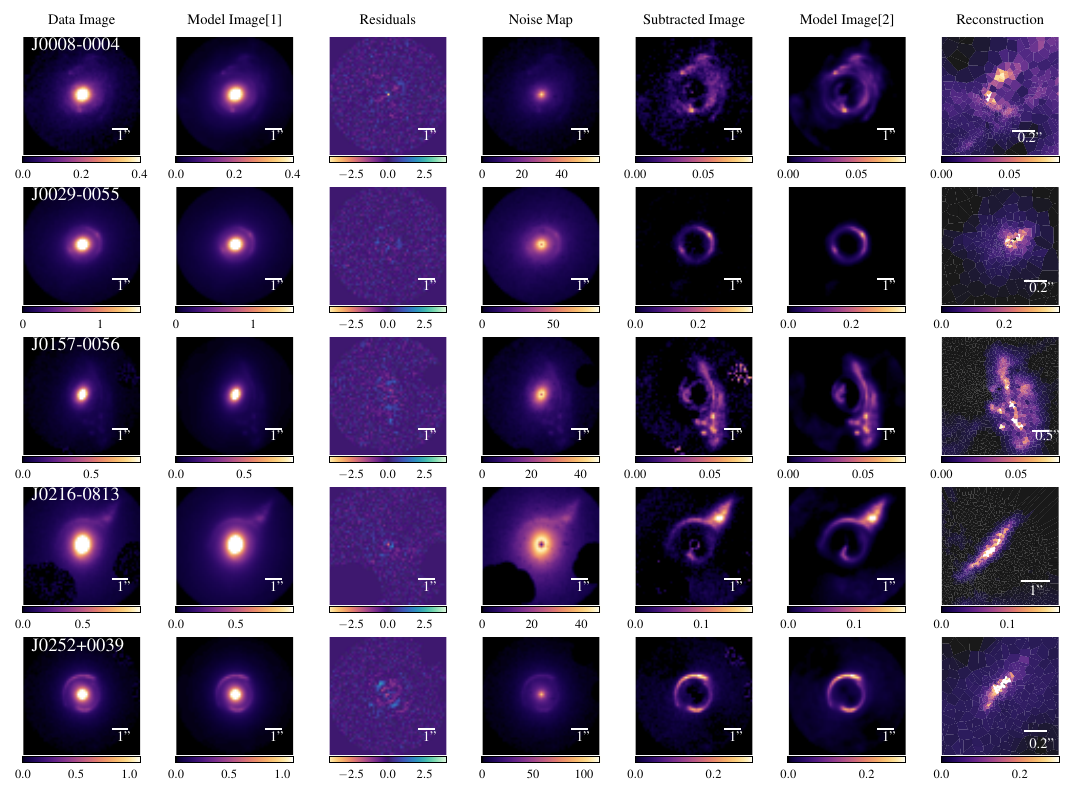}
    \caption{Model fits for the first five SLACS lenses in the ``Gold'' sample in order of Right Ascension. Model fits for the full sample of lenses are available online. Residuals are the normalised residuals.}
    \label{Figure: SLACS modelfits}
\end{figure*}

\begin{figure*}
    \centering
    \includegraphics[width=0.67\linewidth]{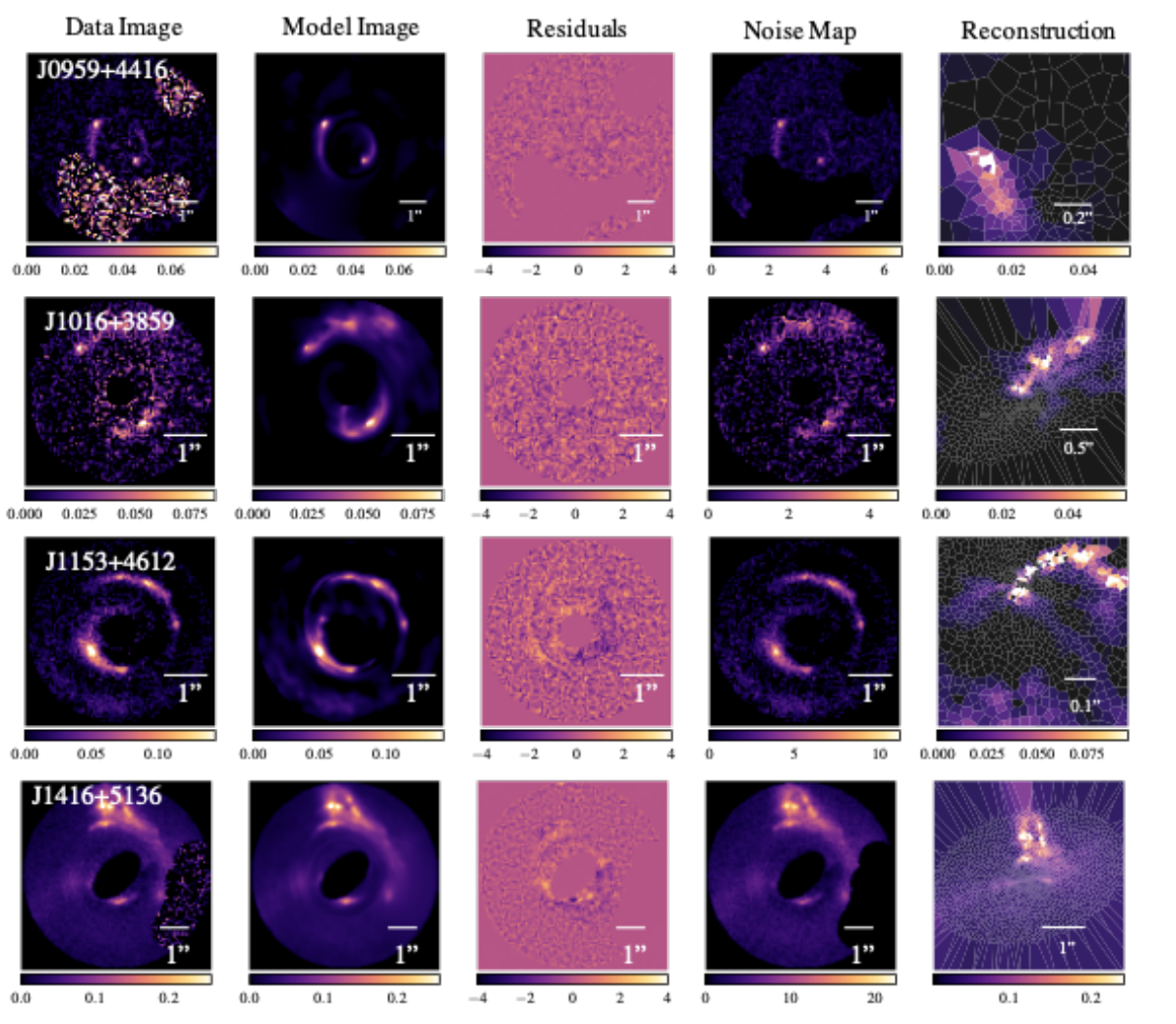}
    \caption{SLACS ``Silver'' model fits. Residuals are the normalised residuals.}
    \label{Figure: SLACS silver modelfits}
\end{figure*}

\begin{figure*}
    \centering
    \includegraphics[width=0.92\linewidth]{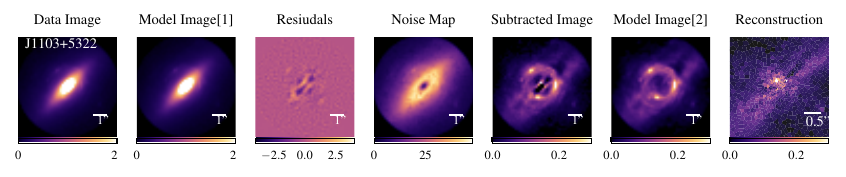}
    \caption{SLACS ``Bronze'' model fit. Residuals are the normalised residuals.}
    \label{Figure: SLACS bronze modelfits}
\end{figure*}

In this study we categorised the model fits into ``Gold'', ``Silver'', and ``Bronze'' depending on the quality of the model fit. The ``Gold'' fits are presented in Figure~\ref{Figure: SLACS modelfits} for SLACS lenses and Figure~\ref{Figure: GALLERY modelfits} for GALLERY lenses. The ``Silver'' lenses are then presented in Figure~\ref{Figure: SLACS silver modelfits} and the ``Bronze'' lens in Figure~\ref{Figure: SLACS bronze modelfits}.

\label{lastpage}

\end{document}